\documentclass[12pt]{article}
\pdfoutput=1
\usepackage{graphicx,epstopdf,amssymb,amsfonts,amsmath,amsthm,array,
mathrsfs,amscd}
\usepackage{todonotes} 
\newcommand{\pbs}[1]{\let\temp=\\#1\let\\=\temp}
\numberwithin{equation}{section}
\def\be{\begin{equation}}\def\ee{\end{equation}}
\def\cvp{\raise 2pt\hbox{,}}

\def\re{\mathop{\rm Re}\nolimits}  \def\res{\mathop{\rm res}\nolimits}

 \def\d{{\rm d}}

\def\la{\lambda}\def\La{\Lambda}

\def\Seff{S_{\text{eff}}}\def\A{\mathscr A}
\def\dia{\mathscr D}
\def\diagrama{\begin{picture}(6,6)(0,0)
\put(3,2){\circle{6}}
\end{picture}}
\def\diagramabig{\begin{picture}(10,10)(0,0)
\put(5,3.5){\circle{10}}
\end{picture}}
\def\diagramb{\begin{picture}(6,6)(0,0)
\put(3,2){\circle{6}}
\put(3,-1){\line(0,1){6}}
\end{picture}}
\def\diagrambbig{\begin{picture}(10,10)(0,0)
\put(5,3.5){\circle{10}}
\put(5,-1.5){\line(0,1){10}}
\end{picture}}
\def\diagramc{\begin{picture}(12,6)(0,0)
\put(3,2){\circle{6}}\put(9,2){\circle{6}}
\end{picture}}
\def\diagramcbig{\begin{picture}(20,10)(0,0)
\put(5,3.5){\circle{10}}\put(15,3.5){\circle{10}}
\end{picture}}
\def\diagramd{\begin{picture}(15,6)(0,0)
\put(3,2){\circle{6}}\put(6,2){\line(1,0){3}}\put(12,2){\circle{6}}
\end{picture}}
\def\diagramdbig{\begin{picture}(26,10)(0,0)
\put(5,3.5){\circle{10}}\put(10,3.5){\line(1,0){5}}\put(20,3.5){\circle{10}}
\end{picture}}

\def\Re{\mathop{\rm Re}}
\def\d{\partial}

\def\d{{\rm d}}

\def\a{\alpha}
\def\b{\beta}
\def\g{\gamma}
\def\G{\Gamma}

\def\m{\mu}

\def\k{\kappa}
\def\l{\lambda}

\def\p{\psi}

\def\vf{\varphi}
\def\L{\Lambda}

\def\wh{\widehat}
\def\wt{\widetilde}

\def\l{\lambda}

\def\del{\partial}

\def\ba{\begin{eqnarray}}
\def\ea{\end{eqnarray}}

\theoremstyle{plain}

\theoremstyle{definition}
\theoremstyle{remark}

\def\plb#1#2#3{{\it Phys.\ Lett.\ }{\bf B #1} (#2) #3}
\def\npb#1#2#3{{\it Nucl.\ Phys.\ }{\bf B #1} (#2) #3}

\def\prl#1#2#3{{\it Phys.\ Rev.\ Lett.\ }{\bf #1} (#2) #3}
\def\jhep#1#2#3{{\it J. High Energy Phys.\ }{\bf #1} (#2) #3}
\def\prd#1#2#3{{\it Phys.\ Rev.\ }{\bf D #1} (#2) #3}

\def\cmp#1#2#3{{\it Comm.\ Math.\ Phys.\ }{\bf #1} (#2) #3}

\def\ap#1#2#3{{\it Ann.\ of Phys.\ }{\bf #1} (#2) #3}

\def\imath#1#2#3{{\it Invent math }{\bf #1} (#2) #3}

\def\jdiffgeo#1#2#3{{\it J.\ Diff.\ Geom.\ }{\bf #1} (#2) #3}
\def\am#1#2#3{{\it Advances in Math.\ }{\bf #1} (#2) #3}
\def\jstat#1#2#3{{\it J.\ Stat.\ Mech.\ }{\bf #1} (#2) #3}
\def\cqg#1#2#3{\emph{Class.\ Quant.\ Grav.\ }{\textbf{#1}} (#2) #3}
\begin{document}
%
%
{\pagestyle{empty}
\parskip 0in
\

\vfill
\begin{center}




{\LARGE Multi-Loop Zeta Function Regularization}

\medskip

{\LARGE and Spectral Cutoff in Curved Spacetime}



\vspace{0.4in}

Adel B{\scshape ilal}$^{*}$ and Frank F{\scshape errari}$^{\dagger}$
\\
\medskip
{$^{*}$\it Centre National de la Recherche Scientifique\\
Laboratoire de Physique Th\'eorique de l'\'Ecole Normale Sup\'erieure\\
24 rue Lhomond, F-75231 Paris Cedex 05, France}

\smallskip

{$^{\dagger}$\it Service de Physique Th\'eorique et Math\'ematique\\
Universit\'e Libre de Bruxelles and International Solvay Institutes\\
Campus de la Plaine, CP 231, B-1050 Bruxelles, Belgique}

\smallskip

{\tt adel.bilal@lpt.ens.fr, frank.ferrari@ulb.ac.be}
\end{center}
\vfill\noindent

We emphasize the close relationship between zeta function methods and arbitrary spectral 
cutoff regularizations in curved spacetime. This yields, on the one hand, a physically sound and mathematically rigorous justification of the standard zeta function regularization at one loop and, on the other hand, a natural generalization of this method to higher loops. In particular, to any Feynman diagram is associated a generalized meromorphic zeta function. For the one-loop vacuum diagram, it is directly related to the usual spectral zeta function. To any loop order, the renormalized amplitudes can be read off from the pole structure of the generalized zeta functions. We focus on scalar field theories and illustrate the general formalism by explicit calculations at one-loop and two-loop orders, including a two-loop evaluation of the conformal anomaly.

\bigskip


\medskip
%
\begin{flushleft}
\today
\end{flushleft}
\newpage\pagestyle{plain}
\baselineskip 16pt
\setcounter{footnote}{0}
\tableofcontents\pagenumbering{arabic}

}

\section{\label{s1} General presentation}

\subsubsection*{Introduction}

Quantum field theory in curved spacetime is a mature area of research with many outstanding applications, including particle creation in time-dependent background and black hole evaporation (see e.g.\ \cite{QFTcurved} and references therein). Interesting applications to the calculation of the leading quantum corrections to the area law for the black hole entropy have also appeared recently \cite{BHSen}. However, the subject has been almost entirely focusing on free fields or, equivalently, on the one-loop vacuum energy. One of the difficulties to compute at higher loops is to define an appropriate reparameterization invariant regularization scheme. In principle, one may use dimensional regularization, but this scheme it is not very natural in curved spacetime because there is no canonical way to generalize a general $d$-dimensional spacetime manifold $\mathscr M_{d}$ to arbitrary $d+\epsilon$ dimensions. A much preferred and powerful regularization method is the zeta function scheme \cite{hawking}. This approach is very elegant and manifestly reparameterization invariant. However, it is only defined at the one-loop level. The main goal of the present work is to show that zeta function methods are also very natural at higher loop order, by highlighting a close relationship between the zeta function scheme and the general physical spectral cutoff regularization. 

\subsubsection*{On the zeta function regularization} 

As a simple illustration of the zeta function method, let us consider a massless scalar field on a two dimensional spacetime of the form $\mathbb R\times\text{S}^{1}$, the length of the circle being $a$. Its momentum is quantized in units of $2\pi/a$ and the vacuum energy is formally given by an infinite sum,
\be\label{zetex1} E = \frac{2\pi}{a}\sum_{n>0} n\, .\ee
The zeta function prescription amounts to replacing the above ill-defined sum by the analytic continuation of the Riemann $\zeta$ function
\be\label{zetaRdef}\zeta_{\text R}(s) = \sum_{n>0}\frac{1}{n^{s}}\ee
at the physically relevant value $s=-1$. $\zeta_{\text R}$ is a meromorphic function with a single pole at $s=1$ with unit residue and  $\zeta_{\text R}(-1)=-\frac{1}{12}$. Hence
\be\label{zetex2} E_{\zeta} = \frac{2\pi}{a}\zeta_{\text R}(-1) = -\frac{\pi}{6 a}\,\cdotp
\ee

Much more generally, a typical one-loop calculation in curved spacetime involves the computation of a Gaussian path integral which yields the determinant of some wave operator $D$. For example, in the case of a scalar field on a Euclidean Riemannian manifold endowed with a metric $g$, 
\be\label{Dscalgen} D = \Delta + m^{2} + \xi R\, ,\ee
where $\Delta$ is the positive Laplace-Beltrami operator, $m$ the mass parameter, $\xi$ an arbitrary dimensionless constant and $R$ the Ricci scalar. If we denote the eigenvalues of $D$ by $\la_{r}$, the determinant of $D$ is formally given by an infinite product
\be\label{DetDform} \det D = \prod_{r} \la_{r}\, .\ee
In the zeta function scheme, this infinite product is defined by introducing the spectral zeta function associated with the wave operator $D$,
\be\label{zetaDdef} \zeta_{D}(s) = \sum_{r}\frac{1}{\la_{r}^{s}}\,\cdotp\ee
It can be shown that $\zeta_{D}$ is a meromorphic function on the complex $s$-plane and that $s=0$ is a regular point. Motivated by the identity
\be\label{zetader} \zeta_{D}'(s) = -\sum_{r}\frac{\ln \la_{r}}{\la_{r}^{s}}\, \cvp\ee
one then sets
\be\label{zetaex3} \det D = e^{-\zeta'_{D}(0)}\, .\ee

In the case of the two-dimensional massless scalar field considered above, the integration over the scalar field produces the effective action
\be\label{Seff0} \Seff = \frac{1}{2}\ln{\det}'\Delta\, ,\ee
where the prime indicates that the zero eigenvalue is not included.
The corresponding zeta function per unit length of the cylindrical Euclidean spacetime is
\be\label{zetaex4} \zeta (s) = \int_{-\infty}^{+\infty}\frac{\d k}{2\pi}\sum_{n\in\mathbb Z^{*}}\frac{1}{\bigl(k^{2} + (2\pi n/a)^{2}\bigr)^{s}}
= \frac{1}{\sqrt{\pi}}\frac{\Gamma(s-1/2)}{\Gamma(s)}\Bigl(\frac{a}{2\pi}\Bigr)^{2s -1}\zeta_{\text R}(2s -1)\, .\ee
The effective potential is then given by
\be\label{Seff} V_{\text{eff}} = -\frac{1}{2}\zeta'(0) = \frac{2\pi}{a}\zeta_{\text R}(-1) = - \frac{\pi}{6 a}\,\cvp\ee
consistently with \eqref{zetex2}. 

The zeta function method is not limited to the effective action. For example, and as we shall review later, it also provides a definition of the stress-energy tensor, avoiding the ambiguities of the point-splitting method, and of other operators of the same type. Actually, virtually all one-loop effects can be consistently computed in this scheme. The results are manifestly reparameterization invariant, since $\zeta_{D}$ depends only on the spectrum of the operator $D$.

In spite of its power and elegance, the zeta function approach suffers from two important drawbacks. The first drawback, shared with dimensional regularization, is the absence of any obvious reason for why precisely it works. Even though replacing sums like $\sum_{n>0}n$ by $\zeta_{\text R}(-1) = -1/12$ is a perfectly well-defined procedure in the mathematical sense, it is abstract and unphysical. It is clear that the analytic continuation subtracts the divergence, as required, but it is very unclear how it does so explicitly and why the remaining finite part is the actual correct physical value. Physically, the renormalization (group) theory implies that subtractions must always correspond to adding local counterterms to the microscopic action. In a renormalizable theory, there are only a finite number of such terms, constrained by power counting. For example, for the massless scalar on the cylinder, the only available counterterm is the cosmological constant, which produces a term linear in $a$ in the energy. The sum \eqref{zetex1} should thus be of the form
\be\label{zetex1bis} E = c_{\infty} a -\frac{\pi}{6 a}\,\cvp\ee
for an infinite, but $a$-independent, constant $c_{\infty}$. The finite physical energy, obtained after subtraction of the infinite local counterterm, will be
\be\label{Efiniteex1} E_{\text{finite}} = c a -\frac{\pi}{6 a}\,\cvp\ee
for an arbitrary finite constant $c$ corresponding to the a priori arbitrary physical cosmological constant. Making the statement \eqref{zetex1bis} precise is essential in understanding the validity of the $\zeta$ function procedure.

One of the upshots of the present paper will be to make the consistency of the zeta function approach with the renormalization group ideas and the subtraction of local counterterms completely explicit, in the most general cases. This yields a streamlined and pedagogical derivation of all the known one-loop results in curved spacetime in which the r\^ole of the zeta function method is shifted, from an abstract trick to provide finite alternatives to 
otherwise ill-defined expressions, to a powerful mathematical tool allowing to compute rigorously physically sound and mathematically well-defined observables. We believe that this point of view puts the theory of quantum fields in curved spacetime on firmer grounds and should be of great help for teaching the subject to students, eliminating once and for all the need to call upon wisecrack statements like \eqref{zetex2} without justification.

The second and, for practical purposes, most important drawback of the 
zeta function scheme is that it only works at one loop. 
This is so because the quantum effects take the form of a functional determinant only at one loop. Guessing a generalization of the method to any loop order has proven to be rather difficult. As we have said, the prescription for the finite parts of amplitudes should correspond to adding local counterterms to the microscopic action. Any abstract mathematical proposal to extract these finite parts from complicated multi-loop diverging amplitudes is unlikely to be consistent with this requirement and, in particular, will violate unitarity. Such difficulties are seen, for example, in the operator regularization method \cite{ORscheme}, which could be viewed as an interesting attempt to generalize the zeta function scheme.

\subsubsection*{The all-loop zeta function scheme}

A central part of the present paper is to illustrate the fact that zeta function methods apply naturally to any loop order. We are mainly going to study vacuum diagrams, which compute the gravitational effective action, and focus on the case of the scalar field, but it will be clear from our presentation that our analysis can be generalized to arbitrary correlation functions and more general field content. To any Feynman diagram, we shall associate a generalized zeta function $Z(s)$ with the following properties:

i) $Z$ is a meromorphic function on the complex $s$-plane, with poles at integer values on the real $s$-axis.

ii) If the amplitude $\A$ associated with the Feynman diagram is finite, then $Z$ does not have poles with $\re s>0$ and has a simple pole at $s=0$ with residue $\A$.

iii) For the one-loop diagram $\diagramabig\,$, the function $Z=Z_{\diagrama}$ is expressed in terms of the standard spectral $\zeta$ function,
\be\label{Zoneloop} Z_{\diagrama}(s) = \frac{\zeta(s+1)}{s}\,\cdotp\ee

iv) To any given loop order, the renormalized effective action can be derived from the pole structure of the functions $Z(s)$ associated to all the contributing Feynman diagrams.

\noindent These properties will be discussed in Section \ref{allloopsec} and illustrated by explicit calculations at two loops in Section \ref{appsec}. 

\subsubsection*{The spectral cutoff}

The construction of the generalized zeta functions $Z$ will actually follow straightforwardly from the general analysis of the much more concrete physical cutoff scheme. Such a scheme is usually thought to be difficult and cumbersome to use, particularly beyond the one-loop order. Moreover, even at one loop, we shall explain below that the simplest flat spacetime cutoff procedure, which amounts to cutting sharply all momenta greater than some fixed energy scale, cannot be generalized to curved spacetime! However, these difficulties are only superficial. It turns out that the general smooth cutoff can actually be used very elegantly and that the right mathematical tool to deal with it is precisely the zeta function. 

A simple reparameterization invariant cutoff scheme generalizing the flat spacetime sharp cutoff procedure can be set up by putting a cutoff $\La$ on the spectrum of the wave operator $D$. For example, a regularized version of the sum \eqref{zetex1} can be defined by cutting off sharply all the modes having frequencies greater than $\La$,
\be\label{sharpex1} E_{\La} =\frac{2\pi}{a}\sum_{n\geq 1}\theta\Bigl(\La - \frac{2\pi n}{a}\Bigr) n = \frac{2\pi}{a}\sum_{n=1}^{\lfloor\frac{a\La}{2\pi}\rfloor} n\, .\ee
The symbol $\lfloor x\rfloor$ denotes the floor function, the largest integer smaller than or equal to $x$, and $\theta$ is the Heaviside step function, defined for convenience such that $\theta(0)=1$. Of course, the sum \eqref{sharpex1} can be easily computed, see \eqref{sharpE4} and \eqref{Efloordef}. However, it does not have a well-defined asymptotic expansion at large $\La$! All we can say is that
\be\label{sharpex2} E_{\La} = \frac{a\La^{2}}{4\pi} + O(\La)\, .\ee
The leading divergence is, as expected from \eqref{zetex1bis}, linear in $a$, but the discontinuities of the floor function $\lfloor \frac{a\La}{2\pi}\rfloor$ make the reminder a discontinuous function of order $\La$. In the very simple case of the sum \eqref{sharpex1}, one might propose an averaging procedure over the discontinuities to try to extract a finite piece, but this would be an unjustified ad-hoc prescription that could not be generalized to more complicated situations. This problem with the sharp cutoff does not occur in infinite flat spacetime but is generic in non-trivial geometries. It is associated with very interesting mathematics, which we shall very briefly describe later. It makes the use of a sharp cutoff inconsistent in curved spacetime.

The situation is much more favorable if one uses a smooth spectral cutoff regularization, characterized by a smooth cutoff function $f$. The only conditions to impose on $f$ are
\be\label{fzero} f(0)=1\ee
and a vanishing condition at infinity, for example that $f$ should be a Schwartz function. In this scheme, the sum \eqref{zetex1} is replaced by
\be\label{smoothex1} E_{f,\La} = \frac{2\pi}{a}\sum_{n\geq 1}
f\Bigr(\frac{2\pi n}{a\La}\Bigr)\, n\, .\ee
Unlike with a sharp cutoff, the regularized energy $E_{f,\La}$ does have a well-defined large $\La$ expansion. This expansion can be found by using the Euler-MacLaurin formula, which yields
\be\label{smoothex2} E_{\La} = \frac{a\La^{2}}{2\pi}\int_{0}^{\infty}\!\d x \, x f(x) - \frac{\pi}{6 a} f(0) + o(1)\, .\ee
The result is in beautiful agreement with the expected formula \eqref{zetex1bis}: all the dependence in the cutoff function can be absorbed in the local counterterm $c_{\infty}a$ and the finite part correctly matches the zeta function prescription by taking into account \eqref{fzero}. Actually, \eqref{smoothex2} provides the rigorous justification of the abstract zeta function result \eqref{zetex2}.

The use of a general cutoff function, as presented above, is of course standard and appears, for example, in rigorous textbook treatment of the Casimir effect, see e.g.\ \cite{IZbook}. At first sight, it may seem to be rather limited in scope, because the Euler-MacLaurin formula, which is instrumental in deriving the large $\La$ expansion, can be used only for a rather limited class of sums like \eqref{smoothex1}. For example, the smooth cutoff version of the logarithm of the determinant of the wave operator \eqref{Dscalgen}, a quantity directly related to the one-loop effective action, is
\be\label{logDsmooth1} \bigl(\ln\det D\bigr)_{f,\La} = \sum_{r}f\bigl(\la_{r}/\La^{2}\bigr)\ln \la_{r}\, .\ee
The Euler-MacLaurin formula is powerless in evaluating the large $\La$ asymptotics of this sum, except in very special cases, because, in general, the eigenvalues $\la_{r}$ are not known explicitly. A central guideline of our work is that the zeta function technique is precisely the right tool to compute the asymptotics of general sums like \eqref{logDsmooth1}, without referring to the Euler-MacLaurin formula. A direct link between the general smooth cutoff scheme and the zeta function prescription can thus be established. The justification of why the zeta function prescription is correct stems from this connection.

This brings us very near the punch line. The smooth cutoff regularization can be straightforwardly defined to any loop order in perturbation theory and even non-perturbatively. The mathematical analysis performed at one loop generalizes effortlessly and produces the simple all-loop generalization of the zeta function scheme which is a central result of our work.

\subsubsection*{A note on some of our original motivations}

Recently, a non-perturbative definition of the path integral over the K\"ahler metrics on a compact complex manifold of arbitrary dimension was proposed in \cite{2dgrav} (see also \cite{2dgrav2,2dgrav3} for related works). The main ingredient in the construction of the path integral is to approximate the infinite dimensional space of K\"ahler metrics $\mathscr M$ at fixed K\"ahler class by finite dimensional subspaces $\mathscr M_{n}$ of so-called Bergman metrics. These subspaces are characterized by an integer $n$ such that $\lim_{n\rightarrow\infty}\mathscr M_{n}=\mathscr M$ in a very precise sense \cite{2dgrav,math}.
The path integral over $\mathscr M$ is then regularized by a finite dimensional integral 
over $\mathscr M_{n}$. In two dimensions, since all metrics are K\"ahler, the construction yields a non-perturbative definition of the full quantum gravity path integral in the continuum formalism.

The non-perturbative nature of the regulator $n$ introduced in \cite{2dgrav} makes it very different from the standard schemes. Mathematically, it is related to the degree of a certain line bundle used in the construction of the spaces $\mathscr M_{n}$. Physically, it is best thought of 
as a sort of cutoff. On Riemann surfaces of fixed area $A$, the physical cutoff $\La$ is given in terms of $n$ by a relation of the form $\La^{2}\sim n/A$ and corresponds to the order of magnitude of the highest scalar curvatures for the metrics in $\mathscr M_{n}$. It is satisfying that a nice physical cutoff regulator emerges from the mathematical constructions in \cite{2dgrav,math}, but it also raises non-trivial questions about how the infinite cutoff limit is to be taken. This question and our will to perform explicit two-loop quantum gravity calculations, which will be presented in separate publications \cite{BFtoappear}, led us to the present investigations. 

\subsubsection*{The plan of the paper}

Some of our main ideas are introduced pedagogically in Section \ref{cutoffsec}, by studying in details a few basic examples. This allows us to discuss, in a simple set-up, the general properties of the cutoff regularization schemes, sharp and smooth, and make the link with the zeta function techniques explicit. In Section \ref{oneloopsec}, we revisit some of the pivotal ingredients of curved spacetime quantum field theory at one loop (the gravitational effective action, the Green's functions at coinciding point, the definition of the stress-energy tensor and the conformal anomaly), offering streamlined and simple derivations in our framework. Since Sections 2 and 3 do not contain any fundamentally new result compared to the existing literature, the expert reader may wish to skip directly to Section \ref{allloopsec}, which is the heart of the paper. It contains the construction of the all-loop generalization of the zeta function scheme and a discussion of its main properties. We define in particular the meromorphic function $Z$ associated with any given Feynman diagram and explain how the divergent and finite parts of the corresponding amplitude are related to its pole structure. Section \ref{appsec} is devoted to the presentation of explicit two-loop calculations illustrating the general framework. We compute in particular the two-loop gravitational effective action for the $\phi^{3}$ and $\phi^{4}$ scalar field theory in dimension four, providing a detailed discussion of the required counterterms and checking our formalism in details. In the case of the conformal $\phi^{4}$ model, this allows us to show explicitly that the two-loop conformal anomaly vanishes. In an effort to make our work self-contained, we have also included an Appendix reviewing the main properties of heat kernels and zeta functions that are used heavily throughout the main text.

\newpage
\section{\label{cutoffsec} Cutoff and zeta function in simple examples}

In this section, after a discussion of the properties of a general cutoff function $f$, we present a peda\-go\-gical introduction to some core ideas of our work on a very simple example: the vacuum energy of the familiar massless two-dimensional scalar field. We are going to discuss the difficulties associated with the use of a sharp cutoff and how these difficulties are resolved by using a smooth cutoff. We are also going to explain the intimate connection between the cutoff schemes and the zeta function formalism, a central idea which will be fully exploited in the later Sections.

\subsection{On the cutoff function}\label{cutpropsec}

The idea of the spectral cutoff is to weight the contribution of modes of ``energy'' $\epsilon_{r}$ by a factor $f(\epsilon_{r}^{2}/\La^{2})$ (or $f(\epsilon_{r}/\La)$ if it's more convenient), where $f:\mathbb R_{+}\rightarrow \mathbb R_{+}$ is the cutoff function and $\La$ the ultraviolet cutoff. The energies squared $\epsilon_{r}^{2}$ are typically the eigenvalues of a wave operator like \eqref{Dscalgen}. The cutoff procedure must keep untouched the infrared spectrum, that is to say the modes with energies much smaller than $\La$. This clearly implies the condition \eqref{fzero}, $f(0)=1$, together with a smoothness condition for $f$ at $x=0$. It is natural to assume that $f$ is infinitely differentiable at $x=0^{+}$. On the other hand, $f$ must go to zero at infinity in order to eliminate the ultraviolet modes. The decrease of $f$ must be fast enough for all the physical quantities of interest to be properly regularized. It is usually sufficient to consider a Schwartz-like condition,
\be\label{finfinity} f (x) \underset{x\rightarrow\infty}{=} O(1/x^{n}) \ \ \text{for any } n\geq 0\, .\ee
Of course, the simplest cutoff function,
\be\label{fsharp} f(x) = \theta (1-x)\ee
where $\theta$ is the Heaviside step function, satisfies the above conditions. However, we have already mentioned that it is plagued by inconsistencies and that it is necessary to consider smooth cutoff functions. It will be very convenient to restrict ourselves to smooth functions that can be written as a Laplace transform,
\be\label{Laplacef} f(x) = \int_{0}^{\infty}\!\d\alpha\, \varphi(\alpha) e^{-x\alpha}\, .\ee
Working with this very large class of functions is more than enough for our purposes, but it is certainly possible to adapt our work to other classes of cutoff functions, for example to Fourier transforms instead of Laplace transforms. 

On the Laplace transform, the conditions \eqref{fzero} and \eqref{finfinity} correspond to
\begin{align}\label{phicond1} & \int_{0}^{\infty}\!\d\alpha\,\varphi(\alpha) = 1\, ,\\\label{phicond2} &
\varphi(\alpha) \underset{\alpha\rightarrow 0}= O(\alpha^{n})\ \ \text{for any } n\geq 0\, ,
\end{align}
whereas the smoothness behavior of $f$ near zero yields
\be\label{phicond3} \int_{0}^{\infty}\!\d\alpha\, \alpha^{n}\varphi(
\alpha) < \infty\ \ \text{for any } n\geq 0\, .\ee
Technically, these conditions will be used in the following way. We are going to come across integrals over several variables $\underline\alpha = (\alpha_{1},\ldots,\alpha_{p})$ of the form
\be\label{intcoex} \mathscr A(\La;\varphi) = \int_{\mathbb R_{+}^{p}}
\! \d\underline\alpha\,\varphi(\alpha_{1})\cdots\varphi(\alpha_{p})\, \mathsf K
\bigl(\underline\alpha/\Lambda^{2}\bigr)\, .\ee
The functions $\mathsf K$ we shall deal with have  smooth finite limits at infinity and asymptotic expansions around zero of the form
\be\label{Piasympgen} \mathsf K (\underline\alpha t) = \sum_{k,q\geq 0}
\mathsf A_{k,q}(\underline\alpha)\, t^{k-N/2}(\ln t)^{q}\, \cvp\ee
for some integer $N$.
The conditions \eqref{phicond1} and \eqref{phicond2} make the integral \eqref{intcoex} convergent. The condition \eqref{phicond3} ensures that the large $\La$ asymptotic expansion of $\mathscr A (\La;\varphi)$ can be obtained from the small $t$ asymptotic expansion of 
$\mathsf K (\underline\alpha t)$, to any order. 

\subsection{Vacuum energy on the cylinder}

Let us start by revisiting the case of the massless scalar on the cylinder, with vacuum energy \eqref{zetex1}. The sharp and smooth cutoff versions of the energy are given in \eqref{sharpex1} and \eqref{smoothex1} respectively.

\subsubsection*{Sharp cutoff}
\begin{figure}
\centerline{\includegraphics[width=5in]{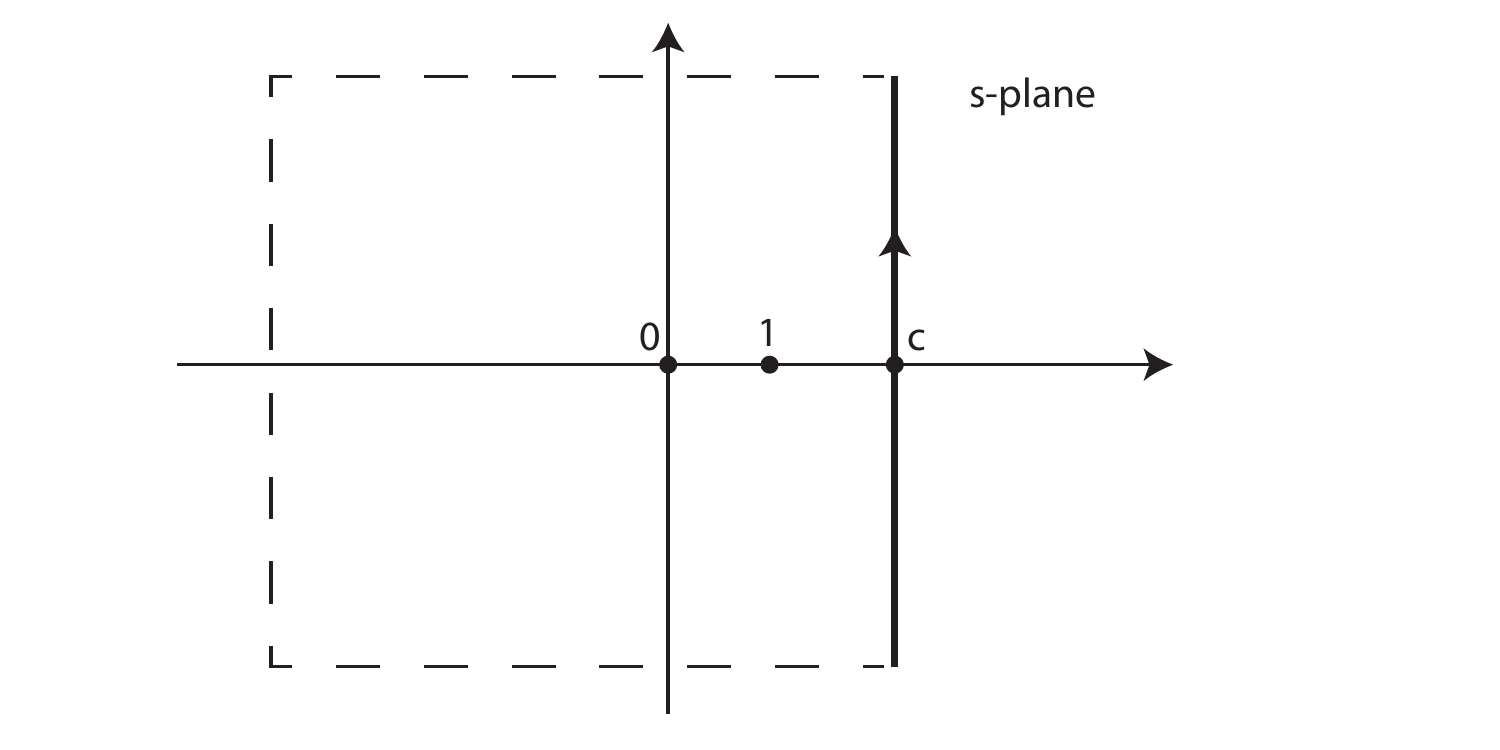}}
\caption{The contour of integration $]c-i\infty,c+i\infty[$ (thick line) used in the main text. In some cases, the large cutoff asymptotic expansion of the corresponding integrals can be obtained by closing the contour on the left by an infinite semi-rectangle (dashed line), e.g.\ in \eqref{smoothE2} and \eqref{S2dsm3}. \label{fig1}}
\end{figure}

Our goal is to make the link between the cutoff scheme and the zeta function. In the case of the sharp cutoff, our starting point is the following Mellin integral representation of the Heaviside step function,
\be\label{stepint} \theta (\la - \la') =\frac{1}{2i\pi} \int_{c-i\infty}^{c+i\infty}
\frac{\d s}{s}\biggl(\frac{\la}{\la'}\biggr)^{s}\, ,\ee
where the integration over $s$ is made along a contour parallel to the imaginary axis, with any constant real part $c>0$, see Fig.\ \ref{fig1}. The idea of the proof of \eqref{stepint} is that, on the one hand, for $\la>\la'$, one can close the contour by an infinite semi-rectangle on the left, the contribution of the integral over the semi-rectangle vanishing due to the fast decrease of the integrand. The integral is then given by the residue at $s=0$, which is one. On the other hand, for $\la<\la'$, one can close the contour by an infinite semi-rectangle on the right. Since the integrand is analytic for $\re s>0$, the integral is then zero. 
Plugging \eqref{stepint} in \eqref{sharpex1}, we obtain
\be\label{sharpE1} E_{\La} = \frac{1}{ia}\sum_{n=1}^{\infty}
\int_{c-i\infty}^{c+i\infty}\frac{\d s}{s}
\biggl(\frac{a\La}{2\pi}\biggr)^{s} n^{1-s}\, .\ee
If we choose $c>2$, in order for the series $\sum_{n\geq 1}n^{1-s}$ to converge, we can commute the sum and integral signs and we find
\be\label{sharpE2} E_{\La} = \frac{1}{ia}
\int_{c-i\infty}^{c+i\infty}\frac{\d s}{s}
\biggl(\frac{a\La}{2\pi}\biggr)^{s} \zeta_{\text R}(s-1)\, .\ee
We thus get a formula for the regularized energy where the Riemann zeta function has appeared naturally.

\begin{figure}
\centerline{\includegraphics[width=4in]{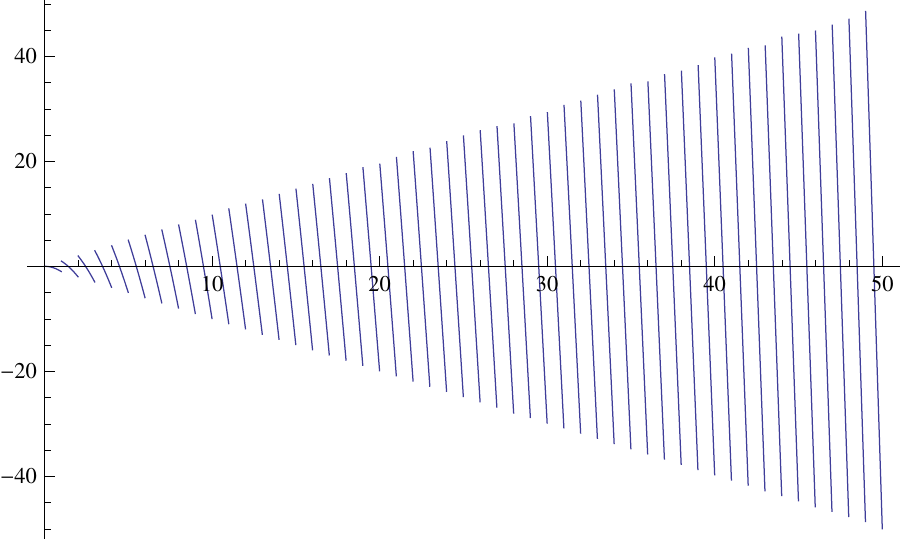}}
\caption{Plot of the function $\mathcal E(x)$ defined in \eqref{Efloordef}. The diverging oscillatory behavior at large $x$ prevents the function from having a smooth large $x$ asymptotic expansion, a typical problem associated with the use of a sharp cutoff.\label{figplotE}}
\end{figure}

In the sum \eqref{sharpE1}, when $a\La/2\pi> n$, we can close the contour of integration on the left to pick the pole at $s=0$ and get back to \eqref{sharpex1}. It is natural to expect that the same may be done in \eqref{sharpE2} when $\La\rightarrow\infty$. If this idea were correct, and since $c>2$, the integral \eqref{sharpE2} would pick a pole at $s=2$ due to the simple pole of the Riemann zeta function and another pole at $s=0$ due to the factor $1/s$. We would thus obtain the following large $\La$ asymptotics of $E_{\La}$,
\be\label{sharpE3} E_{\La} \overset{?}{\underset{\La\rightarrow\infty}{=}}
\frac{a\La^{2}}{4\pi} + \frac{2\pi}{a}\zeta_{\text R}(-1)
= \frac{a\La^{2}}{4\pi} -\frac{\pi}{6a}\,\cdotp
\ee
Unfortunately, this formula is \emph{wrong}. Only the leading term proportional to $\La^{2}$ is correct. As we have discussed in Section \ref{s1} around equation \eqref{sharpex2}, the corrections to the leading term are discontinuous and of order $\La$. These corrections can actually be easily evaluated exactly, because the sum \eqref{sharpex1} is elementary. Denoting by $\lfloor\,\rfloor$ the floor function, we have
\be\label{sharpE4} E_{\La} - \frac{a\La^{2}}{4\pi} = \frac{\pi}{a}
\mathcal E \Bigl(\frac{a\La}{2\pi}\Bigr)\, ,\ee
where
\be\label{Efloordef} \mathcal E (x) = \lfloor x\rfloor \bigl( \lfloor x
\rfloor +1\bigr) - x^{2}\, .\ee
The plot of $\mathcal E(x)$ in Fig.\ \ref{figplotE} illustrates well the oscillatory behavior and the linear divergence of the remainder term. 

The conclusion is that, unlike for \eqref{stepint}, we are not allowed to close the contour by an infinite semi-rectangle on the left in the integral \eqref{sharpE2}, even when $\La\rightarrow\infty$. Technically, the problem comes from the non-trivial large $|s|$ behavior of the Riemann zeta function which makes the contribution of the integral over the infinite semi-rectangle non-trivial, even when $\La\rightarrow\infty$.

Even though \eqref{sharpE3} is not a correct mathematical statement, the reader will have noticed that it does yield the correct finite part in terms of $\zeta_{\text R}(-1)$! Intuitively, closing the contour as we have done correspond to some sort of averaging procedure over the oscillations of the function $\mathcal E(x)$. In the present extremely simple example, we could make this statement more precise, but we are not going to pursue this idea because it cannot be generalized to more complicated and interesting cases.

\subsubsection*{Smooth cutoff}

It is much more fruitful to study how the situation is modified when we use a smooth cutoff instead of the sharp cutoff. Using \eqref{Laplacef}, the regularized energy \eqref{smoothex1} is then given by
\be\label{smoothE1} E_{f,\la} = \frac{2\pi}{a}
\int_{0}^{\infty}\!\d\alpha\,
\varphi(\alpha) \sum_{n\geq 1}ne^{-\frac{2\pi n}{a\La}\alpha}\, .
\ee
The infinite sum in this equation is elementary and we could proceed by computing it exactly. However, this will not be possible in more complicated examples.
Instead, let us try to make the link with the zeta function, as we have done for the sharp cutoff. To do so, we use the Mellin representation of the exponential function,
\be\label{expint} e^{-u} = \frac{1}{2i\pi}\int_{c-i\infty}^{c+i\infty}
\d s\, \Gamma(s)u^{-s}\,\cvp\ee
which is valid as long as $c>0$. The proof of \eqref{expint} is based on the
good large $|s|$ behavior of the $\Gamma$ function, given by the Stirling's formula, which implies that the contour of integration in \eqref{expint} can be closed on a semi-infinite rectangle on the left without changing the value of the integral. One thus picks all the poles of the integrand on the half-plane $\re s\leq 0$. These poles come from the simple poles of $\Gamma$ at integer values $s=-k\leq 0$, 
with residue $(-1)^{k}/k!$, and summing over all the poles yields the series representation of the exponential function $e^{-u}$, as called for. Plugging \eqref{expint} into \eqref{smoothE1} and choosing $c>2$ in order to be able to commute the sum and integral signs then leads to
\be\label{smoothE2} E_{f,\La} = 
\frac{1}{ia}\int_{0}^{\infty}\!\d\alpha\,\varphi(\alpha) 
\int_{c-i\infty}^{c+i\infty}\d s\, \Gamma(s)\biggl(\frac{a\La}{2\pi\alpha}\biggr)^{s}
\zeta_{\text R}(s-1)\, .\ee
This equation is analogue to the sharp cutoff equation \eqref{sharpE2}, the crucial difference being the insertion of the $\Gamma$ function. This insertion improves greatly the large $|s|$ asymptotic behavior of the integrand and the large $\La$ asymptotic expansion of the integral can then be correctly obtained by closing the contour on the left by an infinite semi-rectangle. The diverging and finite terms are given by the poles on the positive $s$-axis. There is one such pole at $s=2$ due to the Riemann zeta function and another such pole at $s=0$ due to the $\Gamma$ function, from which we get
\begin{align}\label{smoothE3} E_{f,\La} &=
\frac{2\pi}{a}\int_{0}^{\infty}\!\d\alpha\,\varphi(\alpha)\biggl[
\Bigl(\frac{a\La}{2\pi}\Bigr)^{2}\frac{1}{\alpha^{2}} + \zeta_{\text R}(-1)
+ O(1/\La)\biggr]\\
&= \frac{a\La^{2}}{2\pi}\int_{0}^{\infty}\!\d\alpha\,\frac{\varphi(\alpha)}{\alpha^{2}} - \frac{\pi}{6 a} + O\bigl(1/\La\bigr)\, .
\end{align} 
By using the simple identity
\be\label{triviallapid} \int_{0}^{\infty}\!\d\alpha\,\frac{\varphi(\alpha)}{\alpha^{2}} = \int_{0}^{\infty}\!\d x\, xf(x)\, ,\ee
we find the correct expansion \eqref{smoothex2}. 

The interest of the derivation we have just presented is that it does not use the Euler-MacLaurin formula and thus can be easily generalized. The link with the zeta function formalism is also made manifest.

\subsection{Vacuum energy on a compact Riemann surface}

Let us now test the power of the method in the much less trivial case of the massless scalar field on an arbitrary compact Riemann surface of genus $h$, endowed with a metric $g$ of total area $A$. The gravitational effective action, obtained after integrating over the scalar field in the path integral, is given by
\be\label{Seff2d} S(g) = \frac{1}{2}\ln\bigl(\mu^{-2} A^{-1}{\det}'(\mu^{-2}\Delta)\bigr)\, .\ee
The factor $A^{-1}$ makes up for the zero eigenvalue that is removed from the determinant and is required by consistency with conformal invariance, see e.g.\ \cite{2dgrav2} and references therein. The scale $\mu$ has been introduced for dimensional reason and can be viewed as an arbitrary renormalization scale. 

\subsubsection*{Sharp cutoff}
\begin{figure}
\centerline{\includegraphics[width=6.25in]{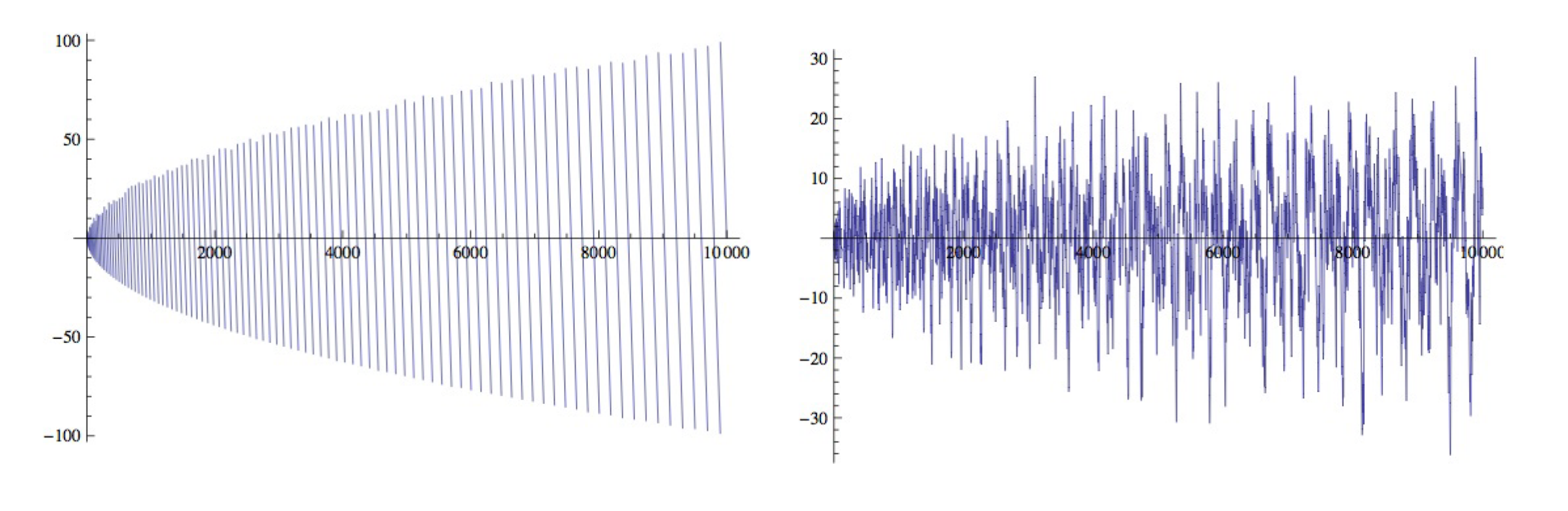}}
\caption{The function $N(\la)-\la$ in the case of the round sphere (left inset) and the flat torus of modulus $\tau=i$ (right inset), with an area $A=4\pi$. The fluctuations are $O(\lambda^{1/2})$ for the sphere and, according to the Hardy's conjecture, $O(\la^{\alpha})$ for any $\alpha>1/4$ for the torus. In all cases, the erratic behavior makes the use of the sharp cutoff inconsistent.\label{torusfig}}
\end{figure}

Let us denote by $0=\lambda_{0}<\lambda_{1}\leq\la_{2}\leq\cdots$ the eigenvalues of the positive Laplacian $\Delta$. The sharp cutoff version of the effective action \eqref{Seff2d} reads
\be\label{S2d1}\begin{split} S_{\La}(g) & = \frac{1}{2}\sum_{r\geq 1}\theta\bigl(\La^{2}-\la_{r}\bigr)\ln\bigl(\mu^{-2}\la_{r}\bigr)-\frac{1}{2}\ln\bigl(\mu^{2}A\bigr)\\ &= \frac{1}{2}\sum_{r=1}^{N(\La^{2})}
\ln\bigl(\mu^{-2}\la_{r}\bigr)-\frac{1}{2}\ln\bigl(\mu^{2}A\bigr)\, ,
\end{split}\ee
where $N(\la)$ is the number of eigenvalues that are less than or equal to $\la$. The celebrated Weyl's law governs the large $\la$ asymptotics of $N(\la)$,
\be\label{Weyl} N(\la)\underset{\la\rightarrow\infty}{\sim} \frac{A\la}{4\pi}\,\cvp\ee
and this yields the leading divergence in \eqref{S2d1},
\be\label{S2dleaddiv} S_{\La}\underset{\La\rightarrow\infty}{\sim}
\frac{A\La^{2}}{8\pi}\ln\frac{\La^{2}}{\mu^{2}}\,\cdotp\ee
However, the remainder term in Weyl's law is an extremely complicated function. In particular, it cannot have a smooth large $\la$ asymptotic expansion, because $N(\la)$ is discontinuous for all the values $\la=\la_{r}$, the amplitude of the discontinuity being equal to the degeneracy of the eigenvalue $\la_{r}$. A smooth asymptotic expansion can thus be valid, at best, up to undetermined bounded terms $O(1)$. Actually, the most general result that one can prove is \cite{hormander}
\be\label{Weyl2} N(\la) = \frac{A\la}{4\pi} + O\bigl(\la^{1/2}\bigr)\, ,\ee
the bound being saturated for the round sphere. The $O(\la^{1/2})$ term is highly oscillatory and we have depicted the cases of the round sphere and a flat torus for illustration in Fig.\ \ref{torusfig}. 
A nice discussion of these issues can be found, for example, in \cite{revWeyl}. The consequence of these facts for the effective action $S_{\La}$ is even more drastic. The remainder term obtained by subtracting the leading divergence \eqref{S2dleaddiv} is a wildly oscillating function of the cutoff, the amplitude of the oscillations growing like $\La\ln\La$. Any attempt to average over these oscillations to extract a finite part would be a very complicated and ambiguous procedure. We see, again, that the sharp cutoff regularization is not appropriate. 

This well-known conclusion is completely general. Unlike in infinite flat space, where it appears to be quite natural and tractable, the sharp cutoff regularization scheme is inconsistent in curved space, or even in flat space with non-trivial topology. We will definitively abandon it from now on. Further discussion may be found in the Ch.\ 5 of \cite{Fursaev} and references therein. 

\subsubsection*{Smooth cutoff}

The smooth cutoff version of the effective action is given by
\be\label{S2dsm1}S_{f,\La}(g) =  \frac{1}{2}\sum_{r\geq 1}
f\bigl(\la_{r}/\La^{2}\bigr)
\ln\bigl(\mu^{-2}\la_{r}\bigr)-\frac{1}{2}\ln\bigl(\mu^{2}A\bigr)\, .
\ee
Finding the large $\La$ asymptotic expansion of such a sum, for a general cutoff function $f$, might naively seem intractable. Except in very special instances, like the round sphere, for which the eigenvalues are explicitly known, the Euler-MacLaurin formula is useless. Trying to use the large $r$ asymptotics for $\la_{r}$ also fails, because the best we can say is that, consistently with \eqref{Weyl2},
\be\label{Weyloneigen} \la_{r}=\frac{4\pi r}{A} + O(\sqrt{r})\, ,\ee
for a very irregular remainder term $O(\sqrt{r})$, and this will fix only the leading quadratic divergence. 

Fortunately, the method that we have used in the previous subsection to evaluate the much simpler sum \eqref{smoothE1} does not suffer from these caveats. Using \eqref{Laplacef} and \eqref{expint}, we can rewrite \eqref{S2dsm1} as
\be\label{S2dsm2} S_{f,\La}(g) = \frac{1}{4i\pi}\int_{0}^{\infty}\!\d\alpha\,\varphi(\alpha)\sum_{r\geq 1}\int_{c-i\infty}^{c+i\infty} \d s\, \Gamma(s)
\frac{\La^{2s}}{\alpha^{s}\la_{r}^{s}}\ln\frac{\la_{r}}{\mu^{2}}
-\frac{1}{2}\ln\bigl(\mu^{2}A\bigr)\, .\ee
To go further, we need to know some basic properties of the spectral zeta function
\be\label{zetaLapdef} \zeta_{\Delta}(s) = \sum_{r\geq 1}\frac{1}{\la_{r}^{s}}\ee
associated with the Laplacian. From Weyl's law \eqref{Weyl}, we see that the series on the right-hand side of \eqref{zetaLapdef} converges for $\re s>1$. It can be shown, along lines reviewed in Sec.\ \ref{heatzetasec} and \ref{analyticsec}, that $\zeta_{\Delta}$ can be analytically continued to a meromorphic function over the whole complex $s$-plane. It has a unique simple pole at $s=1$ with residue
\be\label{zetaDres} \res_{s=1}\zeta_{\Delta} = \frac{A}{4\pi}\,\cdotp\ee
The derivative $\zeta'_{\Delta}$ is thus also a meromorphic function, with a double pole at $s=1$ and a series representation
\be\label{zetapLap} \zeta'_{\Delta}(s) = -\sum_{r\geq 1}\frac{\ln\la_{r}}{\la_{r}^{s}}\ee
which is valid for $\re s>1$. From all these properties, we can deduce that
the sum and integral signs in \eqref{S2dsm2} can be permuted if we choose $c>1$, which yields
\be\label{S2dsm3} S_{f,\La}(g) = -\frac{1}{4i\pi}
\int_{0}^{\infty}\!\d\alpha\,\varphi(\alpha)
\int_{c-i\infty}^{c+i\infty} \d s\, \Gamma(s)
\frac{\La^{2s}}{\alpha^{s}}\bigl(\zeta'_{\Delta}(s) + \zeta_{\Delta}(s)\ln\mu^{2}\bigr)-\frac{1}{2}\ln\bigl(\mu^{2}A\bigr)\, .\ee
We then use the trick of closing the contour of integration on an infinite semi-rectangle on the left to find the large $\La$ asymptotics. 
The diverging and finite pieces come from the poles of the integrand on the positive real axis (including zero). There is a double pole at $s=1$ coming from $\zeta'_{\Delta}$ as well as a simple pole from $\zeta_{\Delta}$ and a simple pole at $s=0$ coming from $\Gamma$. Extracting the residues, using in particular $\Gamma'(1)=-\gamma$ where $\gamma$ is the Euler's constant, and taking into account the constraint \eqref{phicond1}, we finally get
\begin{multline}\label{S2dsm4}
S_{f,\La}(g) = \frac{A\La^{2}}{8\pi}\int_{0}^{\infty}\!\d\alpha\,
\frac{\varphi(\alpha)}{\alpha}\Bigl(-\gamma + \ln\frac{\La^{2}}{\mu^{2}\alpha}\Bigr)\\ -\frac{1}{2}\zeta'_{\Delta}(0) - \frac{1}{2}\zeta_{\Delta}(0)\ln\mu^{2} -\frac{1}{2}\ln\bigl(\mu^{2}A\bigr)
+ O\bigl(1/\La^{2}\bigr)\, .
\end{multline}

Equation \eqref{S2dsm4} is perfectly consistent with the field theory lore. The cutoff-dependent divergent piece in $S_{f,\La}$ is proportional to the area and can thus be canceled by a cosmological constant counterterm. The finite piece is cutoff-independent and is consistent, as hoped, with the zeta-function prescription \eqref{zetaex3} and \eqref{Seff0} for the determinant of the Laplacian. The dependence in the scale $\mu$ is also irrelevant, as expected. 
Part of it comes from a term proportional to the area and can thus be absorbed in the cosmological constant. The other part is proportional to $\zeta_{\Delta}(0) + 1$ which is itself proportional $1-h$ because it can be shown (along lines explained in Sec.\ \ref{heatzetasec}) that
\be\label{zetazeto2d} \zeta_{\Delta}(0) = -\frac{h+2}{3}\,\cdotp\ee
It can thus be absorbed in the Einstein-Hilbert counterterm which, in two dimensions, is topological and proportional to $1-h$.

\subsubsection*{Side remarks}

\noindent\emph{Dimensional analysis}

Equation \eqref{S2dsm4} is not manifestly consistent with dimensional analysis. The situation can be improved by using a dimensionless version of the zeta function,
\be\label{zetadimno} \zeta_{\Delta,\,\mu}(s) = \mu^{2s}\zeta_{\Delta}(s)\, ,\ee
in terms of which
\be\label{S2dsm5}
S_{f,\La}(g) = \frac{A\La^{2}}{8\pi}\int_{0}^{\infty}\!\d\alpha\,
\frac{\varphi(\alpha)}{\alpha}\Bigl(-\gamma + \ln\frac{\La^{2}}{\mu^{2}\alpha}\Bigr) -\frac{1}{2}\zeta'_{\Delta,\,\mu}(0) -\frac{1}{2}\ln\bigl(\mu^{2}A\bigr)+ O\bigl(1/\La^{2}\bigr)\, .
\ee
Using the dimensionless version of the zeta function is in some sense more natural, but it is also unconventional and we shall keep using the standard $\mu$-independent zeta functions in the following. 

\smallskip

\noindent\emph{$\varphi$ versus $f$}

Cutoff-dependent terms are more naturally expressed in terms of the Laplace transform $\varphi$ in our approach, but they can also be expressed in terms of the cutoff function $f$. For example, by using the identity 
\be\label{idflaplap} \int_{0}^{\infty}\!\d x\, f(x)\ln x =
-\int_{0}^{\infty}\!\d\alpha\,\frac{\varphi(\alpha)}{\alpha}\bigl(\gamma
+\ln\alpha\bigr)\, ,\ee
we can put \eqref{S2dsm4} in the form
\begin{multline}\label{S2dsm6}
S_{f,\La}(g) = \frac{A\La^{2}}{8\pi}\ln\frac{\La^{2}}{\mu^{2}}\,\int_{0}^{\infty}\!\d x\, f(x)\, + \frac{A\La^{2}}{8\pi}
\int_{0}^{\infty}\!\d x\, f(x)\ln x\\
-\frac{1}{2}\zeta'_{\Delta}(0) - \frac{1}{2}\zeta_{\Delta}(0)\ln\mu^{2} -\frac{1}{2}\ln\bigl(\mu^{2}A\bigr)+ O\bigl(1/\La^{2}\bigr)\, .
\end{multline}
Such expressions may be of academic interest, but we shall not bother to come back to $f$ systematically.

\smallskip

\noindent\emph{The area dependence}

It is interesting to realize that the field theory lore goes a long way in fixing the area dependence of the effective action in two dimensions, without having to perform any explicit calculation. To understand this point, let us choose the metric to be of the form $g= {A\over A_0}g_{0}$, for some fixed $g_{0}$ of area $A_0$. The effective action can depend on the renormalization scale $\mu$, the cutoff $\La$ and the areas $A_{0}$ and $A$. 
Using the fact that the only available counterterms are the cosmological constant and the topological Einstein-Hilbert terms, the most general formula consistent with dimensional analysis is given by
\be\label{Seffco1} S_{\La,f} = r_{1}(\La/\mu) \mu^{2}A + (1-h) r_{2}(\La/\mu) + S_{\text{finite}}(\mu^{2}A) + o(1)\, .\ee
By locality of the counterterms, the dimensionless functions $r_{1}$ and $r_{2}$ cannot depend on the metric and thus in particular on the area. As for the finite part $S_{\text{finite}}$, it cannot depend on the cutoff (of course, $S_{\text{finite}}$ depends on additional dimensionless parameters, like the moduli of the Riemann surface). Since $S_{\La,f}$ does not depend on the arbitrary scale $\mu$, the function $S_{\text{finite}}$ must be such that any variation of $\mu$ can be balanced by a variation of the counterterms.\footnote{Let us note that we could identify $A_0=\m^{-2}$, in which case the $\mu$-independence is a weak form of background independence.} It is straightforward to check that this condition implies\footnote{Introduce the dimensionless variables $x=\L^2/\m^2$ and $y=\m^2 A$. Then  $S_{\La,f}$ is of the form $ S_{\La,f}=F(x) y + G(x) +S_{\text{finite}}(y)$. The $\m$-independence yields
$-x y F'(x)  + y F(x)  - x G'(x)  + y S_{\text{finite}}'(y) = 0$. Taking $\d/\d y$ yields $xF'(x)-F(x)=(y  S_{\text{finite}}'(y))'$ which must equal some constant $c_3$. Integrating the two equations gives $F(x)=c_1 x - c_3$ and $S_{\text{finite}}(y)=c_2 \ln y + c_3 y + c_4$. Inserting this back into the previous equation yields $G(x)=c_2 \ln x + c_5$, so that 
$S_{\La,f} = c_1 x y + c_2 \ln x + c_2 \ln y + c_4+c_5$.
}
\be\label{Slore} S_{\La,f} = c_1 \L^2 A + (1-h) c_2 \ln{\L^2\over \m^2} + c_2 (1-h) \ln\bigl(\mu^{2}A\bigr)
+S_{\text{finite}}^{(0)}\, ,\ee
where $S_{\text{finite}}^{(0)}$ does not depend on the area $A$ and $c_{1}$ and $c_{2}$ are metric-independent constants. Thus the non-trivial area dependence of the effective action is simply given by
\be\label{SfiniteAdep} c_2 (1-h) \ln A\ee
and is fixed up to a single number $c_{2}$. 
In particular, on a torus, we deduce that there is no non-trivial area dependence at all.

The result \eqref{S2dsm5} is perfectly consistent with \eqref{Slore}. Indeed, the area dependence of the zeta function can be obtained easily by noting that the eigenvalues of the Laplacians for the metrics $g_{0}$ and $g={A\over A_0}g_{0}$ are related according to $\la_{r} = \la_{0,r}{A_0\over A}$ and thus $\zeta_{\Delta}(s) = (A/A_{0})^{s}\zeta_{\Delta_{0}}(s)$. Plugging this result into \eqref{S2dsm5} and using \eqref{zetazeto2d} yields \eqref{Slore} with
\be\label{betavalue} c_2 = -\frac{1}{6}\,\cdotp\ee
%

%
%

\section{\label{oneloopsec} Free fields in curved spacetime revisited}

Let us now extend the ideas of the previous Section to standard important applications of the zeta function scheme in the context of free quantum fields in curved spacetime. In each case we shall see that the general spectral cutoff regularization provides a simple and physical justification of the zeta function prescription. Since we have already explained in details the basic ideas, our style of presentation will be less pedagogical and more succinct. We shall also briefly review some properties of the heat kernel and the zeta functions, which will be useful in later Sections as well. 

Of course, heat kernel and zeta function methods have been used long before in many different ways in quantum field theory on curved spacetimes, see e.g.\ \cite{hawking,Gilkey,Dowker,Avramidi,Toms} and in particular the review \cite{Vassil}, book \cite{Fursaev} and references therein.

\subsection{The model}\label{modelsub}

We shall focus for concreteness on the well-studied scalar field on a $d$  dimensional Euclidean Riemannian manifold, with metric $g$, volume $V$ (which may be infinite) and action
\be\label{Action}
\begin{split}
 S &= \frac{1}{2}\int\!\d^{d}x\sqrt{g}\, \Bigl( g^{ij}\partial_{i}\phi\partial_{j}\phi + m^{2}\phi^{2} + \xi R \phi^{2}\Bigr)\\
 & = \frac{1}{2}\int\!\d^{d}x\sqrt{g}\, \phi D\phi\, ,
 \end{split}
\ee
where the wave operator $D$ was already defined in \eqref{Dscalgen}. The field equations reads
\be\label{fieldeq} D\phi = \bigl(\Delta + m^{2} + \xi R\bigr)\phi = 0\, .\ee
The stress-energy tensor, defined by the variation of the action with respect to the metric,
\be\label{Tdef} \delta S = \frac{1}{2}\int\!\d^{d}x\sqrt{g}\, T^{ij}\delta g_{ij}\, ,\ee
is given by
\begin{multline}\label{Tformula} T^{ij} = \bigl(1/2 - 2\xi\bigr) \partial_{k}\phi\partial^{k}\phi\, g^{ij} -\bigl(1 - 2\xi\bigr) \partial^{i}\phi
\partial^{j}\phi + \frac{1}{2} m^{2}\phi^{2}\, g^{ij}\\
+\xi\Bigl[ \Bigl(\frac{1}{2} R g^{ij} - R^{ij}\Bigr)\phi^{2} + 2\phi\Delta\phi\, g^{ij} + 2 \phi\nabla_{i}\partial_{j}\phi\Bigr]\, . 
\end{multline}
The associated trace
\be\label{traceT} T = T^{i}_{i} = \Bigl(\frac{d-2}{2} - 2(d-1)\xi\Bigr)
\partial_{i}\phi\partial^{i}\phi + \frac{d}{2} m^{2}\phi^{2} +
\frac{d-2}{2}\xi R\phi^{2} + 2(d-1)\xi \phi\Delta\phi
\ee
governs the variation of the action under the Weyl's rescaling
\be\label{Weylscaling} \delta g_{ij} = 2g_{ij}\delta\omega\, .\ee
By using the field equation \eqref{fieldeq} we can recast the trace in the form
\be\label{TwithFE} T = \Bigl(\frac{d-2}{2} - 2(d-1)\xi\Bigr)\Bigl(
\partial_{i}\phi\partial^{i}\phi + \xi R\phi^{2}\Bigr)
+\Bigl(\frac{d}{2} - 2(d-1)\xi\Bigr) m^{2}\phi^{2}\, .\ee
We see that the model is classically Weyl invariant if 
\be\label{xiWeyldef} \xi = \xi_{d} = \frac{d-2}{4(d-1)}\, \cvp\quad m^{2} =0\, .\ee
This can also be understood by noting that the classical action \eqref{Action} is invariant off-shell under the simultaneous transformations
\be\label{Weyltrans} \delta g_{ij} = 2g_{ij}\delta\omega\, ,\quad 
\delta\phi =  -\frac{d-2}{2}\phi\delta\omega\ee
of the metric and the field.

The model \eqref{Action} is well-defined as long as all the eigenvalues of the operator $D$ are strictly positive. The case with zero modes, which occurs in particular for the massless scalar field at $\xi=0$, can also be treated by slightly modifying the formalism presented below. In particular, in two dimensions, or in higher dimensions and finite volume, the zero modes must be removed from the path integral.

%
\subsection{Heat kernel and zeta functions}
\label{heatzetasec}

Let us assume that the volume $V$ is finite for convenience (the required modifications in infinite volume are trivial and thus will not be mentioned). The eigenvalues $\la_{r}$ and associated eigenvectors $\psi_{r}$ of the operator $D$ can then be labelled by a discrete index $r$. 
Let $\psi_{r}$ denote the eigenvector associated with the eigenvalue $\la_{r}$, normalized such that
\be\label{eigenvectors} \int\!\d^{d}x\sqrt{g}\,\psi_{r}(x)\psi_{r'}(x) = \delta_{rr'}\, .\ee
We introduce the generalized kernel
\be\label{Kgen} \mathscr K(t,s,x,y) = \sum_{r\geq 0}
\frac{e^{-\la_{r}t}}{\la_{r}^{s}}\, \psi_{r}(x)\psi_{r}(y)\, ,\ee
together with its coinciding points and integrated versions,
\begin{align}\label{Kgenx} \mathscr K(t,s,x) &= \mathscr K(t,s,x,x)\, ,\\
\label{Kgenint}
\mathscr K(t,s) &= \int\!\d^{d}x\sqrt{g}\, \mathscr K(t,s,x) = \sum_{r\geq 0}\frac{e^{-\la_{r}t}}{\la_{r}^{s}}\,\cdotp
\end{align}
Usual heat kernels and zeta functions are given by
\begin{align}\label{Kdefs} & K(t,x,y)=\mathscr K(t,0,x,y)\, ,\
K(t,x) = \mathscr K(t,0,x)\, ,\ K(t) = \mathscr K(t,0)\, ,\\
\label{zetadefs} & \zeta(s,x,y) = \mathscr K(0,s,x,y)\, ,\
\zeta(s,x)=\mathscr K(0,s,x)\, ,\ \zeta(s) = \mathscr K(0,s)\, .
\end{align}
Relations between these quantities can be found by using
\be\label{Mellin1} \frac{1}{\la^{s}} = \frac{1}{\Gamma(s)}\int_{0}^{\infty}\!\d t\, t^{s-1}e^{-\la t}\ee
or \eqref{expint}. For example,
\begin{align}\label{Ktozeta} \zeta(s,x,y)& = \frac{1}{\Gamma(s)}\int_{0}^{\infty}\!\d t\, t^{s-1} K(t,x,y)\, ,\\\label{zetatoK}
\mathscr K(t,s,x,y) &= \frac{1}{2i\pi}\int_{c-i\infty}^{c+i\infty}\d s'\,\Gamma(s') t^{-s'}\zeta(s+s',x,y)\, .
\end{align}
From Weyl's law, we can deduce that the series representation of the zeta function converges when the real part of its argument is strictly greater than $d/2$. If we choose $c>d/2-\re s$, \eqref{zetatoK} is then valid for any $x$ and $y$, including at $x=y$. 

The heat kernel $K(t,x,y)$ is a well-known standard tool in quantum field theory on curved manifolds \cite{Gilkey,Avramidi,Vassil}. It has a very useful small $t$ asymptotic expansion of the form
\be\label{Kxyexp} K(t,x,y) = \frac{e^{-\ell(x,y)^{2}/(4 t)}}{(4\pi t)^{d/2}}
\biggl(\sum_{k= 0}^{n}a_{k}(x,y) t^{k} + O\bigl(t^{n+1}\bigr)\biggr)\, .\ee
We have denoted by $\ell(x,y)$ the geodesic distance between $x$ and $y$ and the coefficient functions $a_{k}(x,y)$ are bilocal scalars having a smooth expansion around $x=y$. In particular,
\be\label{Kxexp} K(t,x) = \frac{1}{(4\pi t)^{d/2}}
\biggl(\sum_{k= 0}^{n}a_{k}(x) t^{k} + O\bigl(t^{n+1}\bigr)\biggr)\, ,\ee
where the $a_{k}(x)=a_{k}(x,x)$ are local scalar polynomials in the curvature. The overall normalization in \eqref{Kxyexp} and \eqref{Kxexp} has been chosen such that $a_{0}(x)=1$. Explicit formulas for the other coefficients are given in the Appendix \ref{heatapp}, see e.g.~\eqref{ficoeff}, \eqref{ardiag} and \eqref{frmfr}.

The expansions \eqref{Kxyexp} and \eqref{Kxexp} can be used to derive the analytic structure of the zeta functions (see Section \ref{analyticsec} for more details). The function $\zeta(s,x,y)$ is holomorphic over the whole complex $s$-plane if $x\not = y$. As for $\zeta(s,x)$, it can pick poles from the small $t$ region in the integral representation \eqref{Ktozeta}. If $d$ is even, we find simple poles at $s=d/2-k$ for $0\leq k\leq d/2-1$, with 
\be\label{residueeven} \res_{s=\frac{d}{2}-k}\zeta(s,x) = \frac{a_{k}(x)}{(4\pi)^{d/2}\Gamma(d/2-k)}\,\cvp\quad 0\leq k\leq d/2 -1\, .\ee
Moreover, the would-be poles at zero or negative integer values of $s$ are canceled by the poles of the $\Gamma$ function and we find
\be\label{zetaevenneg} \zeta(-k,x) = (-1)^{k}k! \frac{a_{d/2 + k}(x)}{(4\pi)^{d/2}}\, \cvp\quad
k\geq 0\, ,\quad d \text{\ even.}\ee
If $d$ is odd, there are simples poles at $s=d/2-k$ for all $k\geq 0$, with residues given by the same formula as in \eqref{residueeven}. The pole structure of the $\Gamma$ function also yields in this case
\be\label{zetaoddneg} \zeta(-k,x) = 0\, ,\quad k\geq 0\, ,\quad d \text{\ odd.}\ee

It is interesting to note that the heat kernel expansion \eqref{Kxexp} can be derived from the analytic structure of the zeta function that we have just described, by using \eqref{zetatoK}. The reasoning is exactly the same as the one that we have used in Section 2. The small $t$ asymptotic expansion of the integral \eqref{zetatoK}, for $x=y$, can be found by closing the contour on the semi-infinite rectangle on the left, as in Fig.\ \ref{fig1}. Summing up the residues to the desired order, we find back \eqref{Kxexp}.

\subsection{The gravitational effective action}\label{grav1lsec}

The gravitational effective action is formally given by the one-loop formula
\be\label{Sgrav} S(g) = \frac{1}{2}\ln\det D\ee
and is rigorously defined in terms of an arbitrary smooth cutoff function $f$ and renormalization scale $\mu$ by
\be\label{Sgravsmooth} S_{f,\La}(g) = \frac{1}{2}\sum_{r\geq 0}f\bigl(\la_{r}/\La^{2}\bigr) \ln\bigl(\mu^{-2}\la_{r}\bigr)\, .\ee
A small comment on this prescription should be made at this stage. Instead of using the regularizing factor $f(\la_{r}/\La^{2})$, we could also use $f((\la_{r}+\sigma)/\La^{2})$ for any finite parameter $\sigma$ with the dimension of a mass squared. Such a choice may be natural, for example, in the massive theory at $\xi=0$, where one may want to insert $f(\delta_{r}/\La^{2})$, where the $\delta_{r}$ are the eigenvalues of the Laplacian, instead of $f(\la_{r}/\La^{2}) = f((\delta_{r} + m^{2})/\La^{2})$. Our formalism can be straightforwardly adapted to deal with all these cases but, of course, the difference between these prescriptions is immaterial. They are simply related to one another by finite shifts of the infinite local counterterms one must add to the microscopic action.

To obtain the large $\La$ asymptotic expansion of the sum \eqref{Sgravsmooth}, we proceed along the lines explained in Section \ref{cutoffsec}. The formula generalizing \eqref{S2dsm3}, which does not contain a term similar to $\ln A$ because we do not have a zero mode in the present case, reads
\be\label{Sgsmooth2} S_{f,\La}(g) = -\frac{1}{4i\pi}
\int_{0}^{\infty}\!\d\alpha\,\varphi(\alpha)
\int_{c-i\infty}^{c+i\infty} \d s\, \Gamma(s)
\frac{\La^{2s}}{\alpha^{s}}\bigl(\zeta'(s) + \zeta(s)\ln\mu^{2}\bigr)
\, .\ee
The constant $c$ must be such that the series representation of the zeta function converges, that is $c>d/2$. Closing the contour of integration on the semi-infinite rectangle on the left, we pick all the simple and double poles of the integrand. However, the poles at $\re s< 0$ yield vanishing contributions when $\La\rightarrow\infty$. Using the results reviewed in the previous subsection, in particular \eqref{residueeven}, we obtain
\begin{multline}\label{Sgsmooth3} S_{f,\La}(g) = \frac{1}{2 (4\pi)^{d/2}}
\sum_{k=0}^{\lfloor\frac{d-1}{2}\rfloor}a_{k}\Lambda^{d-2 k}\int_{0}^{\infty}\!\d\alpha\,\frac{\varphi(\alpha)}{\alpha^{d/2-k}}\Bigl(\ln\frac{\La^{2}}{\mu^{2}\alpha} + \Psi(d/2-k)\Bigr)\\ - \frac{1}{2}\zeta'(0) - \frac{1}{2}\zeta(0)\ln\mu^{2} + o(1)\, ,
\end{multline}
where $\Psi=\Gamma'/\Gamma$ and the coefficients
\be\label{intakdef} a_{k} = \int\!\d^{d}x\sqrt{g}\, a_{k}(x)\ee
are the integrated versions of the coefficients that appear in the expansion \eqref{Kxexp}. In particular, $a_{0}=V$ and $\zeta(0) = a_{d/2}/ (4\pi)^{d/2}$ or zero depending on whether $d$ is even or odd.

The formula \eqref{Sgsmooth3} is in perfect agreement with the renormalization group ideas and provides a full justification of the zeta function prescription for the finite part of the determinant in \eqref{zetaex3}. As expected, this prescription amounts to subtracting infinite but local counterterms from the action, which are proportional to the coefficients $a_{k}$ in \eqref{Sgsmooth3}. Moreover, both the cutoff function and the arbitrary renormalization scale $\mu$ appear only in these local counterterm and are thus absorbed in the associated renormalized couplings when the cutoff is sent to infinity.

\newpage
\smallskip

\noindent\emph{Side remark}

There exists a commonly used heuristic cutoff procedure at one-loop based on the ``identity''
\be\label{Schwinger} \ln\la = -\int_{\frac{1}{\La^{2}}}^{\infty}\!
\d t\,\frac{e^{-\la t}}{t}
\ee
which is ``justified'' by taking the derivative of both sides with respect to $\la$. This procedure is equivalent to the definition
\be\label{Sgspecial} S_{\La}(g) = -\frac{1}{2}\int_{\frac{1}{\La^{2}}}^{\infty}\!\d t\,\frac{K(t)}{t}\ee
for the regularized gravitational effective action. The divergent piece comes from the small $t$ region in the integral \eqref{Sgspecial} and can thus be derived straightforwardly from the expansion \eqref{Kxexp}. To get the finite piece as well, the simplest method is to use \eqref{zetatoK} in \eqref{Sgspecial} and to evaluate the large $\La$ asymptotics by closing the contour as usual. This yields
\be\label{Sgspediv} S_{\La}(g) = -\frac{1}{(4\pi)^{d/2}}\sum_{k=0}^{\lfloor\frac{d-1}{2}\rfloor} \frac{a_{k}\La^{d-2 k}}{d-2k} - \frac{1}{2}
\bigl( \ln\La^{2} - \gamma\bigr)\zeta(0) - \frac{1}{2}\zeta'(0) + o(1)\, .
\ee
This result is consistent and, if not for its rather dubious starting point \eqref{Schwinger}, could also be seen as a justification of the zeta function prescription. Let us note that, interestingly, it is not a special case of the general cutoff scheme, since \eqref{Sgspediv} is not a special case of \eqref{Sgsmooth3}. Of course, the main drawback of this simple heuristic approach is the lack of a natural higher loop generalization.

\subsection{The Green function at coinciding points}
\label{Greensec}

The Green function $G(x,y)$ of the wave operator $D$ is defined by the condition
\be\label{Greendef} D_{x}G(x,y) = \bigl(\Delta_{x} + m^{2} + \xi R(x)\bigr)G(x,y) = \frac{\delta (x-y)}{\sqrt{g}}\,\cdotp\ee
It can be expressed in terms of the eigenfunctions and eigenvalues of the wave operator as
\be\label{Greendef2} G(x,y) = \sum_{r\geq 0}\frac{\psi_{r}(x)\psi_{r}(y)}{\la_{r}} = \zeta(1,x,y)\, .\ee
The simplest divergences in perturbation theory come from the self-contractions of the scalar field, which yield infinite contributions $G(x,x)$. Making sense of these contributions is crucial, in particular, to construct the quantum stress-energy tensor from the formula \eqref{Tformula} which involves composite operators. 

In flat space, the self-contractions can be suppressed by the normal ordering prescription, which amounts to setting $G(x,x)=0$. This is consistent because it can be shown to be equivalent to the subtraction of local counterterms. However, this simple prescription does not work in curved space because it would violate the reparameterization invariance. ``Normal ordering" in curved space amounts to replacing the ill-defined $G(x,x)$ by a non-vanishing renormalized version  called the Green's function at coinciding points. 

This Green function has been defined in several ways in the literature. The most common approach is to subtract the divergences of $G(x,y)$ when $x\rightarrow y$ to get a renormalized $G_{R}$. For example, for the two-dimensional scalar field, we can define
\be\label{Gdef2d} G_{R}(x) = \lim_{y\rightarrow x}\Bigl[ G(x,y) + \frac{1}{2\pi}\ln\bigl(\mu \ell(x,y)\bigr)\Bigr]\, ,\ee
where $\ell(x,y)$ is the geodesic distance and $\mu$ is an arbitrary renormalization scale. Another approach is to use the zeta function. Equations \eqref{Greendef2} suggests to identify the renormalized version of $G$ with $\zeta(s=1,x)$. This makes perfect sense in odd dimension, because $\zeta(s,x)$ is holomorphic in the vicinity of $s=1$, but in even dimension we have to subtract the pole. This leads the the following ansatz for the Green's function at coinciding points in the zeta function scheme,
\be\label{Gzeta} G_{\zeta}(x) = 
\begin{cases} \zeta(1,x) & \text{if $d$ is odd,}\\
\lim_{s\rightarrow 1}\bigl( \m^{2s-2}\zeta(s,x) - \frac{a_{d/2 -1}(x)}{(4\pi)^{d/2}}
\frac{1}{s-1}\bigr) & \text{if $d$ is even.}\end{cases}\ee
Of course, as usual with the zeta function method, this definition is an abstract mathematical trick. It is not obvious why it works or how it is related to the point-splitting method.

A more physical approach is to use our general cutoff scheme. The regularized Green's function is then given by
\be\label{Greensmooth} G_{f,\La}(x,y) = \sum_{r\geq 0}
f\bigl(\la_{r}/\La^{2}\bigr)\frac{\psi_{r}(x)\psi_{r}(y)}{\la_{r}}\,\cdotp\ee
When $x\not = y$, $G_{f,\La}(x,y)$ has a finite large cutoff limit given by $G(x,y)$. When $x=y$, on the other hand, the large $\La$ asymptotics can be found as usual by using the integral representation \eqref{zetatoK} for $\mathscr K\bigl(\alpha/\La^{2},1,x\bigr)$,
\begin{align}\label{Gsm1} G_{f,\La}(x) & = \int_{0}^{\infty}\!\d\alpha\,\varphi(\alpha)\mathscr K\bigl(\alpha/\La^{2},1,x\bigr)\\\label{Gsm2}
& =\frac{1}{2i\pi}\int_{0}^{\infty}\!\d\alpha\,\varphi(\alpha)
\int_{c-i\infty}^{c+i\infty} \d s\, \Gamma(s)
\frac{\La^{2s}}{\alpha^{s}}\zeta(s+1,x)\, ,
\end{align}
which is here valid for any $c>d/2 -1$, and then closing the contour of integration of the infinite semi-rectangle on the left. Using \eqref{Gzeta}, and taking into account that the integrand has a double pole at $s=0$ when $d$ is even, we obtain
\begin{multline}\label{Gsmooexpeven}
G_{f,\La}(x) \underset{\La\rightarrow\infty}{=}
\frac{1}{(4\pi)^{d/2}}\sum_{\substack{k\geq 0\\k\not = d/2 -1}}
a_{k}(x)\frac{\La^{d-2k-2}}{d/2-k-1}\int_{0}^{\infty}\!\d\alpha\,
\frac{\varphi(\alpha)}{\alpha^{d/2-k-1}}\\
+ \frac{1}{(4\pi)^{d/2}}a_{d/2-1}(x)\int_{0}^{\infty}\!\d\alpha\,\varphi(\alpha)\Bigl(\ln\frac{\La^{2}}{\m^2\alpha} - \gamma\Bigr) + G_{\zeta}(x)\quad\text{if $d$ is even,}
\end{multline}
\begin{multline}\label{Gsmooexpodd}
G_{f,\La}(x) \underset{\La\rightarrow\infty}{=}
\frac{1}{(4\pi)^{d/2}}\sum_{k\geq 0}
a_{k}(x)\frac{\La^{d-2k-2}}{d/2-k-1}\int_{0}^{\infty}\!\d\alpha\,
\frac{\varphi(\alpha)}{\alpha^{d/2-k-1}}\\ + G_{\zeta}(x)\quad
\text{if $d$ is odd.}
\end{multline}
These results provide a neat justification of the zeta function prescription \eqref{Gzeta}, since \eqref{Gsmooexpeven} and \eqref{Gsmooexpodd} imply that replacing the bare self-contraction $G_{f,\La}$ by $G_{\zeta}$ indeed amounts to subtracting local counterterms in the action.

\smallskip

\noindent\emph{Remark}

One can make the link between $G_{\zeta}(x)$ and the point-splitting method as follows. The integral representation \eqref{Ktozeta} shows that singularities in $\zeta(s,x,y)$ must be related to the small $t$ behavior \eqref{Kxyexp} of $K(t,x,y)$. In particular, the function
\be\label{zetaregular} \zeta_{R}(s,x,y) = \zeta(s,x,y) -
\frac{1}{\Gamma(s)}\int_{0}^{1/\mu^{2}}\!\!\frac{\d t}{(4\pi)^{d/2}}
\sum_{0\leq k\leq d/2 -1} a_{k}(x,y) t^{s+k-d/2-1}e^{-\ell(x,y)^{2}/(4t)}\, ,\ee
which is defined using \eqref{Kxyexp} by subtracting all the terms that can yield a singular behavior around $s=1$, must be completely smooth in the vicinity of $s=1$, for any scale $\mu$. In particular,
\be\label{zetaregident} \lim_{s\rightarrow 1}\lim_{y\rightarrow x}
\zeta_{R}(s,x,y) = \lim_{y\rightarrow x}\lim_{s\rightarrow 1}\zeta_{R}(s,x,y)\, .\ee
The limits on both sides of this identity can be straightforwardly evaluated. From \eqref{Gzeta}, the left-hand side is directly related to $G_{\zeta}$. The limits on the right-hand side can be worked out by noting that the integrals over $t$ can be expressed in terms of the exponential integral functions or the incomplete $\Gamma$ functions, defined by
\be\label{Endef} E_{-n}(z) = \int_{1}^{\infty}\!\d u\,  u^{n}e^{-zu}
=z^{-n-1}\G (n+1,z)
\ee
and evaluated at $z = \mu^{2}\ell(x,y)^{2}/4$, for various values of $n\geq - 1$. When $y\rightarrow x$, one then needs the asymptotics expansions
\be\label{Enexpansions} E_{1}(z) = -\gamma - \ln z + O(z)\, ,\quad
E_{-n}(z) = \frac{\Gamma(n+1)}{z^{n+1}}- \frac{1}{n+1} + O(z)\ \text{if $n>-1$.}\ee
Overall, we obtain
\begin{align}\label{Gspliteven}
\begin{split} G_{\zeta}(x) &= \lim_{y\rightarrow x}
\biggl[ G(x,y) - \frac{1}{(4\pi)^{d/2}}\sum_{k=0}^{d/2 -2}2^{d-2k-2}
\Gamma(d/2 - k -1)\frac{a_{k}(x,y)}{\ell(x,y)^{d-2k-2}}\ \\ &\hskip 3cm
+ \frac{1}{(4\pi)^{d/2}}a_{d/2 -1}(x)\Bigl(\ln\frac{\m^2 \ell(x,y)^{2}}{4} + 
2\gamma\Bigr)\biggr]\quad \text{if $d$ is even,}\end{split}\\
\label{Gsplitodd}\begin{split} G_{\zeta}(x) &= \lim_{y\rightarrow x}
\biggl[ G(x,y) - \frac{1}{(4\pi)^{d/2}}\sum_{k=0}^{(d-3)/2}2^{d-2k-2}
\Gamma(d/2 - k -1)\frac{a_{k}(x,y)}{\ell(x,y)^{d-2k-2}}\biggr]
\\ &\hskip 10.45cm \text{if $d$ is odd.}\end{split}
\end{align}
%

\subsection{The conformal anomaly}\label{CAsec}

\subsubsection*{Generalities}

In the present subsection, we assume that the conditions \eqref{xiWeyldef} are met. The model is then classically Weyl invariant. However, the definition of the quantum theory requires to introduce a regulator which always breaks the Weyl symmetry. This is clear in our general cutoff scheme, which depends on an explicit scale $\La$. When the cutoff is removed, the symmetry violation may persist, in which case the original symmetry of the classical theory is anomalous, meaning that it is altogether absent in the quantum theory.

The quantum stress-energy tensor is defined by varying the gravitational effective action \eqref{Sgrav} with respect to the metric via a formula like \eqref{Tdef}. If the variation of the metric is a Weyl transformation \eqref{Weylscaling}, we get the anomalous quantum trace $\mathscr A_{d}(x)$ as
\be\label{anomalydef} \delta_{\omega} S = \int\!\d^{d}x\sqrt{g}\,\mathscr A_{d}(x)\delta\omega\, .\ee
The anomaly is constrained by general consistency conditions. First, being a consequence of the introduction of a symmetry-violating reparameterization invariant regulator, it must be a local scalar functional. Its dimension is fixed to be $d$ by the formula \eqref{anomalydef}. Second, since it is obtained from the Weyl variation of the quantum effective action, it must satisfy the Wess-Zumino consistency conditions
\be\label{WZcc}\frac{\delta\mathscr A_{d}(x)}{\delta\omega(y)} =
\frac{\delta\mathscr A_{d}(y)}{\delta\omega(x)} = \frac{\delta^{2}S}{\delta\omega(x)\delta\omega(y)}\,\cdotp\ee
Third, since the quantum theory is defined modulo the addition of local renormalizable counterterms to the action, the anomaly itself is defined modulo the Weyl variation of the effect of such terms on the effective action. The above conditions restrict significantly the possible form of the anomaly \cite{anomalyrestric}. For example, in four  dimensions, the anomaly is fixed in terms of two dimensionless constants $a$ and $c$,
\be\label{anomaly4d} \mathscr A_{4} (x) \equiv \frac{1}{5760\pi^{2}}\bigl(
 a E_{4}(x) - c W_{4}(x) \bigr)\ee
where
\be\label{WEdef} W_{4} = C^{ijkl}C_{ijkl} = R^{ijkl}R_{ijkl} - 2 R^{ij}R_{ij} + \frac{1}{3}R^{2}\ee
is the square of the Weyl tensor and
\be\label{E4def}E_{4}=R^{ijkl}R_{ijkl} - 4R^{ij}R_{ij} + R^{2}\ee
is proportional to the Euler density. The symbol $\equiv$ in \eqref{anomaly4d} means ``equal up to the variation of local renormalizable counterterms.''

\subsubsection*{Standard computation}

In our case, the computation of the anomaly, in any dimension, can be made as follows. The variation of the wave operator $D$ for the parameters \eqref{xiWeyldef} under a Weyl rescaling \eqref{Weylscaling} is found to be
\be\label{DWeyltrans} \delta_{\omega} D = -2\delta\omega D - (d-2) g^{ij}\partial_{i}\delta\omega\partial_{j} + \frac{1}{2}(d-2)\Delta\delta\omega\, .\ee
The standard quantum mechanical perturbation theory then yields the associated variations of the eigenvalues of $D$. We get
\be\label{eigenWeyltrans} \delta_{\omega}\la_{r} = \int\!\d^{d}x\sqrt{g}\, \psi_{r}(x) \delta_{\omega} D\,\psi_{r}(x) = -2\la_{r} \int\!\d^{d}x\sqrt{g}\,
\psi_{r}^{2}\delta\omega\, ,\ee
where the second equality is obtained by performing an integration by part. We can then use this formula to compute directly the variation of the gravitational effective action \eqref{Sgravsmooth}. The large $\La$ asymptotic expansion of this variation is then obtained by repeating the same steps that yield the asymptotic expansion \eqref{Sgsmooth3}. 

Even  more conveniently, we can compute the variation of the effective action by starting directly from \eqref{Sgsmooth3}. The variation of the diverging pieces may be computed by using the identity
\be\label{akWeyltrans} \delta_{\omega}a_{k} = (d-2k)\int\!\d^{d}x\sqrt{g}\,
a_{k}(x)\delta\omega\, ,\ee
which is itself obtained by looking at the small $t$ asymptotics of the Weyl variation of the heat kernel $K(t)$. By construction, all these scheme-dependent terms do not contribute to the anomaly which is defined modulo the addition of the variation of local counterterms. We thus get the usual zeta function formula for the conformal anomaly,
\be\label{anomalyzeta} \mathscr A_{d}(x) \equiv -\frac{1}{2}\frac{\delta\zeta'(0)}{\delta\omega(x)}\,\cdotp\ee
By plugging \eqref{eigenWeyltrans} into the series representation of the zeta function, we obtain
\be\label{lazetap} \delta_{\omega}\zeta'(0) = 2\int\!\d^{d}x\sqrt{g}\, \zeta(0,x)\delta\omega\ee
and using \eqref{zetaevenneg} and \eqref{zetaoddneg} we finally get
\be\label{Adzeta} \mathscr A_{d}(x) \equiv
\begin{cases} -\frac{a_{d/2}(x)}{(4\pi)^{d/2}} & \text{if $d$ is even,}\\
0 & \text{if $d$ is odd.}\end{cases}\ee
For example, in dimension four,
\be\label{a2dim4} a_{2}(x) = \frac{1}{180}\bigl(R^{ijkl}R_{ijkl}-R^{ij}R_{ij} -\Delta R\bigr)\, .\ee
The term proportional to $\Delta R$ is generated by the Weyl variation of the local functional $\int\!\d^{d}x\sqrt{g}\, R^{2}$ and can thus be eliminated from the anomaly. The result \eqref{Adzeta} is thus consistent with the general form \eqref{anomaly4d}, with $a=1$ and $c=3$, which are the well-known values for a scalar field.

\subsubsection*{The Fujikawa method} 
 
One-loop anomalies can also be understood as coming from the Jacobian of the transformation in the path integral measure. This measure is the volume form for the metric
\be\label{metricFS}\lVert\delta\phi\rVert^{2} = \int\!\d^{d}x\sqrt{g}\,(\delta\phi)^{2}\ee
in field space and a non-trivial Jacobian is generated because \eqref{metricFS} is not invariant under Weyl transformations.

If we perform the transformations \eqref{Weyltrans} on both the metric and the field $\phi$, then the classical action is invariant and the variation of the effective action can be entirely accounted for by the Jacobian. Regularizing according to our general prescription, we get
\be\label{Jacobian} \delta_{\omega}S = -\ln J =\Bigl(-\frac{d}{2} + \frac{d-2}{2}\Bigr)\int\!\d^{d}x\sqrt{g}\,  \sum_{r\geq 0}f\bigl(\la_{r}/\La^{2}\bigr)\psi_{r}^{2}\delta\omega\, ,\ee
where the factor $-d/2$ comes from the variation of the metric and the factor $(d-2)/2$ from the variation of the scalar field. The large cutoff asymptotics can then be straightforwardly evaluated from \eqref{Laplacef} and the asymptotic expansion of the heat kernel. Up to terms which, according to \eqref{akWeyltrans}, can be absorbed in local counterterms, we find \eqref{Adzeta} again.

It is also instructive to make the reasoning by performing the Weyl transformation \eqref{Weylscaling} on the metric only. According to \eqref{traceT}, the variation of the classical action is then given by
\be\label{Sclvar}\frac{d-2}{2} \int\!\d^{d}x\sqrt{g}\, \phi D\phi\,\delta\omega\ee
whereas the Jacobian yields the term 
\be\label{Jac2cont} -\frac{d}{2}\int\!\d^{d}x\sqrt{g}\,  \sum_{r\geq 0}f\bigl(\la_{r}/\La^{2}\bigr)\psi_{r}^{2}\delta\omega\, .\ee
Overall, the anomaly is thus given by
\be\label{An1} \mathscr A(x) \equiv -\frac{d}{2}\sum_{r\geq 0}f\bigl(\la_{r}/\La^{2}\bigr)\psi_{r}^{2}(x)  + \frac{d-2}{2}\bigl\langle\phi (x) D\phi (x)\bigr\rangle\, .\ee
Classically, $D\phi = 0$, but quantum mechanically the expectation value in \eqref{An1} involves a self-contraction which produces an additional anomalous term. According to \eqref{Greensmooth}, the regularized self-contraction is given by
\be\label{An2} \bigl\langle\phi (x) D\phi (x)\bigr\rangle = \lim_{y\rightarrow x}D_{y}G_{f,\La}(x,y) = \sum_{r\geq 0}f\bigl(\la_{r}/\La^{2}\bigr)\psi_{r}^{2}(x)\ee
which, inserted in \eqref{An1}, yields again the correct anomaly \eqref{Adzeta}.
 
\section{\label{allloopsec} The multiloop formalism}
\begin{figure}
\centerline{\includegraphics[width=6in]{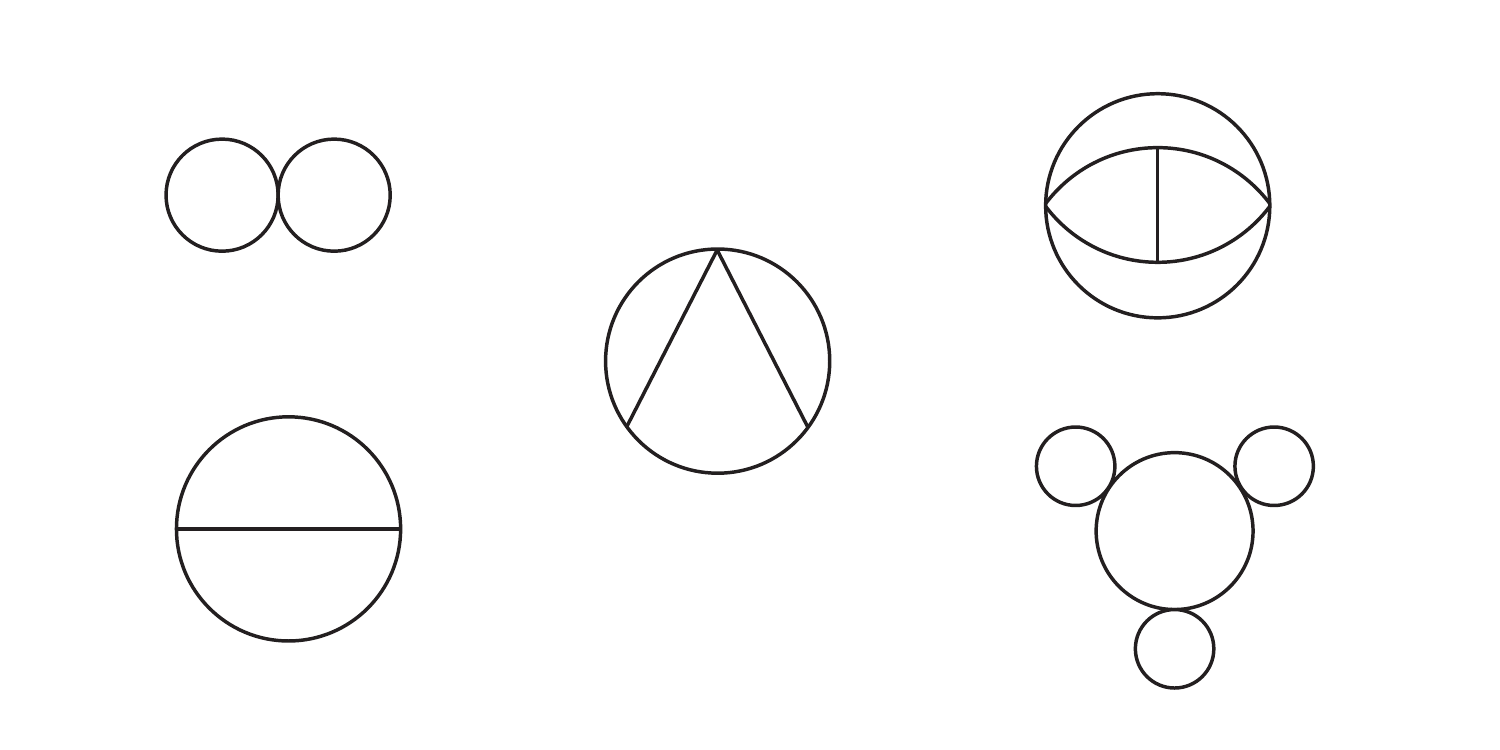}}
\caption{Typical vacuum Feynman diagrams at two, three and four loops. \label{figdiagram}}
\end{figure}

We now consider multi-loop Feynman diagrams. Typical representatives are depicted in Fig.\ \ref{figdiagram}. We shall deal explicitly with vacuum diagrams for a scalar field in $d$ dimensions and wave operator $D$ given by \eqref{Dscalgen}, but the generalization of the basic ideas to arbitrary correlation functions (which actually enter as subdiagrams of the vacuum diagrams) and more general field content are straightforward.

\subsection{Definitions}
\label{defsection}

Let us start with a few useful definitions. To simplify the notation, we denote the integration measure over spacetime by 
\be\label{spacemeasure}
\d\nu(x) = \d^{d}x\sqrt{g(x)}\, .
\ee 
Ordinary $n$-uplet are noted as $\vec a = (a_{1},\ldots,a_{n})$ whereas unordered $n$-uplet are noted as $\underline a = \{a_{1},\ldots,a_{n}\}$. We define
\be\label{sumdef} a=\sum_{i=1}^{n}a_{i}
\ee
and
$\underline a >0$ if $a_{i}\geq 0$ for all $i$ and $a>0$.

Let $\dia$ be an arbitrary connected vacuum Feynman diagram. We shall always denote by $n$ and $v$ the total number of internal lines and vertices, respectively. The $i^{\text{th}}$ internal line connects the spacetime points $x_{i}$ and $y_{i}$. The full set of points $x_{i}$ and $y_{i}$ are denoted collectively by $z_{I}$, $1\leq I\leq 2n$. To each vertex $\mathsf V$ we associate a set of indices $\mathcal I_{\mathsf V}$ corresponding to the spacetime points that are glued at $\mathsf V$ and a privileged index $I_{\mathsf V}$ which can be chosen arbitrarily amongst the elements of $\mathcal I_{\mathsf V}$. The integration measure of the diagram is then defined to be
\be\label{diagrammeasure} \d\Omega_{\dia}(z) = \prod_{J=1}^{2n}\d\nu (z_{J})\prod_{\mathsf V} \prod_{I\in\mathcal I_{\mathsf V}\setminus\{I_{\mathsf V}\}}
\frac{\delta(z_{I}-z_{I_{\mathsf V}})}{\sqrt{g(z_{I_{\mathsf V}})}}\, \cdotp
\ee
After taking into account the constraints from the delta functions, we have $v$ independent integration variables $z_{I_{\mathsf V}}=z_{\mathsf V}$.

\subsubsection*{Barnes spectral zeta functions}

First we introduce generalized spectral zeta functions \`a la Barnes, associated to the wave operator $D$ with eigenvalues $\la_{r}>0$ and eigenfunctions $\psi_{r}$,
\be\label{zetandef}\zeta_{n}(s,\vec\alpha;\vec x,\vec y) =
\frac{\Gamma(s+n-1)}{\Gamma(s)}\sum_{r_{1},\ldots,r_{n}\geq 0}
\frac{\prod_{i=1}^{n}\psi_{r_{i}}(x_{i})\psi_{r_{i}}(y_{i})}{\Bigl(\sum_{k=1}^{n}\alpha_{k}\la_{r_{k}}\Bigr)^{s+n-1}}\,\cvp\ee
where the Feynman parameters $\vec\alpha = (\alpha_{1},\ldots,\alpha_{n})$ are such that $\underline\alpha >0$. For $n=1$,
\be\label{zeta1} \zeta_{1}(s,\alpha;x,y) = \alpha^{-s}\zeta(s,x,y)\ee
is directly related to the standard zeta function introduced in Section \ref{heatzetasec}.

The zeta function $\zeta_{\dia}$ associated with the diagram $\dia$ is then defined by 
\be\label{zetadiadef} \zeta_{\dia}(s,\underline\alpha) = \int\!\d\Omega_{\dia}(\vec x,\vec y)\, \zeta_{n}(s,\vec\alpha;\vec x,\vec y)\, .
\ee
It has the series representation
\be\label{zetaDseries} \zeta_{\dia}(s,\underline\alpha)=
\frac{\Gamma(s+n-1)}{\Gamma(s)}\sum_{r_{1},\ldots,r_{n}\geq 0}
\frac{C_{\{r_{1},\ldots,r_{n}\}}^{\dia}}{\Bigl(\sum_{k=1}^{n}\alpha_{k}\lambda_{r_{k}}\Bigr)^{s+n-1}}\,\cvp\ee
in terms of the coefficients
\be\label{coefdef} C_{\{r_{1},\ldots,r_{n}\}}^{\dia} = \int\!\d\Omega_{\dia}(\vec x,\vec y)\,
\prod_{i=1}^{n}\psi_{r_{i}}(x_{i})\psi_{r_{i}}(y_{i})\, .\ee
Since by Weyl's law $\l_r\sim r^{d/2}$, the series \eqref{zetandef} and \eqref{zetaDseries} always converge for large enough $\re s$; the precise radius of convergence will be determined below. $\zeta_n$ and $\zeta_{\dia}$  are then defined for all values of $s$ by analytic continuation. Let us finally mention the simple scaling relation
\be\label{zetascale} \zeta_{\dia}(s,w\underline\alpha) = w^{1-n-s}\zeta_{\dia}(s,\underline\alpha)\, ,\quad w>0\, .\ee
%


\noindent\emph{Examples}

The zeta function associated with the one-loop diagram $\diagramabig$ is directly related to the usual spectral zeta function
\be\label{zet1l} \zeta_{\diagrama}(s,
\alpha) = \int\!\d\nu(x)\zeta_{1}(s;x,x)=\alpha^{-s}\zeta(s)\, .\ee
For the two-loop diagrams $\diagramcbig$ and $\diagrambbig$ we get
\begin{align}\label{zet2l1} \zeta_{\diagramc}\bigl(s,\{\alpha_{1},\alpha_{2}\}\bigr) &= \int\!\d\nu(x)\,\zeta_{2}(s,\alpha_{1},\alpha_{2};x,x,x,x)\, ,\\\label{zet2l2}\zeta_{\diagramb}\bigl(s,\{\alpha_{1},\alpha_{2},\alpha_{3}\}\bigr) & =
\int\!\d\nu(x)\d\nu(y)\, \zeta_{3}(s,\alpha_{1},\alpha_{2},\alpha_{3};x,x,x,y,y,y)
\end{align}
and for, e.g., the complicated diagram in the upper right corner of Fig.\ \ref{figdiagram},
\be\label{digdiazeta} \int\!\d\nu(w)\d\nu(x)\d\nu(y)\d\nu(z)\,
\zeta_{7}(s,\vec\alpha;w,w,w,w,x,y,x,z,z,x,y,z,z,y)\, .
\ee

\subsubsection*{The $Z$-functions}

The $Z$-function of the connected Feynman diagram $\dia$ is defined for $\underline\alpha >0$ by the series
\be\label{Zdef1} Z_{\dia}(s,\underline\alpha) = 
\frac{1}{s}\sum_{r_{1},\ldots,r_{n}\geq 0}\frac{C_{\{r_{1},\ldots,r_{n}\}}^{\dia}}{\la_{r_{1}}\cdots\la_{r_{n}}\Bigl(\sum_{k=1}^{n}\alpha_{k}\la_{r_{k}}\Bigr)^{s}}\,\cdotp\ee
Again, Weyl's law implies that this series always converges for large enough $\re s$ and thus defines $Z_{\dia}$ for all values of $s$ by analytic continuation. The $Z$ function satisfies the simple scaling relation
\be\label{Zscale} Z_{\dia}(s,w\underline\alpha) = w^{-s}Z_{\dia}(s,\underline\alpha)\, ,\quad w>0\, .\ee
The zeta function can be easily found from the $Z$ function by using the identity
\be\label{partialZZ} \frac{\partial^{n}Z_{\dia}(s,\underline\alpha)}{\partial\alpha_{1}\cdots\partial\alpha_{n}} = (-1)^{n} \zeta_{\dia}(s+1,\underline\alpha)\, ,\ee
which is obvious from the series representations \eqref{Zdef1} and \eqref{zetaDseries}. Conversely, $Z_{\dia}$ can be expressed in terms of $\zeta_{\dia}$ by integrating \eqref{partialZZ}, fixing the integration constants by using the fact that, for large enough $\re s$, the partial derivatives of $Z_{\dia}$ go to zero when some of the $\alpha_{i}$ go to infinity. This yields the interesting relation
\be\label{zetatoZ} Z_{\dia}(s,\underline\alpha) = \int_{\underline\alpha'>\underline\alpha}\!\d\underline\alpha'\, \zeta_{\dia}(s+1,\underline\alpha')\, ,\ee
where the condition $\underline\alpha'>\underline\alpha$ means that we integrate each $\alpha'_{i}$ from $\alpha_{i}$ to infinity.

\smallskip

\noindent\emph{Example}

The simplest $Z$ function is associated with the one-loop diagram $\diagramabig$. In this case, the series representation \eqref{Zdef1} immediately yields
\be\label{Zoneloop2} Z_{\diagrama}(s,\alpha) = \frac{\alpha}{s}\,\zeta_{\diagrama}(s+1,\alpha) = \frac{\alpha^{-s}}{s}\,\zeta(s+1)\,\cdotp\ee
Using \eqref{zet1l}, we can check that this result is consistent with \eqref{zetatoZ}. Let us note that $Z_{\diagrama}$ has a multiple pole, unlike the zeta function. As discussed in Section \ref{analyticsec}, this is a generic feature of the $Z$ functions.

\subsubsection*{Generalized heat kernels}

Finding relations between the zeta and $Z$ functions and generalized heat kernels proves to be extremely useful, both for understanding the analytic structure of the functions and for practical calculations.
We define two versions of generalized heat kernels associated with a Feynman diagram $\dia$,
\begin{align}\label{kgendef}  k_{\dia}(\underline t) &= 
\sum_{r_{1},\ldots,r_{n}\geq 0}C_{\{r_{1},\ldots,r_{n}\}}^{\dia}e^{-\sum_{i=1}^{n}\la_{r_{i}}t_{i}}\, ,\\\label{Kgendef}  
K_{\dia}(\underline t) &= 
\sum_{r_{1},\ldots,r_{n}\geq 0}C_{\{r_{1},\ldots,r_{n}\}}^{\dia}\frac{e^{-\sum_{i=1}^{n}\la_{r_{i}}t_{i}}}{\la_{r_{1}}\cdots\la_{r_{n}}}\, \cvp
\end{align}
where as usual $n$ is the number of internal lines in $\dia$ and $\underline t>0$. These two versions correspond naturally to the Barnes zeta function $\zeta_{\dia}$ and the $Z_{\dia}$ function respectively. At one loop, $k_{\diagrama}(t)=K(t)$ is the usual heat kernel and $K_{\diagrama}(t) = \mathscr K(t,1)$, see \eqref{Kgenint}. Even more generally, we may define
\be\label{Kgengen} \mathscr K_{\dia}(\underline t,\underline s) = 
\sum_{r_{1},\ldots,r_{n}\geq 0}C_{\{r_{1},\ldots,r_{n}\}}^{\dia}\frac{e^{-\sum_{i=1}^{n}\la_{r_{i}}t_{i}}}{\la_{r_{1}}^{s_{1}}\cdots\la_{r_{n}}^{s_{n}}}\, \cvp
\ee
the kernels \eqref{kgendef} and \eqref{Kgendef} being simple special cases. 
All these kernels are directly related to the more standard kernels defined in Section \ref{heatzetasec} via integral formulas, e.g.
\be\label{KggentoK} \mathscr K_{\dia}(\underline t,\underline s) = 
\int\!\d\Omega_{\dia}(\vec x,\vec y) \prod_{i=1}^{n}\mathscr K(t_{i},s_{i},x_{i},y_{i})\, .\ee

Using \eqref{Mellin1}, we can relate the heat kernels to the zeta and $Z$ functions,
\begin{align}\label{ktozeta}  \zeta_{\dia}(s,\underline\alpha)& =
\frac{1}{\Gamma(s)}\int_{0}^{\infty}\!\d t\, t^{s+n-2}k_{\dia}(\underline\alpha t)\, ,\\\label{KtoZ} Z_{\dia}(s,\underline\alpha) &=
\frac{1}{\Gamma(s+1)}\int_{0}^{\infty}\!\d t\, t^{s-1}K_{\dia}(\underline\alpha t)\, .
\end{align}
Conversely, the inverse Mellin transforms derived by using \eqref{expint} reads
\begin{align}\label{invMellinkdia} k_{\dia}(\underline\alpha t) &= \frac{1}{2i\pi}\int_{c-i\infty}^{c+i\infty}\!\d s\,\Gamma(s-n+1)t^{-s}\zeta_{\dia}(s-n+1,\underline\alpha)\, ,\\\label{invMellinKdia}
K_{\dia}(\underline\alpha t) &= \frac{1}{2i\pi}\int_{c-i\infty}^{c+i\infty}\!\d s\,\Gamma(s)s\,t^{-s}Z_{\dia}(s,\underline\alpha)\, .
\end{align}
The constant $c$ must be chosen is each case in such a way that the series representations \eqref{zetaDseries} and \eqref{Zdef1} of the zeta and $Z$ functions in the integrand converge.

The kernels $k_{\dia}$ and $K_{\dia}$ are related to each other via formulas that mimic the relations \eqref{partialZZ} and \eqref{zetatoZ} between the zeta and $Z$ functions,
\begin{align}\label{partialkk}& \frac{\partial^{n}K_{\dia}(\underline t)}{\partial t_{1}\cdots\partial t_{n}}  = (-1)^{n} k_{\dia}(\underline t)\, ,\\\label{ktoK} & K_{\dia}(\underline t)  = \int_{\underline t'>\underline t}\!\d\underline t'\, k_{\dia}(\underline t')\, .
\end{align}
These fundamental identities may be derived from the series representations \eqref{kgendef} and \eqref{Kgendef} or directly from \eqref{partialZZ} and \eqref{zetatoZ} by using \eqref{ktozeta} and \eqref{KtoZ}. The integral in \eqref{ktoK} is traditionally interpreted as an integral over the moduli space of the Feynman diagram $\dia$. The exponential factor in \eqref{Kxyexp} shows that the parameter $\sqrt{t'_{i}}$ can be associated with the spacetime length of the $i^{\text{th}}$ internal line in the diagram. The parameters $\underline t$ in $K_{\dia}$, which bound from below the integral over $\underline t'$, play the r\^ole of regulators. Of course, a similar interpretation could also be given to \eqref{zetatoZ}. 

\subsection{Analyticity properties}\label{analyticsec}

The analyticity properties of the functions $\zeta_{\dia}$ and $Z_{\dia}$ defined in the previous Section can be most easily derived from the integral representations \eqref{ktozeta} and \eqref{KtoZ}. Since the heat kernels are well-behaved at large $t$, the only possible singularities in $\zeta_{\dia}$ or $Z_{\dia}$ must come from the small $t$ region in the integrals and are thus determined by the small $t$ asymptotic expansions of $k_{\dia}(\underline\alpha t)$ and $K_{\dia}(\underline\alpha t)$. As we now explain, this implies that $\zeta_{\dia}$ and $Z_{\dia}$ have a simple analytic structure. They are meromorphic functions on the complex $s$-plane, with poles located on the real $s$-axis. 

\subsubsection*{The asymptotic expansion of $k_{\dia}(\underline\alpha t)$}

To compute the small $t$ asymptotic expansion of the kernel $k_{\dia}(\underline\alpha t)$, $\underline\alpha >0$ and $t>0$, we can proceed as follows. We start from the integral representation
\be\label{Knormtok} k_{\dia}(\underline\alpha t) = \int\!\d\Omega_{\dia}(\vec x,\vec y)\,\prod_{i=1}^{n}K(\alpha_{i}t,x_{i},y_{i})\, ,
\ee
which is the special case of \eqref{KggentoK} relevant for our purposes. This formula shows that the expansion we seek can be derived from the expansion \eqref{Kxyexp} of the standard heat kernel. When $t$ is small, the exponential damping factor in \eqref{Kxyexp} implies that the points $x_{i}$ and $y_{i}$ must be very close for all $i$. In the case of a connected diagram, this implies that all the spacetime points corresponding to the independent integration variables in \eqref{Knormtok} must also be very close to each other. These spacetime points are associated with the vertices $\mathsf V$ of the diagram and are denoted by $z_{\mathsf V}$. If we pick any particular vertex, say $\mathsf V_{0}$, and write
\be\label{newvaru}
z_{\mathsf V}=z_{\mathsf V_{0}} + \sqrt{t}\, u_{\mathsf V}\ee
for all the other vertices $\mathsf V\not = \mathsf V_{0}$, we can then expand all the coefficients $a_{k}$ and geodesic distances $\ell$ that enter into \eqref{Knormtok} in powers of the $u_{\mathsf V}$s. This calculation can be done efficiently by using, for example, Riemann normal coordinates around $z_{\mathsf V_{0}}$. The factor 
$\sqrt{t}$ has been inserted in \eqref{newvaru} in order to remove the $t$-dependence in the Gaussian weight coming from the geodesic distances in the exponentials. We then finish the calculation by performing the corresponding Gaussian integrals over the $u_{\mathsf V}$, $\mathsf V\not = \mathsf V_{0}$.

What is the general form of the expansion so obtained? We always get a factor of $t^{-nd/2}$, which comes from the $t^{-d/2}$ factor in each of the $n$ kernels in \eqref{Knormtok}. The change of variables \eqref{newvaru} also produces a factor $t^{(v-1)d/2}$, if $v$ is the total number of vertices in the diagram. Overall, we thus get a leading $t^{-(n-v+1)d/2} = t^{-Ld/2}$ factor, where
\be\label{Ldef} L = n-v +1\ee
is the number of loops in the diagram. The corrections to this leading factor come from the expansion in powers of the $u_{\mathsf V}$ via \eqref{newvaru}. Odd powers of $\sqrt{t}$ come with an odd number of $u_{\mathsf V}$ variables and thus vanish after performing the Gaussian integrals. Finally, we thus get
\be\label{kdiaexp} k_{\dia}(\underline\alpha t) = \frac{1}{(4\pi t)^{L d/2}}
\biggl(\sum_{k= 0}^{p}a^{\dia}_{k}(\underline\alpha) t^{k} + O\bigl(t^{p+1}\bigr)\biggr)\, .\ee
Each $a_{k}^{\dia}(\underline\alpha)$ is a reparameterization invariant spacetime integral (corresponding to the integral over the ``privileged'' vertex coordinate $z_{\mathsf V_{0}}$ in our derivation) of a local scalar polynomial of the components of the curvature tensor, which appear from the expansions around $z_{\mathsf V_{0}}$ of the geodesic distances and the coefficients $a_{k}(x,y)$ in \eqref{Kxyexp}. Moreover,
the coefficients $a_{k}^{\dia}$ depend on $\underline\alpha$ through simple rational functions (when $d$ is even) or square roots of rational functions (when $d$ is odd). They also satisfy
\be\label{scalealitk} a_{k}^{\dia}(w\underline\alpha) =w^{k-Ld/2} a_{k}^{\dia}(\underline\alpha)\, ,\quad w>0 \, ,\ee
a scaling law that follows from the invariance of $k_{\dia}(\underline\alpha t)$ under $\underline\alpha\rightarrow w\underline\alpha$, $t\rightarrow t/w$. Explicit examples of expansions \eqref{kdiaexp} will be given below, in particular in Section \ref{appsec}.
 
\subsubsection*{The analytic structure of $\zeta_{\dia}$}

The analyticity properties of $\zeta_{\dia}$ follow from a standard argument, using the integral representation \eqref{ktozeta} and the expansion \eqref{kdiaexp}. We write
\be\label{ankdia1} \zeta_{\dia}(s,\underline\alpha) = \frac{1}{\Gamma(s)}\int_{0}^{\epsilon}\!\d t \, t^{s+n-2} k_{\dia}(\underline\alpha t) +
\frac{1}{\Gamma(s)}\int_{\epsilon}^{\infty}\!\d t \, t^{s+n-2} k_{\dia}(\underline\alpha t)\, ,\ee
where $\epsilon >0$ is arbitrary. The second term on the right-hand side of \eqref{ankdia1} is non-singular and defines an entire function of $s$. The first term, on the other hand, can have singularities due to the small $t$ region of the integral. Using \eqref{kdiaexp}, we immediately find that $\zeta_{\dia}$ is meromorphic, with simple poles on the real $s$-axis.

Explicitly, if 
\be\label{superficialdef} \text{SDiv}_{\dia} = L d - 2n\ee
is the superficial degree of divergence of the diagram $\dia$, we find the following:\\
--- If $\text{SDiv}_{\dia}$ is even, which occurs if $d$ is even, or if $d$ is odd and $L$ is even, $\zeta_{\dia}$ has simple poles at $s=\frac{1}{2}
\text{SDiv}_{\dia} + 1 -k$ for $0\leq k\leq \frac{1}{2}\text{SDiv}_{\dia}$, with residues
\be\label{residkdia1} \res_{s= \frac{1}{2}\text{SDiv}_{\dia} + 1 -k}\zeta_{\dia}(s,\underline\alpha) = \frac{a_{k}^{\dia}(\underline\alpha)}{(4\pi)^{Ld/2}\Gamma(\frac{1}{2}\text{SDiv}_{\dia} + 1 -k)}\,\cvp\quad
0\leq k\leq \frac{1}{2}\text{SDiv}_{\dia}\, .\ee
Moreover, the would-be poles at negative integer values of $s$ and at $s=0$ are canceled by the poles of the $\Gamma$ function and we find
\be\label{zetadiaevenvalues} \zeta_{\dia}(-k,\underline\alpha) = (-1)^{k}k!
\frac{a_{\frac{1}{2}\text{SDiv}_{\dia} + 1 +k}^{\dia}(\underline\alpha)}{(4\pi)^{Ld/2}}\,\cvp\quad k\geq 0\, ,\quad \text{SDiv}_{\dia}\ \text{even.}
\ee
--- If $\text{SDiv}_{\dia}$ is odd, which occurs if both $d$ and $L$ are odd, there are simple poles at $s=\frac{1}{2}\text{SDiv}_{\dia} + 1 -k$ for all $k\geq 0$, with residues given by the same formula as in \eqref{residkdia1}. The pole structure of the $\Gamma$ function also yields in this case
\be\label{zetadiaoddvalues} \zeta_{\dia}(-k,\underline\alpha) = 0\, ,\quad k\geq 0\, ,\quad \text{SDiv}_{\dia}\ \text{odd.}\ee
In all cases, there is no pole for $\re s>\frac{1}{2}\text{SDiv}_{\dia} + 1$, showing that the series representation \eqref{zetaDseries} must converge in this domain.

These results are simple generalizations of the well-known properties of the standard spectral zeta function mentioned in \eqref{heatzetasec}. The zeta functions $\zeta_{\dia}$ are, in this sense, the simplest and most natural higher loop generalizations that one can consider. They capture, through their pole structure, interesting information associated with the diagram $\dia$, including the superficial diverging properties. However, this information is only partial beyond one loop. The full information is coded in the $Z$ functions, or in the associated heat kernels, to which we now turn.

\subsubsection*{The asymptotic expansion of $K_{\dia}(\underline\alpha t)$}

\noindent\emph{Basic properties}

\noindent It is tempting to try to use \eqref{ktoK}, 
\be\label{newktoK} K_{\dia}(\underline\alpha t) = \int_{\underline t'>\underline\alpha t}\!\d\underline t'\,k_{\dia}(\underline t')\, ,\ee
to derive the small $t$ asymptotic expansion of $K_{\dia}$
from the small $t$ asymptotic expansion of $k_{\dia}$. However, only partial results can be obtained in this way. Indeed, the integration over $\underline t'$ is unbounded from above and thus it is not correct, even at small $t$, to replace the integrand in \eqref{newktoK} by the expansion \eqref{kdiaexp}. 

The essence of the problem is already captured by the one-loop diagram. In this case, $k_{\diagrama}$ coincides with the standard heat kernel $K$ and we can write \eqref{newktoK} as
\be\label{exol1}K_{\diagrama}(\alpha t) = \int_{\alpha t}^{\epsilon}\d t'\,
K(t') + \int_{\epsilon}^{\infty}\d t'\,
K(t')\, ,\ee
for an arbitrary $\alpha t$-independent constant $\epsilon > \alpha t$. When $t$ is small, $\epsilon$ can also be chosen to be small and thus we can use the integrated version of \eqref{Kxexp} in the first term on the right-hand side of \eqref{exol1}. As for the second term, it cannot be computed explicitly in this way, but it is independent of $\alpha t$.
Assuming, for concreteness, that $d$ is even (we let the very similar case of $d$ odd to the reader), we get
\be\label{exol2} K_{\diagrama}(\alpha t) =\frac{1}{(4\pi)^{d/2}}
\sum_{\substack{0\leq k\leq p+\frac{d}{2} -1\\ k\not = \frac{d}{2}-1}}\frac{a_{k}}{\frac{d}{2}-k-1}(
\alpha t)^{k-\frac{d}{2}+1} - \frac{a_{\frac{d}{2}-1}}{(4\pi)^{d/2}}\ln (\alpha t) + C + O\bigl(t^{p+1}\bigr)\, ,\ee
for some $\alpha t$-independent constant $C$. The asymptotic expansion of $K_{\diagrama}$ is thus determined, to all orders, in terms of the similar expansion of $k_{\diagrama}=K$, \emph{except} for the term of order $O(1)$. The constant $C$ in \eqref{exol2} can actually be expressed in terms of the integrated Green's function at coinciding points defined in \eqref{Gzeta},
\be\label{Coneloopf} C = -\bigl(\ln\m^2+\gamma\bigr)\frac{a_{\frac{d}{2}-1}}{(4\pi)^{d/2}} + \int\!\d\nu (x)\, G_{\zeta}(x)\, .\ee
This formula can be derived by starting from \eqref{Zoneloop2} and \eqref{invMellinKdia} and computing the small $t$ asymptotics by closing the contour on the semi-infinite rectangle on the left, as usual.\footnote{Note that the $\ln\m^2$ in $C$ originates from the $\m^{2s-2}$ that we have chosen to include in the definition \eqref{Gzeta} of $G_\zeta$; it correctly combines with the $\ln t$ in \eqref{exol2} to give the logarithm of a dimensionless quantity.}

\smallskip

\noindent\emph{The hatted heat kernel}

\noindent To make a more general analysis, we start from the formula \eqref{ktoK} which, together with \eqref{Knormtok}, yields the integral representation
\be\label{KhattoK} K_{\dia}(\underline t) = 
\int\!\d\Omega_{\dia}(\vec x,\vec y) \prod_{i=1}^{n}\wh K(t_{i},x_{i},y_{i})\, ,\ee
where the hatted heat kernel function $\wh K$ is defined by
\be\label{khattoK} \wh K(t,x,y) = \int_{t}^{\infty}\!\d t'\, K(t',x,y) =\mathscr K(t,1,x,y)\, .\ee
Let us note that $\wh K$ evaluated at $t=0$ is nothing but the Green function \eqref{Greendef2}. Equation \eqref{KhattoK} is a good starting point to study the small $t$ expansion of $K_{\dia}(\underline\alpha t)$. It 
is to be compared with \eqref{Knormtok}, which played an analogous r\^ole for $k_{\dia}(\underline\alpha t)$. A basic tool that is needed is the small $t$ expansion of $\wh K$, which we now discuss.

This expansion can be obtained rather straightforwardly by writing
\be
\label{Khatsmallt} \wh K(t,x,y)  = \wh K(0,x,y) - \int_{0}^{t}\!\d t'\,
K(t',x,y) = G(x,y) - \int_{0}^{t}\!\d t'\,K(t',x,y)\, .\ee
One needs to assume that $x\not = y$ to write this equation, since otherwise both terms on the right-hand side are singular. At small $t$, we can then plug the standard heat kernel expansion \eqref{Kxyexp} into the second term on the right-hand side of \eqref{Khatsmallt} and perform the integrals over $t'$ to get
\begin{multline}\label{Khatexp}
\wh K(t,x,y)=G(x,y)-{1\over (4\pi)^{d/2}}\sum_{k=0}^n a_k(x,y) \,  t^{k-d/2+1} \, E_{k+2-d/2}\left( {\ell(x,y)^2\over 4 t}\right)\\+ 
O\bigl( t^{n+3-d/2} e^{-\ell(x,y)^{2}/(4 t)}\bigr)\, .
\end{multline}
The exponential integral functions were defined in \eqref{Endef}.

The above expansion is actually valid all the way down to $x=y$. To understand this point, we can directly evaluate the small $t$ expansion of $\wh K$ at $x=y$. This is done by using \eqref{zetatoK} for $s=1$ and $x=y$ and closing the contour of integration on the semi-infinite rectangle on the left as usual. Picking all the residues of the integrand, using in particular \eqref{residueeven}, \eqref{zetaevenneg} and \eqref{zetaoddneg}, we get 
\be\label{Khatxx}\wh K(t,x) \underset{t\rightarrow 0}{=}\begin{cases}\displaystyle
\frac{1}{(4\pi)^{d/2}} \sum_{k\geq 1}\frac{a_{k-1}(x)}{d/2-k}\, t^{k-\frac{d}{2}} + G_{\zeta}(x)\, ,\ \text{$d$ odd}\\\displaystyle
\frac{1}{(4\pi)^{d/2}}\sum_{\substack{k\geq 1\\k\not = d/2}}
\frac{a_{k-1}(x)}{d/2-k}\, t^{k-\frac{d}{2}} -\frac{a_{d/2-1}(x)}{(4\pi)^{d/2}}\bigl(\gamma + \ln (\mu^{2}t)\bigr) + G_{\zeta}(x)\, ,\ \text{$d$ even.}
\end{cases}\ee
It is then not difficult to check, by using \eqref{Gspliteven}, \eqref{Gsplitodd} and the expansion \eqref{Enexpansions}, that \eqref{Khatxx} is consistent with the $x\rightarrow y$ limit of \eqref{Khatexp}.

A simple and typical illustration of the use of the expansion \eqref{Khatexp} is to compute the large cutoff asymptotics of the following spacetime integrals involving the regularized Green's function
\be\label{exampleI1} i_{1}(x) = \int\!\d\nu(y)\, G_{f,\La}(x,y)\, ,\quad
i_{2}(x) = \int\!\d\nu(y)\, R(y) G_{f,\La}(x,y)\, .\ee
Since
\be\label{GfandKhat} G_{f,\La}(x,y) = \int_{0}^{\infty}\!\d\alpha\,\varphi(\alpha)\wh K\bigl(t=\alpha/\La^{2},x,y\bigr)\, ,\ee
we can write at large $\La$, using \eqref{Khatexp} and $t=\alpha/\La^{2}$,
\begin{multline}\label{i1calcuex1} i_{1}(x) = \int\!\d\nu(y) G(x,y) -
\frac{1}{(4\pi)^{d/2}}\int_{0}^{\infty}\!\d\alpha\,\varphi(\alpha)\\
\int\!\d\nu(y)\sum_{k\geq 0}a_{k}(x,y) t^{k-d/2+1}
E_{k+2-d/2}\Bigl(\frac{\ell(x,y)^{2}}{4t}\Bigr)\, .\end{multline}
The integrals involving the exponential integral functions can be done explicitly at small $t$ by using Riemann normal coordinates around $x$ and making the change of variables $y = x + \sqrt{2t}\, y'$. Using \eqref{ardiag}, \eqref{frmfr} from the Appendix, we find in this way
\be\label{i1expI1} i_{1}(x) = \int\!\d\nu(y)\ G(x,y) + \La^{-2}f'(0) + \frac{m^{2} + \xi R(x)}{2}\La^{-4} f''(0) + O\bigl(\La^{-6}\bigr)\ee
and similarly
\be\label{i2expI2} i_{2} (x) = \int\!\d\nu(y)\ R(y) G(x,y) + R(x)\La^{-2}f'(0) + O\bigl(\La^{-4}\bigr)\, .\ee

\smallskip

\noindent\emph{The expansion of $K_{\dia}$}

The small $t$ expansion of $K_{\dia}(\underline\alpha t)$ can be studied systematically by using \eqref{KhattoK} and \eqref{Khatexp}, for any higher loop diagram $\dia$. The divergences can be understood as coming from various regions of the moduli space of the diagram, where some internal lines are short while others are large. The analysis of the contributions of these various regions is, not surprisingly, reminiscent of the standard analysis of divergences in Feynman diagrams due to divergent subgraphs. Each shrinking subgraph yields an expansion with local coefficients which may be analyzed from the standard heat kernel expansion \eqref{Kxyexp}, whereas the long internal lines yield contributions which can be analyzed with the help of \eqref{Khatexp} and which contain non-local coefficients. Two-loop diagrams will be studied in full details in Section \ref{appsec}.

The general structure that emerges is not difficult to work out. The region where all the internal lines are short yields a small $t$ behavior $\sim 
t^{-\frac{1}{2}\text{SDiv}_{\dia}}$ governed by the superficial degree of divergence \eqref{superficialdef} of the diagram $\dia$. The coefficients of this piece of the expansion are integrals of local polynomials in the curvature. The contributions from the regions where some internal lines are short while others are large can change the leading small $t$ behavior to $\sim t^{-\frac{1}{2}\text{Div}_{\dia}}$, where $\text{Div}_{\dia}\geq \text{SDiv}_{\dia}$ is the genuine degree of divergence of the diagram, defined such that $\text{Div}_{\dia}=0$ if the diagram is convergent (a typical diagram having $\text{Div}_{\dia}>\text{SDiv}_{\dia}$ is depicted on the lower right corner of Fig.\ \ref{figdiagram}). This piece of the expansion can also contain logarithms, that come from the integration over the moduli $t'_{i}$ and from the short distance logarithmic singularities of the Green functions, see \eqref{Gspliteven}. All these features will appear explicitly in the examples studied in Section \ref{appsec}. 
Overall, we can write
\be\label{Kdiaexp} K_{\dia}(\underline\alpha t) = \frac{1}{(4\pi t)^{\frac{1}{2}\text{Div}_{\dia}}}\sum_{k,q\geq 0} A_{k,q}^{\dia}(\underline\alpha) t^{k}(\ln t)^{q}\, .\ee
The coefficients $A_{k,q}$ are in general integrals of non-local functionals of the metric. Another useful basis of coefficients is defined by rewriting \eqref{Kdiaexp} as
\be\label{Kdiaexpbis} K_{\dia}(\underline\alpha t) = \frac{1}{(4\pi t)^{\frac{1}{2}\text{Div}_{\dia}}}\sum_{k,q\geq 0}
\tilde A_{k,q}^{\dia}(\underline\alpha) t^{k}\ln^{q} (\alpha t) \, .\ee
The factor $\alpha=\sum_{i}\alpha_{i}$ has been inserted in the argument of the logarithm in such a way that the coefficients so obtained satisfy the simple scaling law
\be\label{scaleA} \tilde A_{k,q}^{\dia}(w\underline\alpha) = w^{k-\frac{1}{2}\text{Div}_{\dia}} \tilde A_{k,q}^{\dia}(\underline\alpha)\, , \quad w>0\, .\ee
The two basis are simply related to each other,
\be\label{AtildeArel} A_{k,q}^{\dia}(\underline\alpha) = \sum_{p=0}^{L-q}
C_{q+p}^{q}\tilde A_{k,q+p}^{\dia}(\underline\alpha)\ln^{p}\alpha\, .\ee

\subsubsection*{The analytic structure of $Z_{\dia}$}

The derivation of the analytic structure of $Z_{\dia}$ from the asymptotic expansion \eqref{Kdiaexp} proceeds along the same lines as the derivation of the analytic structure of $\zeta_{\dia}$ from \eqref{kdiaexp}. The starting point of the analysis is the integral representation \eqref{KtoZ}, which shows that any singularity in $Z_{\dia}$ must come from the small $t$ behavior of $K_{\dia}(\underline\alpha t)$. The singular piece in $Z_{\dia}$ is thus given by
\begin{multline}\label{Zsingular}\frac{1}{(4\pi)^{\frac{1}{2}\text{Div}_{\dia}}}
\sum_{k\geq 0}\sum_{q=0}^{L}A_{k,q}^{\dia}(\underline\alpha)
 \frac{1}{\Gamma (s+1)}\int_{0}^{\epsilon}\!\d t \,
t^{s-1-\frac{1}{2}\text{Div}_{\dia}+k}\ln^{q}t =\\
\frac{1}{(4\pi)^{\frac{1}{2}\text{Div}_{\dia}}}
\sum_{k\geq 0}\sum_{q=0}^{L}A_{k,q}^{\dia}(\underline\alpha)
\frac{1}{\Gamma (s+1)}
\frac{q! (-1)^{q}}{\bigl(s-\frac{1}{2}\text{Div}_{\dia}+k\bigr)^{q+1}}+ \cdots
\end{multline}
where the $\cdots$ represent non-singular terms. We see that multiple poles can occur. They are associated with the logarithms in the expansion \eqref{Kdiaexp}. For example, the logarithm in \eqref{exol2} yields the double pole at $s=0$ of $Z_{\diagrama}$, which is manifest on the formula \eqref{Zoneloop2}. 

In conclusion, $Z_{\dia}$ is a meromorphic function on the complex $s$-plane. It can be defined by the series \eqref{Zdef1}, which converges as long as $\re s > \frac{1}{2}\text{Div}_{\dia}$. Generically, it has multiple poles on the real $s$-axis at $s_{k}=\frac{1}{2}\text{Div}_{\dia}-k$, $k\in\mathbb N$. The precise structure of the Laurent expansion around $s_{k}$ can be straightforwardly read off from \eqref{Zsingular}, by expanding the factor $1/\Gamma(s+1)$. 

\subsection{Feynman amplitudes}\label{ampsec}

\subsubsection*{Feynman amplitudes, $Z$ functions and generalized heat kernels}

Let us now show that the physical amplitude $\mathscr A^{\dia}$ associated with the Feynman diagram $\dia$ can be easily computed from the asymptotic expansion \eqref{Kdiaexp} of $K_{\dia}$ or equivalently from the pole structure of the function $Z_{\dia}$. 

The regularized amplitude $\mathscr A^{\dia}_{f,\La}$ is defined by associating the regularized Green's function \eqref{Greensmooth} to any internal line in the diagram,
\be\label{Amp1}\mathscr A^{\dia}_{f,\La} = \int\!\d\Omega (\vec x,\vec y)\, 
\prod_{i=1}^{n}G_{f,\La}(x_{i},y_{i})\, .\ee
Note that, for convenience, we do not include numerical combinatorial factors in our definition of the amplitude. Using \eqref{Laplacef} and the definition \eqref{Kgendef}, we can rewrite \eqref{Amp1} as
\be\label{Amp2} \mathscr A^{\dia}_{f,\La}  = \int_{\underline\alpha >0}
\!\d\underline\alpha\,\varphi(\alpha_{1})\cdots\varphi(\alpha_{n})
K_{\dia}\bigl(\underline\alpha /\La^{2}\bigr)\, .\ee
In the physical applications, only the large cutoff expansion of the amplitude is needed. Equation \eqref{Amp2} being exactly of the form \eqref{intcoex}, we know, from the discussion of Section \ref{cutpropsec}, that this expansion
can be obtained by plugging \eqref{Kdiaexp} directly into \eqref{Amp2}, with $t=1/\La^{2}$,
\be\label{Ampexp}\mathscr A^{\dia}_{f,\La}  
\underset{\La\rightarrow\infty}{=}
\frac{1}{(4\pi)^{\frac{1}{2}\text{Div}_{\dia}}}\sum_{k,q\geq 0}
\La^{\text{Div}_{\dia}-2k}\bigl(\ln\La^{-2}\bigr)^{q}
\int_{\underline\alpha>0}\!\d\underline\alpha\,\varphi(\alpha_{1})\cdots
\varphi(\alpha_{n})\, A_{k,q}^{\dia}(\underline\alpha)\, .
\ee
In particular, if the diagram is divergent, the leading divergence will be proportional to $\La^{\text{Div}_{\dia}}$ times a power of $\ln\La$. This is precisely the definition of the degree of divergence of the diagram and justifies a posteriori the leading small $t$ behavior of $K_{\dia}(\underline\alpha t)$ given in \eqref{Kdiaexp}.

Another elegant representation of the asymptotic expansion is obtained by using the integral representation \eqref{invMellinKdia} with $t=1/\La^{2}$,
\be\label{Amp3} \mathscr A^{\dia}_{f,\La} = \frac{1}{2i\pi}\int_{c-i\infty}^{c+i\infty}\!\d s\,\Gamma(s)s \La^{2s}\,
\int_{\underline\alpha>0}\!\varphi(\alpha_{1})\cdots\varphi(\alpha_{n})\, Z_{\dia}(s,\underline\alpha)\, .\ee
We can use the standard trick of closing the contour of integration on the semi-infinite rectangle on the left to obtain the large $\La$ asymptotic expansion of \eqref{Amp3}. The poles of $Z_{\dia}$ on the positive real axis yield the diverging piece of the expansion whereas the poles on the negative real axis yield subleading contributions. 

\subsubsection*{Finite amplitude, physical amplitude and $\varphi$-dependence}

If the amplitude is finite, we have
\be\label{Aiffinite}\mathscr A^{\dia} = A_{0,0}^{\dia}=K_{\dia}(0)\, .\ee
Equivalently, to a finite diagram is associated a function $Z_{\dia}$ with no pole at $s>0$ and a simple pole at $s=0$ of residue $\mathscr A^{\dia}$.

If the amplitude is diverging, but $\text{Div}_{\dia}$ is even, we may define the finite part by the equation
\be\label{Afindef} \mathscr A^{\dia}_{\text{finite}} \overset{?}{=} 
\int_{\underline\alpha>0}\!\d\underline\alpha\,\varphi(\alpha_{1})\cdots
\varphi(\alpha_{n})\, A_{\text{Div}_{\dia}/2,0}^{\dia}(\underline\alpha)\, .\ee
Another natural definition may be to set it equal to the residue of $Z_{\dia}$ at $s=0$,
\be\label{Afindefbis} \mathscr A^{\dia}_{\text{finite}} \overset{?}{=}
\int_{\underline\alpha>0}\!\d\underline\alpha\,\varphi(\alpha_{1})\cdots
\varphi(\alpha_{n})\,\res_{s=0}Z_{\dia}(s,\underline\alpha)\, .\ee
The two definitions can be easily related by using \eqref{Zsingular}. Let us emphasize, however, that there is no a priori direct relationship between the finite parts defined in this way and the physical amplitudes. For example, the finite parts \eqref{Afindef} or \eqref{Afindefbis} depend on the regulator $\varphi$, because $A_{\text{Div}_{\dia}/2,0}^{\dia}$ or $\res_{s=0}Z_{\dia}$ depend on $\underline\alpha$.

When computing a physical observable, for example the gravitational effective action, one must sum up various contributions, including from infinite counterterms in the action that multiply the Feynman amplitudes. In particular, even poles of the $Z$ functions on the negative real axis will play a r\^ole. The end result must always be finite and cutoff-independent, if the theory is renormalizable.

In curved space, renormalizability has even more depth than in flat space. The infinities must be absorbed in counterterms that are local not only with respect to the dynamical field but also with respect to the background metric. Also, new counterterms are needed in curved space because more relevant local operators can be built by using the metric on top of the dynamical fields. In particular, the renormalizability in flat space does not imply the renormalizability in curved space and rather few rigorous results seem to be established in this case \cite{curvedrenorm}.
 
Moreover, in our general formalism, the dependence on the arbitrary cutoff function $\varphi$ yields an additional constraint (or consistency check) that is usually not available: the physical amplitudes must not only be made finite by the addition of the local counterterms, but all the dependence on $\varphi$ must also drop from the final result. In some sense, $\varphi$ plays the r\^ole of a ``sliding function'' generalizing the sliding scale of the standard renormalization group. It would be interesting to investigate in details the $\varphi$-dependence and the associated renormalization group equations and, even more generally, the general consequences of the renormalizability properties of field theory on the analytic functions $Z$.

We shall illustrate all these features in details in the next Section at two loops. At this order, one can show that the simple definitions \eqref{Afindef} or \eqref{Afindefbis} actually do correspond to the finite physical piece in the amplitudes. All the cutoff dependence in these definitions can be directly canceled by finite shifts in the local counterterms. This yields a very simple prescription for the two-loop gravitational effective action in terms of the $Z$ functions, see Eq.\ \eqref{finalformula}, which naturally generalizes the traditional one-loop result $-\frac{1}{2}\zeta_{D}'(0)$.

\subsection{Summary}\label{summarySec}

The regularized Feynman amplitudes associated to a Feynman diagram $\dia$ can be expressed equivalently either in terms of $Z_{\dia}$ by \eqref{Amp3} or in terms of $K_{\dia}$ by \eqref{Amp2}.
The large cutoff expansion of the amplitudes is governed by the pole structure \eqref{Zsingular} of $Z_{\dia}$ or equivalently by the asymptotic expansion \eqref{Kdiaexp} of $K_{\dia}$. This asymptotic expansion can often be conveniently studied from the integral representation \eqref{KhattoK} by using \eqref{Khatexp}.

\subsection{Flat space examples}\label{flatexamplessec}

Before we turn to the full-fledged applications in curved space in the next Section, it is instructive to illustrate the general formalism for a few one and two-loop diagrams in the simple case of the massive scalar field in flat space. 
Up to two loops, we have to consider only the four diagrams $\diagramabig$, $\diagramcbig$, $\diagramdbig$ and $\diagrambbig$. 
Calculations in flat space are greatly simplified by the fact that a simple formula is available for the standard heat kernel, valid for all $t$,
\be\label{Kflat} K(t,x,y) = \frac{e^{-m^{2}t-\frac{|x-y|^{2}}{4 t}}}{(4\pi t)^{d/2}}\,\cdotp\ee
Simple explicit formulas can then be found for many of the quantities introduced previously. For instance, using \eqref{Ktozeta}, \eqref{khattoK}, \eqref{Endef} and \eqref{Gzeta}, we get
\begin{align}\label{zetaxflat} \zeta(s,x) &= \frac{m^{d-2s}}{(4\pi)^{d/2}}\frac{\Gamma(s-d/2)}{\Gamma(s)}\,\cvp\\
\label{Khatflata}\wh K(t,x) &={t^{1-d/2}\over (4\pi)^{d/2}} E_{d/2}(tm^{2})\, ,\\\label{Gzetaflat}
G_{\zeta}(x) & = \begin{cases}\displaystyle
\frac{m^{d-2}}{(4\pi)^{d/2}\Gamma(1-d/2)}\ \text{if $d$ is odd,}\\ 
\displaystyle
\frac{(-1)^{\frac{d}{2}-1}m^{d-2}}{(4\pi)^{d/2}(d/2 -1)!}
\Bigl(\gamma + \Psi(d/2) + \ln\frac{\mu^{2}}{m^{2}}\Bigr)\ \text{if $d$ is even,}
\end{cases}
\end{align}
and more examples are given below.

We note $V$ the (infinite) volume of spacetime.

\subsubsection*{The functions $k_{\dia}$ and $\zeta_{\dia}$ in flat space}

The generalized heat kernels $k_{\dia}$ and Barnes spectral zeta functions $\zeta_{\dia}$ can always be found explicitly in flat space, for any diagram $\dia$. Indeed, 
\begin{align} \label{kfl1} k_{\dia}(\underline t)& = \int\!\d\Omega_{\dia}
\,\prod_{i=1}^{n} K(t_{i},x_{i},y_{i})\\\label{kfl2} & =
\frac{1}{(4\pi)^{nd/2}}\frac{e^{-m^{2} t}}{(t_{1}\cdots t_{n})^{d/2}}
\int\!\d\Omega_{\dia}\, e^{-\sum_{i=1}^{n} \frac{|x_{i}-y_{i}|^{2}}{4 t_{i}}}
\end{align}
is given by a simple Gaussian integral. For example,
\begin{align}\label{kflex0} k_{\diagrama}(t) & = \frac{V}{(4\pi)^{d/2}}
\frac{e^{-m^{2}t}}{t^{d/2}}\,\cvp\\
\label{kflex1} k_{\diagramc}(\underline t) &= \frac{V}{(4\pi)^{d}}\frac{e^{-m^{2}t}}{(t_{1}t_{2})^{d/2}}\,\cvp\\\label{kflex1bis}
k_{\diagramd}(\underline t) &= \frac{V}{(4\pi)^{d}}\frac{e^{-m^{2}t}}{(t_{2}t_{3})^{d/2}}\,\cvp\\
\label{kflex2}
k_{\diagramb}(\underline t) & = \frac{V}{(4\pi)^{d}}\frac{e^{-m^{2} t}}{(t_{1}t_{2}+t_{2}t_{3}+t_{3}t_{1})^{d/2}}\, ,\ \text{etc...}
\end{align}
The associated zeta functions are then obtained effortlessly from \eqref{ktozeta},
\begin{align}\label{zetaflex0} \zeta_{\diagrama}(s,\underline\alpha) & =
\frac{V}{(4\pi)^{d/2}}\alpha^{-s}m^{d-2s}\frac{\Gamma(s-d/2)}{\Gamma(s)}\,\cvp\\
\label{zetaflex1} \zeta_{\diagramc}(s,\underline\alpha) & =
\frac{V}{(4\pi)^{d}}\frac{(m^{2}\alpha)^{d-s-1}}{(\alpha_{1}\alpha_{2})^{d/2}}\frac{\Gamma(s+1-d)}{\Gamma(s)}\,\cvp\\\label{zetaflex1b}
\zeta_{\diagramd}(s,\underline\alpha) & =
\frac{V}{(4\pi)^{d}}\frac{(m^{2}\alpha)^{d-s-2}}{(\alpha_{2}\alpha_{3})^{d/2}}\frac{\Gamma(s+2-d)}{\Gamma(s)}\, ,\\
\label{zetaflex2}
\zeta_{\diagramb}(s,\underline\alpha) & =
\frac{V}{(4\pi)^{d}}\frac{(m^{2}\alpha)^{d-s-2}}{(\alpha_{1}\alpha_{2} + 
\alpha_{2}\alpha_{3}+\alpha_{3}\alpha_{1})^{d/2}}\frac{\Gamma(s+2-d)}{\Gamma(s)}\, ,\ \text{etc...}
\end{align}
More generally, the heat kernel and zeta functions are of the form (recall that $\text{SDiv}_{\dia}=L d-2 n$)
\begin{align}\label{kflgen} k_{\dia}(\underline\alpha t)  &= \frac{V}{(4\pi)^{Ld/2}}\frac{e^{-m^{2}\alpha t}}{t^{Ld/2}} f_{\dia}(\underline\alpha)^{d/2}\, ,\\\label{zetaflgen} \zeta_{\dia}(s,\underline\alpha) &= 
\frac{V}{(4\pi)^{Ld/2}} f_{\dia}(\underline\alpha)^{d/2}
(m^{2}\alpha)^{\frac{1}{2}\text{SDiv}_{\dia}+1-s}\,
\frac{\Gamma\bigl(s-\frac{1}{2}\text{SDiv}_{\dia}-1\bigr)}{\Gamma(s)}\,\cvp
\end{align}
for some rational function $f_{\dia}(\underline\alpha)$.

\subsubsection*{The functions $K_{\dia}$ and $Z_{\dia}$ up to two loops}

At one loop, the formulas \eqref{ktoK} and \eqref{zetatoZ} (or \eqref{Zoneloop2} and \eqref{Khatflata}) immediately yield
\begin{align}\label{Konelfl} K_{\diagrama}(t) & = \int\!\d x\,  \wh K(t,x)= {V\over (4\pi)^{d/2}}\,  t^{1-d/2}E_{d/2}(tm^{2})\, ,\\\label{Zonelfl}
Z_{\diagrama}(s,\alpha) & = \frac{V}{(4\pi)^{d/2}}\alpha^{-s}m^{d-2s-2}\frac{\Gamma(s+1-d/2)}{s\Gamma(s+1)}\,\cdotp
\end{align}
The asymptotic expansion \eqref{exol2} can then be computed explicitly from the well-known expansion of $E_n(z)$ or equivalently from the pole structure of \eqref{Zonelfl}. Either way, we find
\begin{align}\label{Amponelfld=2} \mathscr A^{\diagrama}_{f,\La}\big\vert_{d=2}  
& = \frac{V}{4\pi} \Bigl( \ln{\Lambda^2\over m^2}-\g - 
\int_{0}^{\infty}\!\!\d\a\, \vf(\a) \ln\a\Bigr)\, ,\\\label{Amponelfld=4}
\mathscr A^{\diagrama}_{f,\La}\big\vert_{d=4} & = \frac{V\L^2}{(4\pi)^2} \, \int_{0}^{\infty}\!\!\d\alpha\,\frac{\varphi(\alpha)}{\alpha} 
-  \frac{V m^2}{(4\pi)^2} \Bigl( \ln{\Lambda^2\over m^2}+1-\g -\int_{0}^{\infty}\!\d\a\, \vf(\a) \ln\a\Bigr)
\ .
\end{align}
Of course, in the present flat space case, all terms are necessarily proportional to the volume $V$.

The two-loop diagram $\diagramcbig$ factorizes,
\begin{align}\label{k1lto2l} k_{\diagramc}(t_{1},t_{2}) &= \frac{1}{V}k_{\diagrama}(t_{1})k_{\diagrama}(t_{2})\, ,\\
\label{K1lto2l} K_{\diagramc}(t_{1},t_{2}) &= \frac{1}{V}K_{\diagrama}(t_{1})K_{\diagrama}(t_{2}) \, ,\\\label{Amp2lspe}
\mathscr A^{\,\diagramc}_{f,\La} & =\frac{1}{V} \bigl(\mathscr A^{\,\diagrama}_{f,\La}\bigr)^{2}\, ,
\end{align}
and is thus trivially determined in terms of the one-loop diagram in flat space. There is no such simple factorization formula for $Z_{\diagramc}$, but all we need is the amplitude, or the pole structure of $Z_{\diagramc}$, and these are completely fixed by \eqref{k1lto2l}--\eqref{Amp2lspe} (if nonetheless desired, an explicit formula in terms of special functions can easily be found for $Z_{\diagramc}$ by using either \eqref{KtoZ} or \eqref{zetatoZ}). Let us note that in curved space, this simple factorization property does not hold and an independent analysis, which will be performed in the next section, is then required. The same remarks can be made for $\diagramdbig$, for which
\begin{align}\label{kdd1to2}
& k_{\diagramd}(t_{1},t_{2},t_{3})  = \frac{1}{V}
e^{-m^{2}t_{1}}k_{\diagrama}(t_{2})k_{\diagrama}(t_{3})\, ,\\
\label{Kdd1to2} & K_{\diagramd}(t_{1},t_{2},t_{3})  = \frac{1}{V}
\frac{e^{-m^{2}t_{1}}}{m^{2}}K_{\diagrama}(t_{2})K_{\diagrama}(t_{3})\, ,\\
\label{Amp3lspe}&\mathscr A^{\diagramd}_{f,\La}  = \frac{1}{Vm^{2}}
\bigl(\mathscr A^{\,\diagrama}_{f,\La}\bigr)^{2}\, \int_{0}^{\infty}\!\d\alpha\,\varphi(\alpha)e^{-m^{2}\alpha/\La^{2}}\, .
\end{align}
Note that, despite their appearance, the amplitudes \eqref{Amp2lspe} and \eqref{Amp3lspe} are again proportional to the volume $V$.

The most interesting two-loop diagram is $\diagrambbig$. It has some of the generic properties of higher loop diagrams, including an overlapping divergence. We focus on the $d=4$ case, for which
\be\label{Zdiagrambd4} Z_{\diagramb}(s,\underline \a)\big\vert_{d=4}= \frac{V}{(4\pi)^4} \, 
\frac{m^{2-2s}}{s(s-1)}\ I_4(s,\underline \a)\, ,\ee
where the integral $I_{4}$ is defined and studied in App.\ \ref{a4app}.
In particular, for $s=1$, we have the explicit expression
\be\label{Iintsequal1}
I_4(1,\underline \a)=\sum_{i>j=1}^3 \left({1\over \a_i}+{1\over \a_j}\right) \ln(\a_i+\a_j) - \left(\sum_{k=1}^3 {1\over \a_k}\right)\ln\left(\sum_{i>j=1}^3 \a_i\a_j\right) \ .
\ee
Combining the results \eqref{Iint4b}-\eqref{Iint5} from the appendix, we get
\begin{align}\label{Zdiagrambd4bis}
Z_{\diagramb}(s,\underline \a)\big\vert_{d=4}= \frac{V}{(4\pi)^4} \, \Bigg[&
 m^{2-2s} (1+s+s^2)\left( -{3\over s^3} + {\sum_{i>j}\ln(\a_i+\a_j) \over s^2} - {I_4^{\rm reg}(0,\underline\a)\over s}\right) 
\nonumber\\
& + \frac{I_4(1,\underline\a)}{s-1}\Bigg] 
 + \text{poles at $s\le -1$}+ \ \text{regular.}
\end{align}
To get the large $\Lambda$ expansion of the amplitude, we have to insert this into \eqref{Amp3} and use $s \Gamma(s)= 1-\g s +\left({\g^2\over 2}+{\pi^2\over 12}\right) s^2 +O(s^3)$,
\begin{multline}\label{Ampdiagbd4}
\mathscr A^{\,\diagramb}_{f,\Lambda}\big\vert_{d=4}= \frac{V m^2}{(4\pi)^4}  \int_{\underline\a\ge 0} \vf(\a_1)\vf(\a_2)\vf(\a_3)\ \Biggl[
I_4(1,\underline\a) \frac{\Lambda^2}{m^2} 
- {3\over 2} \left(\ln{\Lambda^2\over m^2}\right)^2
\\ +\left(  \sum_{i>j}\ln(\a_i+\a_j)-3+3\g\right) \ln{\Lambda^2\over m^2}
\\+(1-\g)  \sum_{i>j}\ln(\a_i+\a_j) - I_4^{\rm reg}(0,\underline\a)
- 3+3\g -{3\g^2\over 2} -{\pi^2\over 4}
\Biggr] +O(1/\Lambda^2)\ .
\end{multline}
As expected, we find a quadratic divergence associated with the simple pole at $s=1$, as well as a $(\ln\Lambda^2)^2$ and a $\ln\Lambda^2$ divergence from the triple pole at $s=0$ of $Z_{\diagramb}$. 

It is also instructive to check \eqref{Ampdiagbd4} from the small $t$ expansion of $K_{\diagramb}$, which can itself be obtained by using the integral representation \eqref{KhattoK}
\be\label{Khatflat33} K_{\diagramb}(\underline t) = V\int\!\d^{4} x\, 
\wh K(t_{1},x,0) \wh K(t_{2},x,0) \wh K(t_{3},x,0) \ee
and the expansion \eqref{Khatexp} of $\wh K$ which, using the fact that the flat space Green function can be expressed in terms of the Bessel function $K_{1}$, reads
\be\label{Khatflat3}
\wh K(t,x,0)={1\over 4 \pi^2}\biggl( \sqrt{m^2\over x^2} K_1\bigl(m\sqrt{x^2}\bigr)-{e^{-x^2/(4t)}\over x^2}+{m^2\over 4} E_1\bigl(x^2/(4t)\bigr)+O(t)\Bigr) \ .\ee
We have not tried to analytically perform the integral \eqref{Khatflat33} with \eqref{Khatflat3} inserted, but we have checked numerically the perfect agreement with the square bracket in \eqref{Ampdiagbd4}.

\section{\label{appsec} The gravitational effective action at two loops}

We are now going to apply the formalism developed in the previous sections to compute in full details the two-loop gravitational effective action of the general scalar field theory with cubic and quartic interactions. When the value of the spacetime dimension matters, we will focus on $d=4$, but adapting our computations to other dimensions is completely straightforward.
In the case of the classically Weyl invariant model, the result will allow us to compute the two-loop conformal anomaly. Our main goal is to illustrate our general framework on a simple, natural example. Let us mention that two-loop calculations for the scalar field theory on a curved background have been done before using entirely different methods, e.g.\ in \cite{Toms} using dimensional regularization.

\subsection{General analysis}\label{twoLgenSec}

\subsubsection*{The set-up}

We consider the free model \eqref{Action} to which we add $\phi^{3}$ and $\phi^{4}$ interactions, on a Riemannian manifold $\mathscr M$ endowed with a fixed background metric $g$. For concreteness, we assume that $\mathscr M$ is compact, even though most of our discussion does not depend on this assumption.

The most general power-counting renormalizable and reparameterization invariant action of this type reads, in four dimensions, 
\begin{multline}\label{Actionphi} S_{\phi} 
=\int\!\d^{4}x\sqrt{g}\,\Bigl( \frac{1}{2}Z\phi\Delta\phi + \frac{1}{2}Z Z_{m}m^{2}\phi^{2} + \frac{1}{2}Z Z_{\xi}\xi R\phi^{2}  \\ 
+ Z^{1/2} \, (c_1+ \tilde c_1 R)\, \phi
+ \frac{1}{6} Z^{3/2} Z_3 \k_3\phi^{3} + \frac{1}{24} Z^{2} Z_4\k_4\phi^{4}\Bigr)\, ,
\end{multline}
where we called $\k_3$ and $\k_4$ the cubic and quartic couplings. Note that the $\phi^3$ coupling leads to diverging tadpole diagrams, making it necessary to include corresponding counterterms proportional to $c_1\phi$ and $\tilde c_1 R \phi$.
There is also a purely gravitational part $S_{g}$ in the action. If $C$ is the Weyl tensor and $\chi_{\text E}$ the Euler characteristic of the spacetime manifold $\mathscr M$, which is proportional to the integral of the local  
density \eqref{E4def}, we can write
\be\label{Actiong} S_{g} =\frac{1}{16\pi G} \int\!\d^{4}x\sqrt{g}\,
\Bigl( -2 Z_{\la}\la^{2} + Z_G R + Z_2 G\k_2 R^{2} + Z_C G\kappa_C C^{ijkl}C_{ijkl}\Bigr) + Z_E\kappa_E \chi_{\text E}\, .
\ee
The mass $m$, the couplings $\xi$, $\k_2$, $\k_{3}$, $\k_4$, $\k_C$, $\k_E$, Newton's constant $G$ and cosmological constant $\la^{2}$ are the minimal set of physical parameters of the model. Let us stress that considering a quantum interacting scalar field in curved space without, e.g., the $R\phi^{2}$ term, or with a classical gravitational action containing only the usual Einstein's terms, is not allowed quantum mechanically. All the terms in \eqref{Actionphi} and \eqref{Actiong} are required to make sense of the quantum corrections (on a non-compact manifold, an additional term proportional to $\Delta R$ must be added to the action). 

The dimensionless $Z$, $Z_{m}$, $Z_{\xi}$, $Z_G$, $Z_2$, $Z_{3}$, $Z_4$, $Z_{\la}$, $Z_C$ and $Z_E$ are cutoff-dependent renormalization constants, as are $c_1$ and $\wt c_1$. Power counting indicates that $Z$, $Z_{\xi}$, $Z_3$, $Z_4$, $Z_2$, $Z_G$, $Z_C$, $Z_E$ and $\wt c_1$ can be logarithmically divergent when the cutoff goes to infinity, whereas $c_1$ and $Z_{m}$ can be quadratically and $Z_{\lambda}$ quartically divergent. 

The gravitational effective action we seek is formally defined by the path integral
\be\label{Sgravdef} e^{-\frac{1}{\hbar}S(g)} =\int\!\mathscr D\phi\, e^{-\frac{1}{\hbar}(S_{\phi} + S_{g})}\, .\ee
As usual, $\hbar$  provides a loop-counting parameter, as can be seen by rescaling $\phi\rightarrow\sqrt{\hbar}\,\phi$.

The renormalization constants, or, more conveniently, the dimensionless coefficients $c_{\phi}$, $c_{m}$, $c_{\xi}$, $c_2$, $c_{3}$, $c_4$, $c_\l$, $c_C$ and $c_E$ defined by the relations
\begin{multline}\label{csdefs}  
Z = 1 +  c_{\phi}\, ,\ 
ZZ_{m} = 1 +  c_{m}\, ,\
ZZ_{\xi} = 1 +  c_{\xi}\, ,\ 
Z^{3/2}  Z_3 = 1 +  c_3\, ,\
Z^{2}Z_4 = 1 + c_4\, , \\ 
Z_{\la} = 1 + c_{\la}\, ,\ 
Z_G = 1 + c_G\, ,\ Z_2 = 1 +  c_2\, ,\
Z_C = 1 + c_C\, ,\ Z_E = 1 + c_E\, ,
\end{multline}
are adjusted in terms of the cutoff $\La$ to make $S(g)$ finite. Of course, the $c$s vanish at tree-level and are non-vanishing only at one or higher loop order. They are thus $O(\hbar)$. Any term in $S(g)$, possibly divergent, which is the integral of a local polynomial in the curvature of dimension four or less can be absorbed in the coefficients $c_{\la}$, $c_G$, $c_2$, $c_C$ and $c_E$. These contributions are thus ambiguous and this ambiguity in the gravitational action is parameterized by the constants $G$, $\la^{2}$, $\kappa_2$, $\kappa_C$ and $\kappa_E$. Their most natural definition is simply to set
the local piece in $S(g)$ to take exactly the classical form
\be\label{Sloca}S_{\text{cl}}(g)= \frac{1}{16\pi G}\int\!\d^{4}x\sqrt{g}\Bigl(
-2\la^{2} + R + G\kappa_2 R^{2} + G\kappa_C C^{ijkl}C_{ijkl}\Bigr) + \kappa_E \chi_{\text E}\, ,\ee
to any loop order. This being said, we shall work from now on modulo local terms of the form \eqref{Sloca} and use the sign $\equiv$ to indicate equality modulo terms of this sort.

If we wish to consider the two dimensional case, we can still work with the action \eqref{Actionphi},
but the purely gravitational local action contains only the cosmological constant and Euler terms.
Apart from that and the fact that $Z_m$ and $Z_1$  can now be at most logarithmically, and $Z_{\la}$ quadraticaly divergent, the discussion is strictly parallel with the discussion in four dimensions.

The gravitational action $S(g)$ has a small $\hbar$ asymptotic expansion of the form
\be\label{Sloopexp} S(g) = S_{\text{cl}}(g) + \sum_{L\geq 1} \hbar^{L}S^{(L)}(g)\, .\ee
The one-loop approximation $S^{(1)}$ was studied in Sec.\ \ref{grav1lsec} and Eq.\ \eqref{Sgspediv} yields
\be\label{S1form} S^{(1)}(g) \equiv -\frac{1}{2}\zeta'(0)\, .\ee
Our goal in the following is to compute $S^{(2)}$. Expanding the path integral on the right-hand side of \eqref{Sgravdef} to order $\hbar^{2}$, we get
\be\label{S2genform} S^{(2)}(g) = \frac{\k_4}{8}\mathscr A^{\diagramc}
- \frac{\k_3^{2}}{8}\mathscr A^{\diagramd} - \frac{\k_3^{2}}{12}\mathscr A^{\diagramb} + S^{(2)}_{\text{c.t.}}\, ,\ee
where the amplitudes of the various two-loop diagrams were defined in Sec.~\ref{ampsec} and $S^{(2)}_{\text{c.t.}}$ represents the contributions from the counterterms, i.e.\ from the corrections to the renormalization constants \eqref{csdefs}. The diagrams contributing to $S^{(2)}_{\text{c.t.}}$ have less than two loops but include vertices which are at least $O(\hbar)$.
The action $S^{(2)}_{\text{c.t.}}$ is thus easy to find and is discussed in the next subsection. To simplify the formulas we set $\hbar=1$ from now on.


\subsubsection*{Contributions from the one-loop counterterms}

The one-loop diagrams contributing to $S^{(2)}$  can be straightforwardly evaluated from the discussion of Section \ref{oneloopsec}.

--- The counterterm $c_{\phi} \phi\Delta\phi$ yields a contribution which, in terms of the regularized Green's function $G_{f,\Lambda}(x,y)$ defined in \eqref{Greensmooth}, reads
\begin{multline}\label{cphicont}\frac{1}{2}c_{\phi} \int\!\d\nu(x)\,
\lim_{y\rightarrow x}\Delta_{y}G_{f,\La}(x,y) =\\
\frac{1}{2}c_{\phi} \int\!\d\nu(x)\Bigl( \lim_{y\rightarrow x}D_{y}G_{f,\La}(x,y) - \bigl( m^{2} + \xi R(x)\bigr)G_{f,\La}(x,x)\Bigr)\, .
\end{multline}
Since $c_{\phi}$ is at most logarithmically divergent, we can plug the expansion \eqref{Gsmooexpeven} for $G_{f,\La}(x)$ into the second term on the right-hand side of \eqref{cphicont}. If not for the Green's function at coinciding points $G_{\zeta}$, all the terms in this expansion that have a non-zero $\La\rightarrow\infty$ limit are local polynomial in the curvature of dimension less than four. As for
\begin{align}\label{DGexp} \lim_{y\rightarrow x}D_{y}G_{f,\La}(x,y) & =
\sum_{r\geq 0}f\bigl(\la_{r}/\La^{2}\bigr)\psi_{r}(x)^{2}
= \int_{0}^{\infty}\!\d\alpha\, \varphi(\alpha) K\bigl(\alpha/\La^{2}\bigr)
\\\label{DGexp2} & = \frac{1}{(4\pi)^{d/2}}\sum_{k\geq 0}\La^{d-2 k}
a_{k}(x)\int_{0}^{\infty}\!\d\alpha\, \alpha^{k-\frac{d}{2}}
\varphi(\alpha)\, ,
\end{align}
it is also expressed in terms of local polynomials in the curvature of dimension less than four, up to terms of order $\La^{-2}$. Overall, we thus get
\begin{multline}\label{cphicontf} \frac{1}{2}c_{\phi}
\int\!\d\nu(x)\,
\lim_{y\rightarrow x}\Delta_{y}G_{f,\La}(x,y) 
\\ \equiv -\frac{1}{2} c_{\phi}m^{2}
\int\!\d\nu(x)\, G_{\zeta} (x) -\frac{1}{2}c_{\phi}\xi\int\!\d\nu(x)\, R(x)G_{\zeta}(x)\, .\end{multline}

--- Exactly the same analysis for the counterterm $c_{\xi}\xi R\phi^{2}$, using the fact that $c_{\xi}$ is at most logarithmically divergent, yields the contribution
\be\label{cxicontf} \frac{1}{2}c_{\xi}\xi\int\!\d\nu(x)\, R(x) G_{f,\La}(x,x) \equiv \frac{1}{2}c_{\xi}\xi\int\!\d\nu(x)\, R(x) G_{\zeta}(x)\, .\ee

--- To evaluate the contribution coming from the counterterm $c_{m}m^{2}\phi^{2}$, we need in principle the expansion of $G_{f,\La}$ up to and including the term of order $\La^{-2}$, because $c_{m}$ can be quadratically divergent. From \eqref{Gsmooexpeven}, we see that this term is proportional to $a_{d/2}$ and is thus a local polynomial in the curvature of dimension $d$ that can be discarded as usual. We thus find a contribution
\be\label{cmcontf}\frac{1}{2}c_{m}m^{2}\int\!\d\nu(x)\, G_{f,\La}(x,x) \equiv \frac{1}{2}c_{m}m^{2}\int\!\d\nu(x)\, G_{\zeta}(x)\, .\ee

--- The counterterms proportional to $c_3 \k_3\phi^{3}$ and $c_4 \k_4\phi^{4}$ do not contribute at order $\hbar^{2}$.

--- The counterterms linear in $\phi$ yield a contribution
\be\label{ctlinear}-\frac{1}{2}\int\!\d\nu(x)\d\nu(y)\,
\bigl(c_{1} + \tilde c_{1}R(x)\bigr)G_{f,\La}(x,y)\bigl(c_{1}+\tilde c_{1}
R(y) + \kappa_{3} G_{f,\La}(y)\bigr)\, .\ee
Using \eqref{i1expI1} and \eqref{i2expI2}, it is straightforward to show that, up to an integral of local polynomials in the curvature of dimension less than four, this is equivalent to
\begin{multline}\label{ctlinear2} -\frac{1}{2}\int\!\d\nu(x)\d\nu(y)\,
\bigl(c_{1} + \tilde c_{1}R(x)\bigr)G(x,y)\bigl(c_{1}+\tilde c_{1}
R(y) + \kappa_{3} G_{f,\La}(y)\bigr)\\ -\frac{1}{2}\kappa_{3}c_{1}\La^{-2}f'(0)\int\!\d\nu(x)\, G_{\zeta}(x)\, .\end{multline}
In the first line of this equation, we can plug in the expansion \eqref{Gsmooexpeven} for $G_{f,\La}(y)$, up to and including the term of order $\La^{-4}$.


Putting everything together,
\begin{multline}\label{Sctfinal} 
S_{\text{c.t.}}^{(2)} \equiv
\frac{1}{2}\bigl((c_{m}-c_{\phi}) m^{2}-\kappa_{3}c_{1}\La^{-2}f'(0)\bigr)\hskip-1.mm\int\!\d\nu(x)\, G_{\zeta}(x) + \frac{1}{2}\bigl(c_{\xi}-c_{\phi}\bigr) \xi\hskip-1.mm\int\!\d\nu(x)\, R(x) G_{\zeta}(x) \\
-\frac{1}{2}\int\!\d\nu(x)\d\nu(y)\,
\bigl(c_{1} + \tilde c_{1}R(x)\bigr)G(x,y)\bigl(c_{1}+\tilde c_{1}
R(y) + \kappa_{3} G_{f,\La}(y)\bigr)\, .
\end{multline}
We see that the field renormalization constant $c_{\phi}$ is redundant in the calculation of the gravitational effective action, since it can be shifted into $c_m$ and $c_\xi$. We also see that terms of the form $\int\! G_{\zeta}$ and $\int\! RG_{\zeta}$ in the gravitational action, albeit non-local in the metric, are ambiguous, in the same sense as the integrals of local polynomials in the curvature of dimensions less than $d$ are ambiguous: they can be absorbed in the \emph{local} mass and $\xi$-counterterms in the microscopic action. In particular, it is always possible to define $m$ and $\xi$ in order to absorb these terms in the one-loop gravitational action. 

\subsection{The two-loop diagrams in details}
\subsubsection*{Warming up: the functions $k_{\dia}$ and $\zeta_{\dia}$}



As explained at the beginning of Sec.\ \ref{analyticsec}, the asymptotic expansions of the functions $k_{\dia}$ can be derived straightforwardly from the asymptotic expansion \eqref{Kxyexp} of the heat kernel, by using the integral representation \eqref{Knormtok}. For example, at two loops, we find in this way
\begin{align}\label{kdiagccurved}
k_{\diagramc}(t_1,t_2)&= \int\!\d\nu(x)\, K(t_1,x,x) K(t_2,x,x)\nonumber\\
&={1\over (4\pi)^d  (t_1 t_2)^{d/2}} \int\! \d\nu(x)\, \Big[ 1 + (t_1+t_2)a_1(x) +(t_1^2+t_2^2) a_2(x) 
\nonumber\\
&\hskip 8cm
+ t_1 t_2 a_1^2(x) +O(t^3)\Big] \, ,
\end{align}
\begin{align}\label{kdiagbcurved}
k_{\diagramb}(t_1,t_2,t_3) &= \int\!\d\nu(x)\d\nu(y)\, K(t_1,x,y) K(t_2,x,y) K(t_3,x,y)\nonumber\\
&=  {1\over (4\pi)^d (t_1 t_2+t_1 t_3+t_2 t_3)^{d/2}}\nonumber\\ &
\hskip 2cm
\int\!\d\nu(x)\,\Big[ 1 +{1\over 6 \sum_i t_i^{-1}} R(x) 
+ \sum_i t_i\ a_1(x) +O(t^2)\Big]
\end{align}
and
\begin{align}
k_{\diagramd}(t_1,t_2,t_3) &= \int\!\d\nu(x)\d\nu(y)\, K(t_1,x,y) K(t_2,x,x) K(t_3,y,y)\nonumber\\ \label{kdiagdcurved}&
= {1\over (4\pi)^d (t_2  t_3)^{d/2}} \int\d^d y\, \sqrt{g(y)} \Big[ 1 +\sum_i t_i \, a_1(y) -{t_1\over 6}R(y) +O(t^2)\Big]\, .
\end{align}
These formulas should be compared with their flat space versions \eqref{kflex1}--\eqref{kflex2}. Note that the diagonal heat kernel coefficients are explicitly given in the appendix \ref{heatapp} by Eq.\ \eqref{ardiag} and \eqref{frmfr}, e.g.\ $a_1(x)={1-6\xi\over 6} R(x)-m^2$. If desired, higher order terms can be straightforwardly computed, using e.g.\ \eqref{Knint}. Of course, the expansions so obtained are consistent with the general formula \eqref{kdiaexp}.

The analytic structure of the various $\zeta_{\dia}$ functions follow immediately from the above expansions, by using \eqref{ankdia1}. For example,
\begin{multline}\label{zetadiagccurved}
\zeta_{\diagramc}(s,\underline\alpha)={1\over (4\pi)^d  (\a_1 \a_2)^{d/2}} \int \!\d\nu(x)\,\biggl[ {1\over \G(d-1)\, (s-d+1)} + {(\a_1+\a_2)\, a_1(x)\over \G(d-2)\, (s-d+2)}  \\
+{(\a_1^2+\a_2^2) \, a_2(x) 
+ \a_1 \a_2 \, a_1^2(x) \over \G(d-3)\, (s-d+3)} + \cdots \biggr] +\text{analytic in $s$}\, ,
\end{multline}
where $+\cdots$ indicates further pole terms at $s=d-4$, $s=d-5$, etc. Note that the would-be pole at $s=0$ is canceled by the $1/\G(s)$ in  \eqref{ankdia1}. Thus, in $d=4$, all the pole terms at $s\geq 0$ are explicitly written in \eqref{zetadiagccurved}. The leading pole is at $s=d-1$, while the superficial degree of divergence of this diagram is ${\rm SDiv}_{\diagramc}=2(d-2)$, in agreement with the relation \eqref{residkdia1}, which gives the leading pole at $s={\rm SDiv}_{\diagramc}/2+1$. Similarly,
\begin{multline}\label{zetadiagbcurved}
\zeta_{\diagramb}(s,\a_1,\a_2,\a_3)={1\over (4\pi)^d  (\a_1 \a_2+\a_1\a_3+\a_2\a_3)^{d/2}} \int\!\d\nu(x)\\\biggl[ {1\over \G(d-2)\, (s-d+2)} 
+ {\sum_i\a_i a_1(x)+R(x)/(6\sum_i \a_i^{-1})\over \G(d-3)\, (s-d+3)} 
+ \cdots \biggr] +\text{analytic in $s$.}
\end{multline}
The leading pole is at $s=d-2$, in agreement with a superficial degree of divergence ${\rm SDiv}_{\diagramb}=2d-6$, except in $d=2$, where it is canceled by the $\G(d-2)$ (of course, in $d=2$ the diagram is finite.)
Finally,
\begin{multline}\label{zetadiagdcurved}
\zeta_{\diagramd}(s,\a_1,\a_2,\a_3)={1\over (4\pi)^d  (\a_1 \a_3)^{d/2}} \int \!\d\nu(x)\\ \biggl[ {1\over \G(d-2)\, (s-d+2)}
+ {\sum_i\a_i a_1(x)- \a_2 R(x)/6\over \G(d-3)\, (s-d+3)} 
+ \cdots \biggr] +\text{analytic in $s$.}
\end{multline}
The leading pole is again at $s=d-2$, consistently with the superficial degree of divergence being again $2d-6$, although the true degree of divergence of this diagram is $2(d-2)$.

\subsubsection*{The functions $K_{\diagramd}$ and $K_{\diagramc}$ in $d=4$}

We now proceed to the discussion of the functions $K_{\dia}$, which are directly related to the Feynman amplitudes, see e.g.\ \eqref{Ampexp} and \eqref{Kdiaexp}. The basic tools at our disposal are the integral representation \eqref{KhattoK} together with the asymptotic expansion \eqref{Khatexp}. 

To present the calculations, it is actually very convenient to include right away, for any divergent diagram, the relevant counterterm contributions that make the diagram finite. The form of these contributions were discussed in Section \ref{twoLgenSec}. We will thus actually focus on the functions
\be\label{Kpluscounter}
K_\dia^{\text{plus ct}}(\underline t)=K_\dia(\underline t)\ +\ \text{counterterm contributions relevant for diagram}\,  \ \dia
\ee
which have a smooth limit when $\underline t\rightarrow 0$.

\smallskip

\noindent\emph{The diagram $\diagramdbig$}

\noindent We have
\be\label{Kdiad}
K_{\diagramd}(\underline t)=\int\!\d\nu(x)\d\nu(y) \, \wh K(t_2,x) \wh K(t_1,x,y) \wh K(t_3,y) \ .
\ee
At small $t$, \eqref{Khatxx} yields, for $d=4$,
\be\label{Khat4dexp} \wh K(t, x) = \frac{1}{(4\pi)^{2}}\Bigl[ \frac{1}{t} - \bigl(\gamma + \ln (\mu^{2}t)\bigr) a_{1}(x)\Bigr] + G_{\zeta}(x) + O(t)\, .\ee
This strongly suggests to consider
\begin{multline}\label{Kdiadct}
K_{\diagramd}^{\text{plus ct}}(\underline t)=\int\!\d\nu(x)\d\nu(y) \,
\wh K(t_{1},x,y)\\
\Bigl[ \wh K(t_2,x) 
-{1\over (4\pi)^2}\Big(\, {1\over t_2}  -a_1(x)  \bigl(\ln (\m^2 t_2) +\g\bigr)\Bigr)\Bigr]\\
\Bigl[ \wh K(t_3,y,y) 
-{1\over (4\pi)^2}\Big(\, {1\over t_3}  -a_1(y)  \big(\ln (\m^2 t_3) +\g\bigr)\Bigr)\Bigr]\ .
\end{multline}
Using also the expansion \eqref{Khatexp}, it is straightforward to check that \eqref{Kdiadct} does have a finite small $\underline t$ limit given by
\be\label{Kdiag3} K_{\diagramd}^{\text{plus ct}}(0) = \int\!\d\nu(x)\d\nu(y)\, G_{\zeta}(x)G(x,y)G_{\zeta}(y)\, .\ee
Carefully taking into account the fact that the contribution of our diagram to the amplitude should be $-\frac{\kappa_{3}^{2}}{8}K_{\diagramd}^{\text{plus ct}}(0)$ and using the formula
\be\label{a1form} a_{1}(x) = \frac{1-6\xi}{6} R(x) - m^{2}\ee
derived in the appendix \ref{heatapp}  (Eqs.\ \eqref{ardiag} and \eqref{frmfr}), it is straightforward to check that the subtractions included in \eqref{Kdiadct} are equivalent to adding the counterterm \eqref{ctlinear}, with
\begin{align}\label{c1direct} c_{1}& = -\frac{\kappa_{3}}{2 (4\pi)^{2}}
\int\!\d\alpha\,\varphi(\alpha)\biggl[\frac{\La^{2}}{\alpha} + m^{2}\Bigl(
\gamma + \ln \frac{\mu^{2}\alpha}{\La^{2}}\Bigr)\biggr]\, ,\\
\label{ct1direct} \tilde c_{1}& = \frac{\kappa_{3}}{2 (4\pi)^{2}}\Bigl(\frac{1}{6}-\xi\Bigr)
\int\!\d\alpha\,\varphi(\alpha)\Bigl(
\gamma + \ln \frac{\mu^{2}\alpha}{\La^{2}}\Bigr)\, .
\end{align}
Overall, we thus obtain a contribution
\be\label{conttoamp1} -\frac{\kappa_{3}^{2}}{8}\int\!\d\nu(x)\d\nu(y)\, G_{\zeta}(x)G(x,y)G_{\zeta}(y)\ee
to the gravitational effective action.

\smallskip

\noindent\emph{The diagram $\diagramcbig$}

\noindent A similar analysis can be done for
\be\label{Kdiac}
K_{\diagramc}(\underline t)=\int\!\d\nu(x) \, \wh K(t_1,x) \wh K(t_2,y) \ .
\ee
The expansion \eqref{Khat4dexp} suggests to consider
\begin{multline}\label{Kdiacplusct}
K_{\diagramc}^{\text{plus ct}}(\underline t)
=\int\!\d\nu(x)  \, \Bigl[\wh K(t_1,x) -{1\over (4\pi)^2} \Big( {1\over t_1}-a_1(x) \bigl(\ln(\m^2 t_1)+\g\bigr)\Bigr)    \Bigr]\\
\Bigl[\wh K(t_2,x) -{1\over (4\pi)^2} \Bigl( {1\over t_2}-a_1(x) \bigl(\ln(\m^2 t_2)+\g\bigr)\Bigr)     \Bigr]\, ,\end{multline}
which has a finite small $\underline t$ limit given by
\be\label{Kdiag3bis} K_{\diagramc}^{\text{plus ct}}(0) = \int\!\d\nu(x)\, G_{\zeta}(x)^{2}\, .\ee
Using e.g.\ \eqref{Gsmooexpeven}, it is easy to check that the subtractions included in \eqref{Kdiacplusct} correspond to adding terms which are integrals of local polynomials in the metric to the action together with terms of the form \eqref{Sctfinal}, with precise contributions to the coefficients $(c_{m}-c_{\phi})m^{2}$ and $(c_{\xi}-c_{\phi})\xi$ given by
\be\label{cmcont1} -\frac{\kappa_{4}}{2 (4\pi)^{2}}
\int\!\d\alpha\,\varphi(\alpha)\biggl[\frac{\La^{2}}{\alpha} + m^{2}\Bigl(
\gamma + \ln \frac{\mu^{2}\alpha}{\La^{2}}\Bigr)\biggr]\ee
and
\be\label{cxicont1} 
\frac{\kappa_{4}}{2 (4\pi)^{2}}\Bigl(\frac{1}{6}-\xi\Bigr)
\int\!\d\alpha\,\varphi(\alpha)\Bigl(
\gamma + \ln \frac{\mu^{2}\alpha}{\La^{2}}\Bigr)\ee
respectively. Overall, the diagram thus contributes a term
\be\label{conttoamp2} \frac{\kappa_{4}}{8}\int\!\d\nu(x)\, G_{\zeta}(x)^{2}
\ee
to the gravitational effective action.

\subsubsection*{The function $K_{\diagramb}$ in $d=4$}

The last diagram at two loops, given by
\be\label{Kdiab} K_{\diagramb}(\underline t) = \int\!\d\nu(x)\d\nu(y)\, \wh K(t_{1},x,y)\wh K(t_{2},x,y)\wh K(t_{3},x,y)\, ,\ee
yields the most interesting contribution. Unlike the previous diagrams we considered, it is a genuine two-loop diagram, in the sense that is cannot be related to quantities appearing already at one-loop, and it has an overlapping divergence. We are going to study it in some details, using a strategy which can be straightforwardly generalized to any higher loop diagram.

A fundamental tool to study the small $t=t_{1}+t_{2}+t_{3}$ behavior is the expansion \eqref{Khatexp} for the hatted heat kernel. In four dimensions, it reads
\be\label{Khatexpd4} \wh K(t,x,y) = G(x,y) - \frac{a_{0}(x,y)}{(2\pi)^{2}\ell(x,y)^{2}}e^{-\frac{\ell(x,y)^{2}}{4 t}} - \frac{1}{(4\pi)^{2}}
\sum_{k\geq 1}a_{k}(x,y) t^{k-1}E_{k}\Bigl( \frac{\ell(x,y)^{2}}{4t}\Bigr)\, .\ee
The first two terms on the right-hand side of \eqref{Khatexpd4} define a sort of regularized Green's function, which has a simple integrable logarithmic singularity when $y\rightarrow x$. The other terms are exponentially suppressed when $t\rightarrow 0$, since
\be\label{Ekasbis} E_{k}(z)\underset{z\rightarrow\infty}{\sim} \frac{e^{-z}}{z}\, \cvp\ee
except if $\ell(x,y)\sim\sqrt{t}$. 

We can use directly \eqref{Khatexpd4} to show that
%
\be\label{Kdiabnice2} K_{\diagramb}(\underline t) \equiv \int\!\d\nu(x)\d\nu(y)\,
\prod_{i=1}^{3}\biggl( G(x,y) - \frac{a_{0}(x,y)}{(2\pi)^{2}\ell(x,y)^{2}}e^{-\frac{\ell(x,y)^{2}}{4 t_{i}}}\biggr) \, ,\ee
where the symbol $\equiv$ means, as usual, equality up to terms that are integrals of local polynomials in the metric of dimension less than or equal to four. However, this simple-looking formula is not very convenient to study the small $t$ asymptotic expansion, and we shall use a more systematic method.

\smallskip

\noindent\emph{The structure of the divergences}

\noindent As for any Feynman diagram (see the discussion in Sec.\ \ref{analyticsec}), the small $t$ (or equivalently large cutoff $\La$) divergences come from the fact that the internal lines in the diagram can be arbitrarily short. In the integral representation
\be\label{wkintnew} \wh K(t,x,y) = \int_{t}^{\infty}\!\d\beta\, K(\beta,x,y)\, ,\ee
this corresponds to the region of small $\beta$. It is thus natural to decompose
\be\label{wKhdecompose} \wh K(t,x,y) = \wh K_{-}(t,x,y) + \wh K_{+}(t,x,y)\, ,\ee
with
\be\label{wKplusminus} \wh K_{-}(t,x,y) = \int_{t}^{T}\!\d\beta\, K(\beta,x,y)\, ,\quad \wh K_{+}(t,x,y) = \int_{T}^{\infty}\!\d\beta\, K(\beta,x,y)\, ,\ee
where $T>t$ is a fixed scale which can be taken to be arbitrarily small when $t\rightarrow 0$. The diagram \eqref{Kdiab} can thus be written as a sum of four terms $K_{\diagramb}^{(i)}$, $0\leq i\leq 3$, containing products of $i$ factors $\wh K_{-}$ which are associated with $i$ small internal lines,
\be\label{Kbdecaccordlines} K_{\diagramb}(\underline t) = 
K_{\diagramb}^{(0)}(\underline t) + K_{\diagramb}^{(1)}(\underline t) + K_{\diagramb}^{(2)}(\underline t) + K_{\diagramb}^{(3)}(\underline t)\, .\ee

The diverging contribution coming from $K_{\diagramb}^{(3)}(\underline t)$, with all three internal lines very short, can be immediately obtained from the expansion \eqref{Kxyexp} of the standard heat kernel. Using  \eqref{Knint} for $n=3$, we get
\begin{multline}\label{F123-1}
K_{\diagramb}^{(3)}(\underline t)={1\over (4\pi)^4}
\int\!\d\nu(x)
\left[\prod_{i=1}^3 \int_{t_i}^T \d\b_i\right] {1\over (\b_1\b_2+\b_1\b_3+\b_2\b_3)^2}\\
\Bigg\{ 1+ a_1(y)\sum_i\b_i
+\,{R(y)\over 6} {\b_1\b_2\b_3\over \b_1\b_2+\b_1\b_3+\b_2\b_3} + O(\b^2) \Bigg\} \, ,
\end{multline}
which yields
\begin{multline}\label{F123res}
K_{\diagramb}^{(3)}\bigl(\underline t=\underline\alpha/\La^{2}\bigr)={1\over (4\pi)^4}\int\!\d\nu(x)\, \biggl[
I_{4}(1,\underline\a) \L^2+{3\over 2} a_1(x) \Big(\ln {\L^2\over \m^2}\Big)^2
\\+\Big({R(x)\over 12} -{3\over T}\Big)\ln  {\L^2\over\m^2}  
- a_1(x) \Bigl( \sum_{i>j}\ln(\a_i+\a_j)-3\ln\bigl( \m^2 T\bigr)\Bigr) \, \ln {\L^2\over \m^2}\biggr] + \text{finite.}
\end{multline}
The integral $I_{4}(1,\underline\alpha)$ was defined in \eqref{I4int} and computed in \eqref{Iintsequal1}.

The contribution from
\be\label{K2explfor} K_{\diagramb}^{(2)}(\underline t) = 
\int\!\d\nu(x)\d\nu(y)\, \bigl( \wh K_{-}(t_{1},x,y) \wh K_{-}(t_{2},x,y) \wh K_{+}(t_{3},x,y) + \text{two terms}\bigr)\ee
can be most efficiently evaluated by writing $\wh K_{+}=\wh K - \wh K_{-}$, which shows that
\be\label{K2explfor2} K_{\diagramb}^{(2)} = 
\int\!\d\nu(x)\d\nu(y)\, \bigl( \wh K_{-}(t_{1},x,y) \wh K_{-}(t_{2},x,y) \wh K(t_{3},x,y) + \text{two terms}\bigr) - 3 K_{\diagramb}^{(3)}\, .\ee
At small $\underline t$, we can use the standard heat kernel expansion to evaluate the factors $\wh K_{-}$ and the expansion \eqref{Khatexpd4} to evaluate $\wh K$. Moreover, the factors of $\wh K_{-}$ imply that $\ell (x,y)$ is at most $\sim\sqrt{T}$, and thus small, in the integrals \eqref{K2explfor2}. We can thus use the short distance expansion  
of the Green function $G(x,y)$ which follows from \eqref{Gspliteven},
\be\label{Greenshortd4}
G(x,y)={1\over (4\pi)^2} \left[ {4 a_0(x,y)\over \ell^2(x,y)}-a_1(x,y)\Big(\ln {\m^2\ell(x,y)^{2}\over 4}+2\g\Big)\right] +G_\zeta(x)+O(\ell(x,y)) \ ,\ee
which yields
\begin{multline}\label{Khatexpd4bis}
\wh K(t,x,y)=G_\zeta(y)-{1\over (4\pi)^2}\biggl[ {4 a_0(x,y)\over \ell^2(x,y)} \Big(e^{-\frac{\ell(x,y)^{2}}{4 t}} -1\Big)\\
+ a_1(x,y) \Big(E_1\Bigl({\ell(x,y)^{2}\over 4 t}\Bigr)+ \ln {\m^2\ell(x,y)^{2}\over 4}+2\g   \Bigr)\biggr] +O\bigl(t,\ell(x,y)\bigr) \ .
\end{multline}
Let us emphasize that this expansion assumes that $t$ and $\ell(x,y)$ are small, but the ratio $\ell(x,y)^{2}/t$ can be arbitrary. It is then very natural to introduce Riemann normal coordinates around $x$ and to define
\be\label{normalexp} y = x + \sqrt{2t}\,  z\, .\ee
The expansion \eqref{Khatexpd4bis} then reads
\begin{multline}\label{Khatxclosey4dim}
\wh K(t,x,y)
=G_\zeta(x)-{2\over (4\pi)^2 t} {1+{t\over 6} R_{kl}(x)z^k z^l\over z^2} \Bigl(e^{-\frac{1}{2}z^2} - 1 \Bigr)
\\
-{a_1(x)\over (4\pi)^2} \Bigl[ E_1 \bigl(z^{2}/2\bigr)+\ln{\m^2 t z^2\over 2}+2\g \Big]+O(t)\, .
\end{multline}
Inserting this result in \eqref{K2explfor2}, together with \eqref{Kn2} for $n=2$, we can perform explicitly the integral over $y$ (or $z$), using in particular
\begin{align}\label{intelem1} \int\! \d^4 x\, e^{-a x^2/4}E_1\bigl(b x^2/4\bigr) &={(4\pi)^2\over a^2} \Bigl( \ln\bigl(1+ a/b\bigr)-{1\over 1+b/a} \Bigr)\\
\int\! \d^4 x\, e^{-a x^2/4} \ln\bigl( b x^2/4\bigr) &={(4\pi)^2\over a^2} \Bigl( 1 - \g + \ln{b\over a}\Bigr)\, .
\end{align}
This yields, e.g.,
\begin{multline}\label{F12int}
\int\!\d\nu(y)\,  K(\b_1,x,y) K(\b_2,x,y) \wh K(t_3,x,y)
\\
={1\over (4\pi)^4} \Biggl[
{(4\pi)^2 G_\zeta(x) -(\g+1) a_1(x) \over (\b_1+\b_2)^2}
+{R(x)\over 12} {\b_1\b_2\over (\b_1+\b_2)^2} {\b_1\b_2 + 2(\b_1+\b_2)t_3\over [\b_1\b_2 + (\b_1+\b_2)t_3]^2}
\\
 -{a_1(x)\over (\b_1+\b_2)^2}\ln{\m^2[\b_1\b_2 + (\b_1+\b_2)t_3]\over  \b_1+\b_2}
+{1+a_1(x)(\b_1+\b_2+t_3)\over (\b_1+\b_2)[\b_1\b_2 + (\b_1+\b_2) t_3]}
+O({1\over \b})
\Biggr] \, .
\end{multline}
The last step in the calculation is to perform the integrals over $\beta_{i}\in[t_{i},T]$. The terms $O(1/\beta)$ that we have discarded in \eqref{F12int} yield finite contributions at small $t$. The other integrals can be expressed explicitly in terms of elementary or special functions (poly-logarithms) from which the singular behavior as $t_i\to 0$ can be extracted. Finally, putting everything together in \eqref{K2explfor2} yields
\begin{equation}\label{F12res2}
K_{\diagramb}^{(2)}\bigl(\underline t = \underline\alpha/\La^{2}\bigr) 
={3\over (4\pi)^4}\ln\frac{\La^{2}}{\mu^{2}}\hskip-1.mm
\int\!\d\nu(x) \Bigl[ (4\pi)^2 G_\zeta(x)+{1\over T}  
 - a_1(x) \bigl(\g +\ln (\m^2 T) \bigr) \Bigr]  + \text{finite.}
\end{equation}

From a similar analysis, it is easy to check that $K_{\diagramb}^{(1)}(\underline t)$ has a finite small $t$ limit, which is also obviously the case for $K_{\diagramb}^{(0)}(\underline t)$. Summing up all the contributions in \eqref{Kbdecaccordlines}, we thus obtain
\begin{multline}\label{Fsingres}
K_{\diagramb}\bigl(\underline t = \La^{2}/\underline\alpha\bigr)
={1\over (4\pi)^4}\int\!\d\nu(x)\,\Biggl[
I_{4}(1,\underline\a) \L^2+{3\over 2} a_1(x)\Bigl(\ln {\L^2\over\m^2}\Bigr)^2
\\+\biggl[ 3(4\pi)^2 G_\zeta(x)+{R(x)\over 12}  
- a_1(x) \Bigl( \sum_{i>j}\ln(\a_i+\a_j)+3\g \Bigr)\biggr] \ln {\L^2\over\m^2} \Biggr] + \text{finite.}
\end{multline}
Nicely, all dependence on the arbitrary scale $T$ has canceled, providing a basic consistency check of the method. The dependence in $\mu$ also cancels: the $\mu$-dependence in the first line of \eqref{Fsingres} is absorbed in the $\mu$-dependence of $G_{\zeta}$ (see \eqref{Gzeta}) and in the finite piece, as is the $\mu$-dependence in the second line.
Let us finally note that in flat space, $a_{1}=-m^{2}$ and $G_{\zeta} = -m^{2}/(4\pi)^{2}$ if $\mu=m$; we then find again the result \eqref{Ampdiagbd4}.

The divergences can of course be absorbed in the local counterterms. Apart from the local polynomials in the metric, \eqref{Fsingres} contains a non-local divergence proportional to $G_{\zeta}$. According to \eqref{Sctfinal} and carefully taking into account the factor $-\kappa_{3}^{2}/12$ in front of the amplitude $\mathscr A^{\diagramb}$ in \eqref{S2genform}, this divergence is absorbed with the help of a contribution
\be\label{cmcont2}
\frac{\kappa_{3}^{2}}{2(4\pi)^{2}}\ln\frac{\La^{2}}{\mu^{2}}\ee
to the coefficient $(c_{m}-c_{\phi})m^{2}$.

\smallskip

\noindent\emph{Cutoff independence of the physical finite part}

\noindent We have shown explicitly above that the diverging part of the amplitude can be canceled by appropriate local counterterms.
The finite part is obtained by subtracting \eqref{Fsingres} from \eqref{Kdiab} (or \eqref{Kdiabnice2}) and taking the $\underline t =\underline\alpha/\La^{2}\rightarrow 0$ limit. However, and as discussed at the end of Sec.\ \ref{ampsec}, renormalization implies more than the simple cancellation of divergences; there is also a non-trivial constraint on the physical \emph{finite} part, which must be cutoff-independent. The traditional cutoff procedures do not allow to check explicitly this important feature, since all the explicit cutoff dependence is usually in the divergent pieces. In our case, the cutoff dependence is not only in $\La$ but also in the cutoff function $\varphi$. The only constraint on the integrals of $\varphi$ is \eqref{phicond1}. Cutoff independence of the physical amplitudes is thus equivalent, in our framework, to the $\underline\alpha$-independence of the physical finite parts of the $K_{\dia}$ functions, which is a rather non-trivial constraint. 
At two loops, we thus have to check that the $\underline\alpha$-dependence of the \emph{finite} piece in $K_{\diagramb}$ can be canceled by appropriate finite shifts of the local counterterms.

To do this, we are going to study the partial derivatives $\partial K_{\diagramb}/\partial\alpha_{i}$. Since the three internal lines play a symmetric r\^ole in the diagram $\diagrambbig$, we can focus, for example, on the derivatives with respect to $\alpha_{3}$. From $\underline t = \underline\alpha/\La^{2}$ and the basic identity
\be\label{basicderkhat}\frac{\partial\wh K}{\partial t} = - K\, ,\ee
we get
\be\label{Kdiagbderiv}\frac{\partial K_{\diagramb}}{\partial\alpha_{3}}
=-{1\over \Lambda^2}\int\!\d\nu(x)\, \wh K(t_1,x,y) \wh K(t_2,x,y)  K(t_3,x,y)\ .\ee
The small $t$ behavior can then be studied by using exactly the same method as before. Since $t_{3}$ is small, we can use the standard heat kernel expansion for $K(t_{3},x,y)$, which also implies that $\ell(x,y)$ will be no larger than $\sim\sqrt{t}$ in \eqref{Kdiagbderiv}. We can thus use Riemann normal coordinates and the expansion \eqref{Khatxclosey4dim} for $\wh K(t_1,x,y)$ and for $\wh K(t_2,x,y)$ in \eqref{Kdiagbderiv}, together 
with \eqref{Kn2} for $n=1$. All the resulting spacetime integrals can be performed explicitly and we obtain
%
\begin{multline}\label{Kdiagbderiv3}
\frac{\partial K_{\diagramb}}{\partial\alpha_{3}}
=-{1\over (4\pi)^4}\int\!\d\nu(x)\,\biggl[
\Bigl({\L^2 \over \a_3^2}-{a_1(x)\over \a_3}\Bigr) \ln{(\a_1+\a_3)(\a_2+\a_3)\over \a_1\a_2+\a_1\a_3+\a_2\a_3}
\\
+{R(x)\over 12}\Bigl( {1\over \a_1+\a_3}+{1\over \a_2+\a_3}-{\a_1+\a_2\over \a_1\a_2+\a_1\a_3+\a_2\a_3}\Bigr)\\
+\Bigl( (4\pi)^2G_\zeta (x) -\g a_1(x) +a_1(x) \ln{\L^2\over \m^2}\Bigr) \Bigl(   {1\over \a_1 +\a_3}+{1\over \a_2+\a_3}  \Bigr)\\
+{a_1(x)\over \a_1+\a_3}\ln{\a_1+\a_3\over   \a_1\a_2+\a_1\a_3+\a_2\a_3}
+{a_1(x)\over \a_2+\a_3}\ln{\a_2+\a_3\over\a_1\a_2+\a_1\a_3+\a_2\a_3}+O\Big({1\over \L^2}\Big)\biggr]\, .
\end{multline}
Integrating the diverging pieces, we find again the $\underline\alpha$-dependent terms in \eqref{Fsingres}, as required. But we now also have information on the finite piece. It is easy to find explicit formulas for the $\underline\alpha$-dependence of this finite piece by integrating \eqref{Kdiagbderiv3}, but the resulting formulas (involving dilogarithms) are not particularly illuminating. The fundamental point is already visible on \eqref{Kdiagbderiv3}: all the $\underline\alpha$-dependence is in terms that are local in the metric or proportional to $\int G_{\zeta}$ and can thus, as required, be absorbed in the counterterms.

Of course, this will not be the case for the $\underline\alpha$-independent finite piece. This piece contains all the non-trivial physical information and is highly non-local in the metric. It is the analogue, for the diagram $\diagrambbig$, of more familiar quantities like $\zeta'(0)$ or $\int G_{\zeta}^{2}$, which are relevant at one loop and for $\diagramcbig$, for example. As for these quantities, it cannot be computed in closed form, except on very simple Riemannian manifolds for which the spectrum of the operator $D$ is known explicitly.

\subsubsection*{Remark on the counterterm coefficients}

We have determined above the diverging counterterm coefficients that make the gravitational effective action finite at two loops (the finite pieces in these coefficients are of course ambiguous). The coefficients $c_{1}$, $\tilde c_{1}$ and $c_{\xi}-c_{\phi}$ are given by \eqref{c1direct}, \eqref{ct1direct} and \eqref{cxicont1}, whereas $c_{m}-c_{\phi}$ is given by adding up the contributions \eqref{cmcont1} and \eqref{cmcont2}. Of course, all these coefficients (including $c_{\phi}$ independently of the others) can also be determined, in a more traditional way, from the one-loop one and two-point functions.

The coefficients $c_{1}$ and $\tilde c_{1}$ are determined by canceling the divergences in the one-point function $-m\kappa_{3}G_{f,\La}(x)/2$. Using the expansion \eqref{Gsmooexpeven}, we immediately find again \eqref{c1direct} and \eqref{ct1direct}. As for the coefficients $c_{m}$, $c_{\xi}$ and $c_{\phi}$, they are fixed by looking at the divergences of the two-point function. This requires a bit more work, which is presented in App.\ \ref{apptwopt}., providing en passant an example of a correlator computation in our formalism. We find that $c_{\phi}=0$ and $c_{m}$ and $c_{\xi}$ consistently with our previous analysis.
 
%

%
\subsubsection*{Conclusion and the $Z$-functions}

The results obtained above are perfectly in line with the general discussion of Section~\ref{allloopsec}. 

For example, for the diagram $\diagrambbig$, we have obtained an expansion of the form \eqref{Kdiaexp} for the function $K_{\diagramb}$. Explicitly, \eqref{Fsingres} can be rewritten
%
\be\label{Kbexpgeneral} K_{\diagramb}(\underline\alpha t) = \frac{1}{4\pi t}\Bigl( A_{0,0}^{\diagramb}(\underline\alpha) + 
A_{1,2}^{\diagramb}(\underline\alpha) t(\ln t)^{2} + 
A_{1,1}^{\diagramb}(\underline\alpha) t\ln t +
A_{1,0}^{\diagramb}(\underline\alpha) t\Bigr) + O\bigl(t(\ln t)^{2}\bigr)\, ,\ee
with, choosing $\mu=1$ for simplicity,
\begin{align}\label{Af1}
A_{0,0}^{\diagramb}(\underline\alpha) & = \frac{V}{(4\pi)^{3}}
\biggl[\sum_{i>j}\Bigl(\frac{1}{\alpha_{i}} + \frac{1}{\alpha_{j}}\Bigr)\ln\bigl(\alpha_{i}+\alpha_{j}\bigr) - \Bigl(\sum_{k}\frac{1}{\alpha_{k}}\Bigr)\ln\sum_{i>j}\alpha_{i}\alpha_{j}\biggr]\, ,\\\label{Af2}
A_{1,2}^{\diagramb}(\underline\alpha) & =\frac{3}{2(4\pi)^{3}}\int\!\d\nu(x)\, a_{1}(x)\, ,\\\label{Af3}
A_{1,1}^{\diagramb}(\underline\alpha) & = -\frac{1}{(4\pi)^{3}}\int\!\d\nu(x)\, \biggl[ 3(4\pi)^{2}G_{\zeta}(x) + \frac{R(x)}{12} - a_{1}(x)
\Bigl(\sum_{i>j}\ln\bigl(\alpha_{i} + \alpha_{j}\bigr) + 3\gamma\Bigr)\biggr]\, .
\end{align}
The $\underline\alpha$-dependence of the coefficient $A_{1,0}^{\diagramb}$ has also been determined via \eqref{Kdiagbderiv3}.

As explained in Section \ref{analyticsec}, these results can be coded elegantly in the analytic structure of the $Z$-function. From \eqref{Kbexpgeneral} and \eqref{Zsingular} we find for example that, on the positive real axis, $Z_{\diagramb}(s,\underline\alpha)$ has a simple pole at $s=1$ with residue $\frac{1}{4\pi} A_{0,0}^{\diagramb}(\underline\alpha)$ and a triple pole at $s=0$, with
\begin{multline}\label{Zdiabszeroexp} Z_{\diagramb}(s,\underline\alpha) \underset{s\rightarrow 0}{=} \frac{A_{1,2}^{\diagramb}(\underline\alpha)}{4\pi s^{3}} - \frac{A_{1,1}^{\diagramb}(\underline\alpha)- \gamma
A_{1,2}^{\diagramb}(\underline\alpha)}{4\pi s^{2}}\\ +
\frac{A_{1,0}^{\diagramb}(\underline\alpha)-\gamma A_{1,1}^{\diagramb}(\underline\alpha) +\frac{1}{2}\bigl(\gamma^{2}-\frac{\pi^{2}}{6}\bigr)A_{1,2}^{\diagramb}(\underline\alpha)}{4\pi s} + \cdots\end{multline}

Up to ambiguous terms that can be absorbed in finite shifts of the counterterms, we see that the physical finite part of the amplitude $\mathscr A^{\diagramb}$ is simply given by the residue of $Z_{\diagramb}$ at $s=0$. This is of course also true for the other two-loop diagrams. We can thus write a simple and elegant formula for the two-loop gravitational effective action,
\be\label{finalformula} S^{(2)}(g) \equiv \res_{s=0}\Bigl( \frac{\kappa_{4}}{8}Z_{\diagramc} - \frac{\kappa_{3}^{2}}{8} Z_{\diagramd}-\frac{\kappa_{3}^{2}}{12} Z_{\diagramb}\Bigr)\, ,\ee
where, again, the equality sign here is modulo terms that can be absorbed in finite shifts of the local counterterms. In particular, all the $\underline\alpha$-dependence of the right-hand side of \eqref{finalformula} is within such terms. Formula \eqref{finalformula} is the two-loop generalization of the standard one-loop formula $S^{(1)}(g) =-\frac{1}{2}\zeta_{D}'(0)$.

\subsection{The two-loop conformal anomaly}

As a last application of our formalism, let us show that the two-loop conformal anomaly vanishes for the $\phi^{4}$ model in four dimensions. We thus assume that the conditions \eqref{xiWeyldef}, i.e.\ $m^{2}=0$ and $\xi = 1/6$, are satisfied, together with $\kappa_{3}=0$. In particular, the heat kernel coefficient $a_{1}(x)$ vanishes. 

The standard one-loop conformal anomaly was derived in Sec.\ \ref{CAsec}, see Eq.\ \eqref{Adzeta}. At two loops, the gravitational effective action is given by
\be\label{S2conf} S^{(2)}(g) = \frac{\kappa_{4}}{8}\int\!\d^{4}x\sqrt{g}\, G_{\zeta}(x)^{2}\, .\ee
To compute the conformal anomaly, we thus need to know how the four dimensional Green function at coinciding points $G_{\zeta}(x;g)$ transforms under a Weyl rescaling of the metric $g$. 

\subsubsection*{The Weyl transformation of the Green functions}

The Weyl transformation laws for the Green's function $G(x,y)$ and the Green's function at coinciding points $G_{\zeta}(x)$, in the case $m^{2}=0$ and $\xi=1/6$, are given by
\begin{align}\label{GWeyl} G\bigl(x,y;e^{2\omega}g\bigr) &= e^{-\omega(x) -\omega(y)}G(x,y;g)\, ,\\\label{GzetaWeyl}
G_{\zeta}\bigl(x;e^{2\omega}g\bigr) &= e^{-2\omega(x)}\Bigl[ G_{\zeta}(x;g)
+\frac{1}{48\pi^{2}}\bigl(g^{ij}\partial_{i}\omega\partial_{j}\omega - \Delta_{g}\omega\bigr)\Bigr]\, .
\end{align}
Similar transformation laws were studied in \cite{2dgrav2} in two dimensions and the above relations can be proved by using similar methods.

The most straightforward derivation of \eqref{GWeyl} is to note that the Green's function $G(x,y;g)$ is entirely determined by the differential equation \eqref{Greendef}. It is then immediate to check that, if $G(x,y;g)$ satisfies \eqref{Greendef}, for the Laplacian $\Delta_{g}$ and Ricci scalar $R_{g}$ computed in the metric $g$, then $G(x,y;e^{2\omega}g)$ given by \eqref{GWeyl} satisfies the same equation, but now for the Laplacian and Ricci scalar
\begin{align}\label{DeltaWeylt} \Delta_{e^{2\omega}g} &= e^{-2\omega}\bigl(\Delta_{g} - 2 g^{ij}\partial_{i}\omega\partial_{j}\bigr)\, ,\\
\label{RicciWeyl} R_{e^{2\omega}g} &= e^{-2\omega}\bigl(
R_{g} - 6g^{ij}\partial_{i}\omega\partial_{j}\omega + 6\Delta_{g}\omega\bigr)
\end{align}
computed in the metric $e^{2\omega}g$. Another possible derivation is to start from the representation \eqref{Greendef2} in terms of the eigenvalues and eigenfunctions of the operator $D$ and to use standard quantum mechanical perturbation theory to compute the infinitesimal Weyl variation. A little algebra then yields the infinitesimal version of \eqref{GWeyl}. 

To prove \eqref{GzetaWeyl}, we are going to use the representation \eqref{Gspliteven}, which reads in the present case
\be\label{Gzetaconflim}G_\zeta(x)=\lim_{x\to y}\Bigl[ G(x,y)-{a_0(x,y)\over 4\pi^2 \ell^2(x,y)}\Bigr]\, .\ee
The Weyl variation of the first term on the right-hand side of \eqref{Gzetaconflim} is given by \eqref{GWeyl}. To compute the Weyl variation of the second term, we can proceed as follows.

First, the Weyl variation of the geodesic length $\ell(x,y)$, to any order when $y\rightarrow x$, can be found systematically by the following method. The geodesic $z(s)$ joining the points $x$ and $y$ satisfies the geodesic equation
\be\label{geodeq} \ddot z^{i} + \Gamma^{i}_{jk}\dot z^{j}\dot z^{k} = 0\, ,
\ee
with the boundary conditions $z(s=0)=x$ and $z(s=\ell(x,y))=y$. The solution of \eqref{geodeq} can be expanded in a Taylor series around $s=0$,
\be\label{solgeodexp} z^{i}(s) = \sum_{k\geq 0}\frac{s^{k}}{k!}z^{i(k)}(0) = x^{i} + s \dot z^{i}(0) - \sum_{k\geq 2}\frac{s^{k}}{k!}\Gamma^{i}_{i_{1}\cdots i_{k}}\dot z^{i_{1}}(0)\cdots\dot z^{i_{k}}(0)\, .
\ee
The completely symmetric coefficients $\Gamma^{i}_{i_{1}\cdots i_{k}}$ are expressed in terms of the Christoffel symbols as
\be\label{Gammasymformula} \Gamma^{i}_{i_{1}\cdots i_{k}} = \nabla'_{(i_{1}}
\cdots\nabla'_{i_{k-2}}\Gamma^{i}_{i_{k-1}i_{k})}(x)\, ,\ee
where the ``covariant derivatives'' $\nabla'$ act on the lower indices only. Setting $s=\ell$ in \eqref{solgeodexp}, we can solve recursively for $\ell\dot z^{i}(0)$ as a function of $\epsilon =y-x$, to any desired order. For example, we find in this way
\be\label{szexp} \ell\dot z^{i}(0) = \epsilon^{i} + \frac{1}{2}\Gamma^{i}_{i_{1}i_{2}}\epsilon^{i_{1}}\epsilon^{i_{2}} + \frac{1}{6}\bigl(
\partial_{(i_{1}}\Gamma_{i_{2}i_{3})}^{i} + \Gamma^{j}_{(i_{1}i_{2}}\Gamma^{i}_{i_{3})j}\bigr) \epsilon^{i_{1}}\epsilon^{i_{2}}\epsilon^{i_{3}} + O(\epsilon^{4})\, .\ee
Using $g_{ij}(x)\dot z^{i}(0)\dot z^{j}(0) =1$, an explicit formula for $\ell^{2}$ in terms of $g_{ij}(x)$, $\Gamma^{i}_{jk}(x)$ and $\epsilon$ can be derived by taking the square of \eqref{szexp}. The Weyl variation of $\ell^{2}$ is then found immediately from the Weyl variation of the Christoffel symbols,
\be\label{GammaWeyl} \Gamma^{i}_{jk}\bigl(x;e^{2\omega}g\bigr) = \Gamma^{i}_{jk}\bigl(x;g\bigr) + \delta_{j}^{i}\partial_{k}\omega + \delta_{k}^{i}\partial_{j}\omega - g_{jk}\partial^{i}\omega\, .\ee
The calculation can be simplified if we use a locally flat coordinate system around $x$, like the Riemann normal coordinates, because in this case $\Gamma^{i}_{jk}(x)=0$. In such a coordinate system, and for an infinitesimal Weyl variation, we get
\be\label{deltaellsq} \frac{\delta\ell^{2}}{\ell^{2}} = 2\delta\omega +
\epsilon^{i}\partial_{i}\delta\omega + \frac{1}{3}\partial_{i}\partial_{j}\delta\omega\,\epsilon^{i}\epsilon^{j} + O(\epsilon^{3})\, .\ee

Second, we need the Weyl variation of the heat kernel coefficient $a_{0}(x,y)$. In Riemann normal coordinates, \eqref{ficoeff} yields
\be\label{A0Riem} a_{0}(x,y) = 1 + \frac{1}{12}R_{ij}(x)\epsilon^{i}\epsilon^{j} + O(\epsilon^{3})\ee
and thus
\be\label{A02Riem} \delta a_{0}(x,y) = \frac{1}{12}\delta R_{ij}\epsilon^{i}\epsilon^{j} + \frac{1}{6}R_{ij}\epsilon^{i}\delta\epsilon^{j} + O(\epsilon^{3})\, .\ee
One must take into account the fact that after the Weyl transformation the new normal coordinates are different from the old ones. Hence $\delta R_{ij}=\hat R_{ij}-R_{ij}$, where $\hat R_{ij}$ is the Weyl transformed Ricci tensor in the new normal coordinate system. To the desired order in $\epsilon$, we only need the leading transformation law $\delta\epsilon^{i} = \epsilon^{i}\delta\omega$ of the normal coordinates under Weyl rescaling. Combining the effect of this coordinate transformation together with the usual transformation law of the Ricci tensor yields
\be\label{RWscaling} \delta R_{ij} = -2\delta\omega R_{ij} - 2 \partial_{i}\partial_{j}\delta\omega + \delta_{ij}\Delta\delta\omega + O(\epsilon)\ee
and thus
\be\label{deltaa0finalRiem} \delta a_{0}(x,y) = -\frac{1}{6}\partial_{i}\partial_{j}\delta\omega\, \epsilon^{i}\epsilon^{j} + \frac{1}{12}\ell^{2}\Delta\delta\omega + O(\epsilon^{3})\, .\ee
From \eqref{deltaa0finalRiem} and \eqref{deltaellsq}, we finally get the Weyl variation of the second term on the right-hand side in \eqref{Gzetaconflim}. To the desired order, the result can be most elegantly written as 
\be\label{deltaWeylsecondterm} \delta\biggl[\frac{a_{0}(x,y)}{\ell(x,y)^{2}} \biggr]= -\bigl(\delta\omega(x) + \delta\omega(y)\bigr)\frac{a_{0}(x,y)}{\ell(x,y)^{2}} + \frac{1}{12}\Delta\delta\omega + O(\ell)\, .\ee
We are also providing an alternative derivation of this formula in App.\ \ref{riemappsec}.

Putting together \eqref{GWeyl} and \eqref{deltaWeylsecondterm} into \eqref{Gzetaconflim}, we finally find
\be\label{deltaGzetainfi} \delta G_{\zeta} = -2\delta\omega\, G_{\zeta} - \frac{\Delta\delta\omega}{48\pi^{2}}\, \cvp\ee
which is the infinitesimal version of \eqref{GzetaWeyl}. To derive the more general equation \eqref{GzetaWeyl}, one can check that it is consistent with the composition of two finite Weyl transformations. An alternative derivation is to note that \eqref{deltaGzetainfi} implies that 
\be\label{deltacombGzet} \delta\Bigl(G_{\zeta} + \frac{R}{288\pi^{2}}\Bigr) = -2 \Bigl(G_{\zeta} + \frac{R}{288\pi^{2}}\Bigr) \delta\omega\, ,
\ee
which is easily integrated to 
\be\label{deltacombGzet2}G_{\zeta}\bigl(x;e^{2\omega}g\bigr) + \frac{R_{e^{2\omega}g}(x)}{288\pi^{2}} = e^{-2\omega(x)} \Bigl(G_{\zeta}\bigl(x;g\bigr) + \frac{R_{g}(x)}{288\pi^{2}}\Bigr)
\ee
and which yields \eqref{GzetaWeyl} by using \eqref{RicciWeyl}.

\subsubsection*{The vanishing of the two-loop conformal anomaly}

The infinitesimal Weyl variation of the two-loop gravitational effective action \eqref{S2conf} is
\be\label{Weyl2loopvar} \delta S^{(2)}(g) = \frac{\kappa_{4}}{8}\int\!\d^{4}x\sqrt{g}\, \bigl( 2G_{\zeta}\, \delta G_{\zeta} + 4G_{\zeta}^2\, \delta\omega\bigr)\, ,\ee
with $\delta G_\zeta$ given by \eqref{deltaGzetainfi}. Let us now remind ourselves that the anomaly is defined only up to the addition of local counterterms to the action. For example, the one-loop anomaly discussed in Sec.~3.5 was only defined up to the addition of the Weyl variation of the integral of a local scalar of dimension four, like $\int\! R^2$. Similarly, the two-loop anomaly \eqref{Weyl2loopvar} is defined only up to the Weyl variation of such local terms or the Weyl variation of terms of the form \eqref{Sctfinal} generated by finite shifts of the counterterm coefficients, like $\int\! G_\zeta\, R$. Keeping this in mind, let us note that the identity 
\be\label{Weylinv2}
\delta \left[\frac{\kappa_{4}}{8}\int\!\d^{4}x\sqrt{g}\, \Bigl(G_{\zeta} + \frac{R}{288\pi^{2}}\Bigr)^2 \right] = 0 \, ,
\ee
which is obvious from \eqref{deltacombGzet}, allows to rewrite \eqref{Weyl2loopvar} as
\be\label{S2varfinal1} \delta S^{(2)}(g) = -\frac{\kappa_{4}}{1152\pi^{2}}\,\delta\!\int\!\d^{4}x\sqrt{g}\, RG_{\zeta} - \frac{\kappa_{4}}{8 (288\pi^{2})^{2}}\,\delta\!\int\!\d^{4}x\sqrt{g}\, R^{2} \ .
\ee
Both terms on the right-hand side of \eqref{S2varfinal1} can be absorbed by a finite shift of the counterterms, as just discussed. This proves that the two-loop conformal anomaly vanishes.

\vskip8.mm
\subsection*{Acknowledgments}

This work is supported in part by the belgian FRFC (grant 2.4655.07) and IISN (grants 4.4511.06 and 4.4514.08).

\newpage
\begin{appendix}
\section{Appendix}

\subsection{Riemann normal coordinates}\label{riemappsec}

It is often useful to go to a special coordinate system for which the metric in the vicinity of a given point is as close to the Euclidean metric as possible. In general relativity, these are the coordinates of a freely falling observer. More generally, one defines these Riemann normal coordinates as follows.

\subsubsection*{Definition and basic relations}

We fix any point as the origin (which would have coordinates $y$ in a general coordinate system). Riemann normal coordinates $x$ around this point are defined such that straight lines through the origin are geodesics, and scaled such that the square of the geodesic distance is simply
\be\label{geodistnormal}
\ell^2(x,0)=x^i x^i\equiv x^2 \ .
\ee
The Taylor expansion of the metric around $x=0$ is given by (see e.g.\ \cite{Schubert})
\begin{align}\label{metricnormalexp}
g_{ij}(x)&= \delta_{ij}-{1\over 3} R_{ikjl}x^k x^l -{1\over 6}R_{ikjl;m}x^k x^l x^m 
\nonumber\\
&\hskip3.mm+\ \Big[{2\over 45} R_{ikrl}R^r_{\ mjn} 
-{1\over 20}R_{ikjl;mn}\Big] x^k x^l x^m x^n +O(x^5) \ ,
\end{align}
where $R^i_{\ jkl}$  is the Riemann curvature tensor,\footnote{Our sign convention is $R^i_{\ jkl}=\del_k \G^i_{lj}-\del_l \G^i_{kj}+\G^i_{kr}\G^r_{lj}-\G^i_{lr}\G^r_{kj}$.} 
$(\ldots)_{;mn}=\nabla_n\nabla_m(\ldots)$ denotes covariant derivatives, and $R_{kl}=R^j_{\ kjl}$ and $R=R^k_{\ k}$ denote the Ricci tensor and scalar. All curvature tensors are at $x=0$. In particular, since $g_{ij}(0)=\delta_{ij}$ we do not have to distinguish upper and lower indices on these tensors. The inverse metric is 
\begin{align}\label{inversemetricexp}
g^{ij}(x)=&\ \delta_{ij}+{1\over 3} R_{ikjl}x^k x^l +{1\over 6}R_{ikjl;m}x^k x^l x^m 
\nonumber\\
&\hskip-3.mm+\ \Big[{1\over 15} R_{ikrl}R^r_{\ mjn} 
+{1\over 20}R_{ikjl;mn}\Big] x^k x^l x^m x^n +O(x^5) \ .
\end{align}
Note that in these coordinates one has
$g_{ij} x^j=x^i $ and $g^{ij} x^j = x^i$
since all other terms involve symmetric products of the coordinates $x$ contracted with antisymmetric curvature tensors. The square root of the determinant of the metric is
\begin{align}\label{sqrtg}
\sqrt{g}\ =\ &1-{1\over 6} R_{kl}x^k x^l -{1\over 12} R_{kl;m} x^k x^l x^m \nonumber\\
&+ \left[
{1\over 72}  R_{kl} R_{mn} -{1\over 40}R_{kl;mn} -{1\over 180} R^r_{\ kls}R^s_{\ mnr}
\right] x^k x^l x^m x^n +O(x^5) \ ,
\end{align}
Note that the terms $O(x^5)$ are terms $O(R^3, \nabla R^2)$, i.e.\ involving at least three curvature tensors or two curvature tensors and a covariant derivative.

\subsubsection*{Laplace operator}

The scalar Laplace operator is given in general by 
\be\label{Laplace}
-\Delta={1\over \sqrt{g}}\del_i\left(\sqrt{g} g^{ij}\del_j \right)
=g^{ij}(x)\del_i\del_j +r^j(x) \del_j \ ,
\ee
where we defined
\be\label{ffct}
r^j=\del_i g^{ij} + {1\over 2} g^{ij} \del_i \ln  g \ .
\ee
Using \eqref{inversemetricexp} and \eqref{sqrtg}, a straightforward computation yields for the function $r^j(x)$ in the normal coordinates
\ba\label{ffct2}
\hskip-0.mm r^j(x)&=&-{2\over 3} R_{jk}x^k +\Big[ {1\over 12} R_{kl;j}-{1\over 2} R_{kj;l}\Big] x^k x^l 
-
\Big[{1\over 5} R_{jl;mn}+{1\over 40} R_{mn;lj}-{3\over 40} R_{mn;jl}\nonumber\\
&&\hskip0.5cm 
- \ {23\over 180} R_{ls} R^s_{\ mnj}+{4\over 45} R_{rjls}R_{smnr}\Big] x^l x^m x^n  
\ +\ O(x^4) \ .
\ea
Inserting this into (\ref{Laplace}) yields the expansion of the Laplace operator in the vicinity of $x=0$, up to terms $O(x^4)$ or $O(R^3,\nabla R^2)$. In particular, we have 
\be -\Delta\, \ell^2(x,0)=\hskip-2.mm \left[g^{ij}(x)\del_i\del_j + r^j(x)\del_j\right] (x^k x^k)
= 2\left[g^{jj}(x)+r^j(x) x^j\right]\, ,\ee so that
\begin{align}\label{Laplacegeodist2}
\Delta\, \ell^2(x,0)+2d &={2\over 3} R_{kl} x^k x^l +{1\over 2} R_{kl;m}x^k x^l x^m \nonumber\\
& +\left[{1\over 5} R_{kl;mn} +{2\over 45} R^r_{\ kls}R^s_{\ mnr}\right] x^k x^l x^m x^n  
\ +\ O(x^5)\, .
\end{align}

\subsubsection*{Weyl transformation of the geodesic length}

In the main text, we needed  the transformation of the geodesic length $\ell$ under Weyl rescalings of the metric. One way to prove this formula is to use normal coordinates, as we will do now. Consider the (infinitesimal) Weyl transformation of the metric \eqref{Weylscaling}, i.e.
\be\label{Weylmetric}
\wt g_{ij}(x)=e^{2\omega(x)} g_{ij}(x) \ .
\ee
The curvature tensors of the new metric are then
\begin{align}\label{Weylcurvature}
\wt R^i_{\ jkl}&=R^i_{\ jkl} -\delta_k^i \nabla_l\del_j\omega + \delta_l^i\nabla_k\del_j\omega +g_{jk} \nabla_l\del^i\omega -g_{jl}\nabla_k\del^i\omega + O(\omega^2) \ ,
\nonumber\\
\wt R_{jl}&=R_{jl}-(d-2)\nabla_l\del_j\omega +g_{jl} \Delta\omega+O(\omega^2)\ ,
\nonumber\\
\wt R&=e^{-2\omega}\big( R+2(d-1)\Delta\omega\big) +O(\omega^2) \ .
\end{align}
The new metric does not have the form \eqref{metricnormalexp}, i.e.\ the coordinates $x^i$ are not normal coordinates for the Weyl transformed geometry. Instead, let $\hat x^i(x)$ be the new normal coordinates. Then the metric in these coordinates is $\hat g_{ij}={\del x^m\over \del \hat x^i}{\del x^n\over \del \hat x^j}\wt g_{mn}$. Since the $\hat x^i$ are normal coordinates, this metric and corresponding curvature must satisfy \eqref{metricnormalexp}, 
\be\label{xhatnormal}
{\del x^m\over \del \hat x^i}{\del x^n\over \del \hat x^j}\,\wt g_{mn}=\hat g_{ij}(\hat x)=\delta_{ij}-{1\over 3} \hat R_{ikjl}(0)\hat x^k\hat x^l + O(\hat x^3)\ .
\ee
Obviously, to lowest order in $x$, this implies $\hat x^i=e^{\omega(0)} x^i +O(x^2)$, so that \be\hat R_{ikjl}(0)=e^{-4 \omega(0)} \wt R_{ikjl}(0)=e^{-2 \omega(0)} \wt R^i_{\ kjl}(0)\, .\ee Inserting this back into \eqref{xhatnormal} and using \eqref{Weylmetric} and \eqref{Weylcurvature} we get
\begin{align}\label{xhatnormal2}
e^{2\omega(x)}\big(\delta_{mn}-{1\over 3}R_{mknl}x^k x^l\big)={\del \hat x^i\over \del x^m}&{\del \hat x^j\over \del x^n}\Big( \delta_{ij}-{1\over 3}\big(
R_{ikjl} -\delta_{ij}\del_l\del_k\omega + \delta_{il}\del_k\del_j\omega\nonumber\\
& +\delta_{jk} \del_l\del_i\omega -\delta_{kl}\del_i\del_j\omega\big) x^k x^l\Big) + O(x^3,\omega^2)\ .
\end{align}
If we insert
\be\label{xxhat}
\hat x^i=e^{\omega(0)}\big( x^i + f^i(x)\big) \ , \quad f^i=O(x^2) \ ,
\ee
we see that $f^i$ must be $O(\omega)$ and satisfy
\begin{align}\label{fief}
\del_m f^n+\del_n f^m= 2\delta_{mn}\del_k \omega(0) x^k +{1\over 3}& \Big( 2\delta_{mn}\del_l\del_k\omega(0) +\delta_{ml}\del_k\del_n\omega(0) + \delta_{kn}\del_l\del_m\omega(0)\nonumber\\
&-\delta_{kl}\del_n\del_m\omega(0)\Big) x^k x^l +O(x^3,\omega^2)\, .
\end{align}
\vskip-5.mm
\noindent
The solution is
\be\label{Fisol}
f^i=x^i\Big( x^l\del_l\omega(0) +{1\over 3} x^k x^l \del_k\del_l\omega(0)\Big)-{1\over 2}x^m x^m \Big( \del_i\omega(0)+{1\over 3} x^l \del_l \del_i \omega(0)\Big)+O(x^4,\omega^2)\, .
\ee
It follows that the geodesic distance in the Weyl transformed geometry is given by
\begin{align}\label{geodisttilde}
\wt\ell^2(x,0)&\equiv \hat\ell^2(x,0)=\hat x^i \hat x^i=e^{2\omega(0)}\Big( x^i x^i + 2 x^i f^i +O(\omega^2)\Big)\nonumber\\
&=e^{2\omega(0)} x^i x^i \Big( 1+x^l\del_l\omega(0)+{1\over 3} x^k x^l\del_k\del_l\omega(0)\Big) +O(x^5,\omega^2)\nonumber\\
&=\ell^2(x,0) \Big(1+\omega(x)+\omega(0)-{1\over 6} x^k x^l \del_k\del_l \omega(0) + O(x^3,\omega^2)\Big) \, ,
\end{align}
or, in terms of the Weyl variation $\delta_\omega$ of $\ell^2(x,0)=x^2\equiv x^i x^i$,
\be\label{Weylvar1}
\delta_\omega {1\over x^2}= -{\omega(x)+\omega(0)\over x^2} +{x^k x^l\del_k\del_l\omega(0)\over 6 x^2} +O(x,\omega^2)\, .
\ee
One shows similarly (but with less effort) that
\be\label{Weylvar2}
\delta_\omega \left( R_{ij} {x^i x^j\over x^2}\right) = -\big(\omega(x)+\omega(0)\big) R_{ij} {x^ix^j\over x^2} -(d-2) {x^i x^j \del_i\del_j \omega(0)\over x^2} + \Delta\omega(0) +O(x,\omega^2) \, ,
\ee
so that, in $d=4$, the $x^i x^j \del_i\del_j \omega(0)$ terms disappear from the following combination
\be\label{Weylvar3}
\delta_\omega \left( {1\over x^2} +  {R_{ij} x^i x^j\over 12\, x^2}\right) = -\big(\omega(x)+\omega(0)\big) \left( {1\over x^2}+ {R_{ij} x^ix^j\over12 \,x^2}\right) + {1\over 12}\Delta\omega(0) +O(x,\omega^2) \, .
\ee

\subsection{The heat kernel\label{heatapp}}

In this appendix, we will (re)derive or collect useful formulas about the standard heat kernel $K(t,x,y)$.
It is defined in terms of the eigenvalues $\l_r$ and eigenfunctions $\p_r$ of the wave operator $D=\Delta+m^2+\xi R$, see \eqref{fieldeq}, as
\be\label{heatxdiffy1}
K(t,x,y)=\sum_{r} e^{-\l_r t} \p_r(x) \p_r^*(y) \, .
\ee
Since the operator $D$ is real, one may choose real eigenfunctions $\p_r$, which we assume throughout. They satisfy the orthonormality and completeness relations
\be\label{orthocompl}
\int\!\d^d x\, \sqrt{g(x)}\, \p_r(x) \p_s(x)=\delta_{rs}\, , \quad
\sum_r \p_r(x)\p_r(y)=\delta^{(d)}(x-y) \, [g(x) g(y)]^{-1/4} \, .
\ee
It follows from \eqref{heatxdiffy1} that the heat kernel satisfies 
the fundamental differential equation (generalized heat equation)
\be\label{eqdiff}
\left( {\d\over \d t}+D_x\right) K(t,x,y)=\left( {\d\over \d t}+\Delta_x+m^2+\xi R(x)\right) K(t,x,y) =0\ ,
\ee
with the initial condition
\be\label{initial}
\lim_{t\rightarrow 0} K(t,x,y) = \delta^{(d)}(x-y) \, [g(x) g(y)]^{-1/4}\, .
\ee
Note that, as is obvious from \eqref{heatxdiffy1} or \eqref{eqdiff}, the massive and massless heat kernels are simply related by
\be\label{KmtoK}
K(t,x,y)=e^{-m^2 t}K_{m=0}(t,x,y) \ .
\ee
In flat space, the well-known solution is \be K(t,x,y)=(4\pi t)^{-d/2}
e^{ -m^2 t -{(x-y)^2\over 4 t}}\, .\ee
This must also be the leading small $t$, small distance behavior on a curved manifold, if $(x-y)^2$ is replaced by the square of the geodesic distance $\ell^2(x,y)$ between $x$ and $y$. Corrections to this leading behavior can then be obtained as a perturbative expansion in $t$ and $\ell(x,y)$. 

\subsubsection*{Asymptotic solution of the heat equation and recursion relations}

We will now show that the heat kernel $K(t,x,y)$ admits an asymptotic small $t$ expansion of the form repeatedly used in the main text and derive recursion relations between the coefficients $a_k(x,y)$. This will be done in general. Then we use normal coordinates to explicitly solve the recursion relations and obtain the first few coefficients $a_k$. 

We search for a solution of (\ref{eqdiff}), (\ref{initial}) of the form
\be\label{Kseries}
K(t,x,y)= (4\pi t)^{-d/2} e^{ -{\ell^2(x,y)\over 4 t}}
F(t,x,y) \, ,
\ee
where $\ell(x,y)$ is the geodesic distance between $x$ and $y$ and $F$ is expanded in integer powers of $t$,
\be\label{Fexp}
F(t,x,y)=\sum_{k=0}^\infty a_k(x,y) t^k \ .
\ee
Clearly, this generalizes the flat space result.

Recall that the scalar Laplace operator is given in general by 
\eqref{Laplace}. 
This obviously implies for any $F(x)$ and $H(x)$
\be\label{productformula}
\Delta(F H)=(\Delta F) H + F \Delta H +2 g^{ij}\del_i F \del_j H \ .
\ee
Using this formula, we find
\be\label{LaplKF}
-\Delta_x\left( {e^{-\ell^2/(4t)} F\over t^{d/2}} \right)
={e^{-\ell^2/(4t)}\over t^{d/2}}\Biggl[ {g^{ij}(x)\del_i\ell^2\del_j\ell^2\over 16 t^2} F +{(\Delta_x \ell^2)\over 4t} F -{g^{ij}(x)\del_i\ell^2\over 2t}  \del_j F -\Delta_x F \Biggr]\, ,
\ee
where $\del_i= \del/\del x^i$. 
Now, $g^{ij}(x)\del_i\ell(x,y)\del_j\ell(x,y)=1$, so that
\be\label{ellder}
g^{ij}(x)\del_i\ell^2 \del_j \ell^2 = 4 \ell^2 \ ,
\ee 
and adding the pieces with $\xi R(x)+m^2$ we get
\be\label{LaplKF2}
-D_x\left( {e^{-\ell^2/(4t)} F\over t^{d/2}} \right)
={e^{-\ell^2/(4t)}\over t^{d/2}}\Biggl[ {\ell^2\over 4 t^2} F +{(\Delta_x \ell^2)\over 4t} F -{g^{ij}(x)\del_i\ell^2\over 2t}  \del_j F -D_x F \Biggr]\, .
\ee
On the other hand,
\be\label{dtKF}
{\d\over \d t} \left({e^{-\ell^2/(4t)} F\over t^{d/2}}\right)
={e^{-\ell^2/(4t)}\over t^{d/2}}\Biggl[{\ell^2\over 4 t^2}F-{d\over 2 t}F +{\d F\over \d t}\Biggr] \, .
\ee
Upon inserting this into the heat equation \eqref{eqdiff} and using \eqref{Fexp}, one finds  the following system of differential equations for the functions $a_r(x,y)$,
\begin{align}\label{foeq}
-{1\over 4}(\Delta_x \ell^2+2 d) a_0+{1\over 2}g^{ij}(x)\del_i\ell^2 \del_j a_0 &= 0 \ ,
\\
\label{freq}
-{1\over 4}(\Delta_x \ell^2+2 d) a_r+{1\over 2}g^{ij}(x)\del_i\ell^2 \del_j a_r + r a_r&= -D_x a_{r-1} \, , \ r\ge 1 \, .
\end{align}
Defining $a_r=a_0\, \wt a_r$, the latter equations simplify to
\be\label{freqbis}
{1\over 2}g^{ij}(x)\del_i\ell^2  \del_j \wt a_r + r \wt a_r= -{1\over f_0}D_x a_{r-1}
\, , \quad a_r=a_0 \wt a_r \, .
\ee
Note that the heat kernel $K(t,x,y)$, as well as the expansion \eqref{Kseries}, are symmetric under the interchange of $x$ and $y$. Thus all $a_r(x,y)$ are symmetric functions, and they obey also the same equations \eqref{foeq}--\eqref{freqbis} with the differential operators acting on $x$ replaced by those acting on $y$. To explicitly solve these equations it simplifies things to go to the Riemann normal coordinate system which we discussed above.

\subsubsection*{Solution to the recursion relations of the heat equation in normal coordinates}

We will solve the recursion relations \eqref{foeq}--\eqref{freqbis} and obtain the first few coefficients $a_k(x,y)$. This will be done using normal coordinates centered in $y$, i.e.\ we set $y=0$. We also let $m=0$ first, since the $m$-dependence can be trivially restored in the end as is obvious from \eqref{KmtoK}. 
The recursion relations then read
\begin{align}\label{foeqbis}
-{1\over 4}(\Delta \ell^2+2 d) a_0+x^j \del_j a_0 &= 0 \ ,
\\
\label{freqbis2}
x^j \del_j \wt a_r + r \wt a_r&= -{1\over f_0}\big( \Delta+\xi R(x)\big) a_{r-1}\, , \quad a_r=a_0 \wt a_r \, .
\end{align}
Equation \eqref{foeqbis} for $a_0$ is easily solved. Using (\ref{Laplacegeodist2}) and taking into account the initial condition to fix the normalization, we get
\begin{align}\label{f0}
a_0(x,0)&=1+{1\over 12} R_{kl} x^k x^l +{1\over 24} R_{kl;m}x^k x^l x^m  \nonumber\\
&+{1\over 16}\left({1\over 5} R_{kl;mn} +{2\over 45} R^r_{\ kls}R^s_{\ mnr}+{1\over 18}  R_{kl} R_{mn}\right) x^k x^l x^m x^n  
 +\ O(x^5)\, .
\end{align}
Clearly, the unwritten terms $O(x^5)$ are terms $O(R^3, \nabla R^2)$.
Let us insist that all curvature tensors are evaluated at $0$ (which corresponds to $y$), unless otherwise stated.
Comparing with \eqref{sqrtg}, we see that $a_0(x,0)=g(x)^{-1/4}$. Actually, this is the well-known   VanVleck-Morette determinant, which in normal coordinates indeed reduces to $g(x)^{-1/4}$. It is also easy to see directly that $g(x)^{-1/4}$ is a solution of (\ref{foeq}). Finally note that $a_0(x,0)=g(x)^{-1/4}$ is precisely what is required for the initial condition (\ref{initial}).

Next, one solves \eqref{freqbis2} recursively. With our expansion, we will get $\Delta a_0$ up to terms $O(x^3)$ which are still $O(R^3, \nabla R^2)$. One also has to expand \be\xi R(x)=\xi R+\xi R_{;l}x^l + {1\over 2} \xi R_{;kl}x^k x^l + \cdots\ee Thus
\begin{align}\label{Laplf02}
-{1\over a_0}\big(\Delta+\xi\, R(x)\big) a_0&= {1-6\xi\over 6}  R +{1-6\xi\over 6} R_{;l}x^l \nonumber\\
&+ \big(D_{kl}-{1\over 72} R R_{kl}-{\xi\over 2} R_{;kl}\big) x^k x^l 
+ \ O(R^3, \nabla R^2) \, ,
\end{align}
where we defined
\be\label{Dtensor}
D_{kl}={3\over 20} R_{(jj;kl)} +{1\over 30} R^{s\ \ \ r}_{\ (jj}R^{r\ \ \ s}_{\ kl)} +{1\over 24} R_{(jj}R_{kl)} -{1\over 9} R_{kj} R_{lj} +{1\over 18} R_{ij}R_{ikjl} \, .
\ee
The solution $a_1$ is then easily seen to be
\begin{align}\label{f1sol}
a_1(x,0)=&\ {1-6\xi\over 6} R +{1-6\xi\over 12} R_{;l}x^l + {1\over 3}\left[D_{kl} +{1-9\xi\over 36} R R_{kl} -{\xi\over 2} R_{;kl}\right] x^k x^l\nonumber\\
&+O(R^3, \nabla R^2)\, .
\end{align}
(One cannot add  solutions of the homogeneous equation since they are singular at $x=0$.) 

Computing similarly
\be -\Delta a_1={2\over 3} D_{kk} + {1-9\xi\over 54} R^2 -{\xi\over 3}R_{;k}^{\ ;k}+{\cal O}(R^3, \nabla R^2)\, ,\ee we get
\be\label{f2}
a_2(x,0)= {1\over 3}D_{kk} +{1-18\xi+54\xi^2\over 108} R^2 -{\xi\over 6} R_{;k}^{\ ;k}+{\cal O}(R^3, \nabla R^2) \ .
\ee
Clearly then, 
\be\label{f3}
a_r(x,0)={\cal O}(R^3, \nabla R^2) \ , \quad\  \forall r\ge 3 \ .
\ee

Finally, it is trivial to relate these massless heat kernel coefficients to those of the massive heat kernel. It immediately follows from \eqref{KmtoK} that
\be\sum_r a_r^m\,  t^r = \sum_r a_r^{m=0} \, t^r e^{-m^2 t}\, .\ee Also, it will be convenient for referencing to restore $y$: we still use normal coordinates around $y$, but we no longer set $y=0$. Obviously, this is achieved by replacing $x^i$ everywhere by $x^i-y^i$.

Let us summarize: in general, the heat kernel has the  expansion given by \eqref{Kseries}, \eqref{Fexp}.
In Riemann normal coordinates around $y$, we have
$\ell^2(x,y)=(x-y)^2$ and the {\it massless} $a_k$ are given by
\begin{align}\label{ficoeff}
a_0(x,y)=&\ (g(x))^{-1/4}\nonumber\\
=&\ 1+{1\over 12} R_{kl} (x-y)^k (x-y)^l +{1\over 24} R_{kl;m}(x-y)^k (x-y)^l (x-y)^m  
\nonumber\\
&+{1\over 16}\Big[{1\over 5} R_{kl;mn} +{2\over 45} R^r_{\ kls}R^s_{\ mnr}\nonumber\\
& \hskip10.mm+{1\over 18}  R_{kl} R_{mn}\Big] (x-y)^k (x-y)^l (x-y)^m (x-y)^n  +O(R^3, \nabla R^2)\ ,
\nonumber\\
a_1(x,y)=&\ {1-6\xi\over 6} R +{1-6\xi\over 12} R_{;l}(x-y)^l \nonumber\\
&+ {1\over 3}\left[D_{kl} +{1-9\xi\over 36} R R_{kl} -{\xi\over 2} R_{;kl}\right] (x-y)^k (x-y)^l +O(R^3, \nabla R^2)\ ,
\nonumber\\
a_2(x,y)=&\ {1\over 180}\hskip-1.mm\left[ R_{mnkl}R^{mnkl} -R_{mn} R^{mn}+\Big({5\over 2}-30\xi + 90 \xi^2\Big) R^2 + \big(6-30\xi\big) R_{;m}^{\ \ ;m} \right]\nonumber\\
&+O(R^3, \nabla R^2)\, ,
\end{align}
where all curvature tensors are evaluated at $y$. The tensor $D_{kl}$ was given in \eqref{Dtensor} and we have used the Bianchi identities to simplify $D_{kk}$.

The {\it massless diagonal} heat kernel coefficients $a_k(y)=a_k(y,y)$ obviously are 
\begin{align}\label{ardiag}
a_0(y)&=1 \nonumber\\
a_1(y)&={1-6\xi\over 6} R(y)\nonumber\\
a_2(y)&={1\over 180}\hskip-1.mm\left[ R_{mnkl}R^{mnkl} -R_{mn} R^{mn}+\Big({5\over 2}-30\xi + 90 \xi^2\Big) R^2 + \big(6-30\xi\big) R_{;m}^{\ \ ;m} \right] \, .
\end{align}
The massive coefficients are related to the massless ones by
\be\label{frmfr}
a_0^m=a_0^{m=0} \ , \quad
a_1^m=a_1^{m=0}-m^2 a_0^{m=0} \ , \quad
a_2^m=a_2^{m=0} - m^2 a_1^{m=0} +{m^4\over 2} a_0^{m=0} \, .
\ee

As already noted, the coefficients $a_k(x,y)$ must be symmetric in $x$ and $y$. Although this is not manifest on our expressions \eqref{ficoeff}, it is nevertheless easily checked: if we  Taylor expand the curvature tensors at $y$ around $x$, we end up with the same expressions, except that the roles of $x$ and $y$ are switched (at least to the order in $x-y$ that we have kept).

It is clear from our above derivation that the terms $O(t^3)$ are terms $O(R^3, \nabla R^2)$. Note also that the parameter $\xi$ which entered through the combination $\xi R$, can only appear in terms containing the curvature scalar and its derivatives. In particular, $a_0(x,y)$ does not depend on $\xi$. It is reassuring to note that our (diagonal) $a_2(y)$ coincides\footnote{
The only difference is the overall sign of the term $\big(6-30\xi\big) R_{;m}^{\ \ ;m}$. We believe that the sign in the review would be  correct when using a Minkowski signature $+---$ but should be switched when using a Minkowski signature $-+++$  or in Euclidean space.
} 
with the one quoted in \cite{Vassil}. In turn, this also provides an indirect check  for the (non-diagonal) coefficient $a_1(x,y)$.

\subsubsection*{Two formula involving  products of $n$ heat kernels}

A useful formula involving the product of $n$ heat kernels is
\begin{multline}\label{Kn2}
\sqrt{g(x)}\prod_{i=1}^n K(t_i,x,0)={\exp\left( -m^2 \sum_i t_i-{\sum_i t_i^{-1}\over 4} x^k x^k\right)\over (4\pi )^{nd/2}(\prod_i t_i)^{d/2}}
\Bigg\{
\Big[ { 1+{n-2\over 12} R_{kl} x^k x^l }\\
+{n-2\over 24}R_{kl;m}x^k x^l x^m  
 +{n-2\over 16}\Big({1\over 5} R_{kl;mn} +{2\over 45} R^r_{\ kls}R^s_{\ mnr}+{n-2\over 18} R_{kl} R_{mn}\Big) x^k x^l x^m x^n  \Big]
\\
+\Bigl(\sum_i t_i\Bigr)
\Big[ {1-6\xi\over 6}R +{1-6\xi\over 12} R_{;l}x^l 
 + \Big({1\over 3}D_{kl} +{1-9\xi\over 108} R R_{kl}-{\xi\over 6} R_{;kl}\\+{(n-3)(1-6\xi)\over 72} R \, R_{kl} \Big) x^k x^l  \Big]\\
 +{\bigl(\sum_i t_i^2\bigr)\over 180}\left[ R_{mnkl}R^{mnkl} -R_{mn} R^{mn}+\Bigl({5\over 2}-30\xi + 90 \xi^2\Bigr) R^2 + \big(6-30\xi\big) R_{;m}^{\ \ ;m} \right]
\\
+{(\sum_{i>j}t_i t_j)\, \Bigl({1\over 6}-\xi\Bigr)^2 R^2
+O(R^3, \nabla R^2) }\Bigg\} \, .
\end{multline}
Now, as long as at least one of the $t_i$ is small, $\sum_i t_i^{-1}$ is large and the exponential factor in (\ref{Kn2}) provides an exponential suppression, unless the $x^l$ are all small. In particular, this then justifies the expansion in normal coordinates, and  we can safely replace the integral of (\ref{Kn2}) over the manifold by an integral over ${\bf R}^n$. The latter are trivial to perform and we arrive at  
\ba\label{Knint}
\int\! \d^d x\sqrt{g(x)}\prod_{i=1}^n K(t_i,x,0)\hskip-2.mm&=&\hskip-2.mm
{e^{-m^2\sum_i t_i}\over (4\pi)^{(n-1)d/2}(\prod_i t_i)^{d/2} (\sum_i t_i^{-1})^{d/2}} 
\Bigg\{ {\textstyle 1+{n-2\over 6\sum_i t_i^{-1}} R}
\nonumber\\
&&\hskip-5.cm
 {\textstyle +{n-2\over 60(\sum_i t_i^{-1})^2}\left(R_{mnkl} R^{mnkl}+{5n-8\over 3} R_{mn} R^{mn}
 +{5(n-2)\over 6} R^2+6 R_{;m}^{\ \ ;m}\right) }
\nonumber\\
&&\hskip-5.cm
 {\textstyle +(\sum_i t_i)\, \big({1\over 6}-\xi\big)R 
+(\sum_{i>j}t_i t_j)\, \big({1\over 6}-\xi\big)^2 R^2}
\nonumber\\
&&\hskip-5.cm 
 {\textstyle + {\sum_i t_i\over 90 \sum_i t_i^{-1}} 
\left(
R_{mnkl} R^{mnkl}- R_{mn} R^{mn}
 +{5(n-2)\over 2}(1-6\xi)  R^2+(6-30\xi) R_{;m}^{\ \ ;m}
\right) }
\nonumber\\
&&\hskip-5.cm 
 {\textstyle +{(\sum_i t_i^2)\over 180}\hskip-1.mm\left( R_{mnkl}R^{mnkl} -R_{mn} R^{mn}+\Big({5\over 2}-30\xi + 90 \xi^2\Big) R^2 + \big(6-30\xi\big) R_{;m}^{\ \ ;m} \right)}
\nonumber\\
&&\hskip-5.cm
+O(R^3, \nabla^2 R^2) \Bigg\} \ .
\ea

\subsection{Short-distance expansion of the Green's function\label{Greenapp}}

We have shown in the main text how the short-distance expansion of the Green's function $G(x,y)$ is given in terms of the heat kernel coefficient $a_r(x,y)$, see \eqref{Greenshortd4}, \eqref{Gspliteven}, \eqref{Gsplitodd}. We have found it instructive to show in this appendix, that this same short-distance expansion of the Green's function can also be derived in a very elementary way directly from the recursion relations \eqref{foeq}, \eqref{freq} of the  $a_r$. As a bonus, we will obtain this expansion up to and including terms $\sim \ell^2(x,y)$. We will also see how $G_\zeta$ arises in this expansion as a quantity that cannot be determined from the short distance expansion.

The Green's function satisfies
\be\label{Greeneq}
\big(\Delta_x +\xi R(x)+m^2\big) G(x,y) = \delta^{(d)}(x-y)\, [g(x)g(y)]^{-1/4} \ .
\ee
(We assume there is no zero eigenvalue.) The Green's function and the heat kernel are related by $G(x,y)=\int_0^\infty \d t\,  K(t,x,y)$. Both $G(x,y)$ and $K(x,y)$ are symmetric under interchange of $x$ and $y$.

Obviously, knowledge of the Green function requires more than just the small $t$ asymptotic expansion of the heat kernel. However, just as we constructed this small $t$ expansion of $K$ starting from the exact solution in flat space, we are now going to construct a short-distance development of the Green function $G(x,y)$ starting from the exact solution of (\ref{Greeneq}) in flat space. Not surprisingly, this short-distance development involves exactly the heat kernel coefficients $f_r(x,y)$ via their recursion relations (\ref{foeq}) and (\ref{freq}).

In flat $d$-dimensional space, the Green's functions are easily obtained by integrating the corresponding heat kernels and are exactly given by Bessel functions,
\be\label{flatGreenBessel}
G_{\rm flat}(x,y)={m^{d-2}\over (2\pi)^{d/2}} (m^2 \ell^2)^{2-d/4}
K_{\frac{d}{2}-1}\Bigl( \sqrt{m^2 \ell^2}\Bigr) \, ,
\ee
where we wrote $\ell^2=(x-y)^2$. For $d=2$ and $d=4$ their 
short-distance expansions are explicitly given by
\begin{align}\label{flatGreenexp}
G_{\rm flat}^{d=2}(x,y)&={1\over 4\pi}\Biggl[-\ln {m^2\ell^2\over 4} -2\g\ +{m^2\ell^2\over 4}\Big(-\ln {m^2\ell^2\over 4}-2\g+2\Big) +O(\ell^4\ln\ell^2) \Biggr]\, , \nonumber\\
G_{\rm flat}^{d=4}(x,y)&={m^2\over (4\pi)^2}\Biggl[{4\over m^2\ell^2}+\ln {m^2\ell^2\over 4} +2\g-1 \ +{m^2\ell^2\over 8}\Big(\ln {m^2\ell^2\over 4}+2\g-{5\over 2}\Big)\nonumber\\
&\hskip1.6cm +O(\ell^4\ln\ell^2) \Biggr]\, .
\end{align}
These expansions may also be viewed as expansions in $m$. In particular, the leading short-distance singularities are given by the massless Green's functions.

In the sequel, we will construct the generalizations of these expansions in curved space.  We will repeatedly use the following formula for the Laplacian of the product of functions $F(x,y)$ and $H(\ell^2(x,y))$:
\be\label{product1}
\Delta_x\big( H(\ell^2) F(x,y)\big)= H \Delta_x F-2 H'\, g^{ij}(x)\del_j\ell^2\partial_i F+H'(\Delta_x \ell^2) F -4 \ell^2\,  H'' F \, ,
\ee
which is easily established using (\ref{Laplace}) and \eqref{ellder}.
Adding $(\xi R+m^2) H F$ on both sides of the equation, this becomes
\be\label{product2}
D_x\big( H(\ell^2) F(x,y)\big)= H D_x F-2 H'\, g^{ij}(x)\del_j\ell^2\partial_i F+H'(\Delta_x \ell^2) F -4 \ell^2\,  H'' F \, .
\ee

\smallskip

\noindent\emph{The case $d=4$}

\noindent
We now restrict ourselves to $d=4$ dimensions.  As just recalled, in flat space, the leading singularity is given by
\be G_{\rm flat}^{m=0}(x,y)={1\over (4\pi)^2}\, {4\over (x-y)^2}\,\cdotp\ee In curved space, taking into account that  the $\delta^{(4)}(x)$ is accompanied by \be (g(x))^{-1/4}(g(y))^{-1/4}\, ,\ee we see that the first term in a small $\ell(x,y)$ expansion should be
\be\label{Green1}
G(x,y)={1\over (4\pi)^2}\, {4 a_0(x,y)\over \ell^2(x,y)}+\cdots
\ee
To find the corrections to this expression, we use \eqref{product2}, for $\ell^2\ne 0$, with $H(\ell^2)=1/\ell^2$ and $F(x,y)=a_0(x,y)$. All but the first term on the right-hand side of \eqref{product2} then exactly cancel thanks to (\ref{foeq}) satisfied by $a_0$. Thus
\be\label{DelG1}
D_x\left( {4a_0\over \ell^2}\right)={4 D_x a_0\over \ell^2} \,\cdotp
\ee
Next, we use \eqref{product2} with $H(\ell^2)=\ln{\mu^2\ell^2\over 4}+A$, where $A$ is does not depend on $x$, and  $\m$ is some scale introduced to make the argument of the logarithm dimensionless. The terms on the right-hand side of \eqref{product2} then again nicely combine to yield the left-hand side of \eqref{freq} for $r=1$. As a result, we get
\be\label{DelG2}
D_x \left(\Bigl( \ln {\m^2\ell^2\over 4} +A\Bigr)\, a_1\right)=\Bigl( \ln {\m^2\ell^2\over 4} +A\Bigr)\, D_x a_1+{4 D_x a_0\over \ell^2} \, \cdotp
\ee
Combining the last two equations, we get
\be\label{DelG3}
D_x\left[ {4a_0\over \ell^2}-\Bigl(\ln {\m^2\ell^2\over 4}+A\Bigr) a_1\right] = -\Bigl(\ln {\m^2\ell^2\over 4}+A\Bigr) \, D_x a_1 \, .
\ee
The term on the right-hand side is now $O(\ln \ell^2)$. 

To do better, we compute similarly, with an $x$-independent $B$,
\begin{align}\label{DelG4}
D_x \Big(  {\ell^2\over 4} 
\Bigl(\ln {\m^2\ell^2\over 4}+B\Bigr)\, a_2\Big) 
&= {\ell^2\over 4}  \Bigl(\ln {\m^2\ell^2\over 4}+B\Bigr)  D_x a_2 +\Bigl(\ln {\m^2\ell^2\over 4}+1+B\Bigr) D_x a_1 -a_2\nonumber\\
&\hskip-1.8cm= {\ell^2\over 4}  \Bigl(\ln {\m^2\ell^2\over 4}+B\Bigr)\  D_x a_2 +\Big(\ln {\m^2\ell^2\over 4}+A\Big)\ D_x a_1\nonumber\\
&\hskip-1.5cm+(B+1-A)\left({\Delta_x\ell^2+8\over 4}a_2-{1\over 2}g^{ij}\del_i\ell^2\del_j a_2\right)
-(2B-2A+3) a_2\, ,
\end{align}
where we used \eqref{freq} twice. Now, $\Delta_x\ell^2+8$ is $O(\ell^2)$ and $\del_i\ell^2$ is $O(\ell)$. Thus, if we choose $B=A-{3\over 2}$ the whole last line is $O(\ell)$. Hence, adding \eqref{DelG4} to \eqref{DelG3} we find
\be\label{DelG5}
D_x\left[ {4a_0\over \ell^2}-\big(\ln {\m^2\ell^2\over 4}+A\big)\, a_1+{\ell^2\over 4} \big(\ln {\m^2\ell^2\over 4}+A-{3\over 2}\big) a_2
\right] = O(\ell)\, .
\ee
One might now be tempted to conclude that the expression in the square bracket, multiplied by $1/ (4\pi)^2$, is the short-distance expansion of the Green's function. 
In flat space, where $a_0=1$, $a_1=-m^2$ and $a_2=m^4/ 2$, this is indeed the case, as one sees by comparing with \eqref{flatGreenexp}, provided
\be\label{Aflat}
A\Big\vert_{\text{flat space, $d=4$}}=2\g -1 \, .
\ee
In general however, this cannot be the full answer since we know from \eqref{Gspliteven} that this expansion must involve the non-trivial $G_\zeta(y)$.\footnote{
In flat four-dimensional space, choosing $\m=m$ we have $G_\zeta=-m^2/ (4\pi)^2$, and it only shows up via the $-1$ in \eqref{Aflat}.}
Clearly, one cannot identify $G_\zeta$ with $A$ since, by symmetry, if $A$ does not depend on $x$ it cannot depend on $y$ either and must be a true constant. 

The point is that the short-distance expansion
is a local expansion which ignores any global ``boundary" conditions. This is very easily seen on the following example.
Take flat four-dimensional space, but restricted to $x^1\ge 0$ only, and impose the boundary condition that $G(x,y)$ must vanish whenever $x$ or $y$ is on the boundary. The massless Green's function then is \be G(x,y)={1\over 4 \pi^2}\left( {1\over (x-y)^2}-{1\over (x-y_C)^2}\right)\, ,\ee where $y_C$ is the image point of $y$, i.e. $y^1_C=-y^1$ and $y^i_C=y^i$ for $i\ne 1$. Our expansion \eqref{DelG5} only captures the $ 1/ (x-y)^2$ piece. The second term has a short-distance expansion \be{1\over (4 \pi)^2}\left( -{1\over x^1 y^1}+{(x-y)^2\over 4 (x^1 y^1)^2} + \ldots\right)\, ,\ee which is of the form \be C(x,y) +{\Delta_x C(x,y)\over 8}(x-y)^2 +\cdots\ee In this example, one identifies \be G_\zeta(x)=-{1\over (4\pi)^2} {1\over (x^1)^2}=C(x,x)\, .\ee 

Coming back to the general case, we observe that on the right-hand side of  \eqref{DelG5} we neglected terms that are $O(\ell)$, and we can, a priori, add anything that solves the homogenous equation up to terms $O(\ell)$~: $D_x(\ldots)=O(\ell)$. Let $C(x,y)$ and $\wt C(x,y)$ be  symmetric functions of $x$ and $y$ that are regular as $x\to y$. Then, $D_x C$ and $D_x \wt C$ are also regular and, by \eqref{product2}
\be\label{DxC}
D_x \Big(C(x,y) +{\ell^2\over 8} \wt C(x,y) \Big)
=D_x C +{\Delta_x \ell^2\over 8} \wt C +O(\ell)
=D_x C - \wt C +O(\ell)\, .
\ee 
Choosing $\wt C= D_x C$, or rather ${1\over 2}(D_x C+D_y C)$, the whole expression is $O(\ell)$, i.e. it solves the homogeneous equation, up to terms $O(\ell)$.

We conclude that the general form of the short-distance expansion of the Green's function must be
\begin{multline}\label{Greensmallexp}
(4\pi)^2 G(x,y)=  {4a_0(x,y)\over \ell^2}-\Bigl(\ln {\m^2\ell^2\over 4}+A\Bigr)a_1(x,y) +C(x,y)\\
+{\ell^2\over 4} \Bigl(\ln {\m^2\ell^2\over 4}+A-{3\over 2}\Bigr) a_2(x,y)+{\ell^2\over 16} \big(D_x C(x,y)+D_y C(x,y)\big)
 +{\cal O}(\ell^3) \, .
\end{multline}
Of course, $C(x,x)$ is related to the renormalized Green's function at coinciding points, $G_\zeta(x)$. Indeed, by \eqref{Gspliteven} we have
\be\label{GzetaC}
(4\pi)^2 G_\zeta(x)=(2\g-A) a_1(x) +C(x,x) \, .
\ee
In flat space, \eqref{Aflat} and $G_\zeta=-m^2/ (4\pi)^2$ consistently yield $C=0$. In general,
to actually determine $C(x,y)$ or $G_\zeta(x)$ requires global knowledge that goes beyond the local structure captured by the heat kernel coefficients $a_r$.

\smallskip

\noindent\emph{The case $d=2$}

\noindent
In two dimensions we find similarly
\begin{multline}\label{Greensmallexptwo}
4\pi G(x,y)=  -\Bigl(\ln {\m^2\ell^2\over 4}+A\Bigr) a_0(x,y) +C(x,y)+{\ell^2\over 4} \Bigl(\ln {\m^2\ell^2\over 4}+A-2\Bigr) a_1(x,y)\\
+{\ell^2\over 8} \big(D_x C(x,y)+D_y C(x,y)\big)+{\cal O}(\ell^3) \, ,
\end{multline}
with the function $C$ related to $G_\zeta$ by
\be\label{GzetaC2}
4\pi G_\zeta(x)=(2\g-A) +C(x,x) \, .
\ee
Note again, that in flat space, $G_\zeta=0$ implies consistently $C=0$.

\subsection{Some computational details relevant in flat space\label{flatapp}}
\label{a4app}


In flat four-dimensional space, the function $Z_{\diagramb}(s,\underline\a)$ was given by \be {V\over (4\pi)^4} {m^{2-2s}\over s(s-1)} I_4(s,\underline\a)\, ,\ee see \eqref{Zdiagrambd4}, where $I_4$ is defined by
\be\label{I4int}
I_4(s,\underline \a)=
\int_{\underline \b\ge \underline\alpha}
\frac{(\b_1+\b_2+\b_3)^{1-s}}{(\b_{1}\b_{2} + 
\b_{2}\b_{3}+\b_{3}\b_{1})^2}\,\cdotp
\ee
It is clear that  this is a homogeneous function of degree $-s$, 
\be\label{Ishom}
I_4(s,t \underline \a)=t^{-s}\, I_4(s,\underline \a) \, ,
\ee 
in agreement with \eqref{Zscale}. 
Since, in the integrand of \eqref{I4int}, $\sum_i\b_i\ge \sum_i\a_i=\a$, for  $\Re s>1$ one has $|(\b_1+\b_2+\b_3)^{1-s}|\le \a^{1-s}$, and thus $|I_4(s,\underline \a)|< \a^{1-s}I_4(1,\underline \a)$.
Hence, $I_4(s,\underline \a)$ is analytic for $\Re s \ge 1$. We will now show that it has a double and simple pole at $s=0$, as well as various poles at $s\le -1$. As already repeatedly emphasized, the poles at $s\le -1$ give contributions to the amplitude that are ${\cal O}(1/ \L^2)$, and we will not need to consider them further. On the other hand, due to the explicit $1/ s$ factor in \eqref{Zdiagrambd4}, we will also need the finite part of $I_4(s,\underline \a)$, once the pole terms at $s=0$ is subtracted.

To compute $I_4(s,\underline\a)$, we change variables to $u=(\b_1+\b_2+\b_3)^{-1}$, $x=u\b_1$ and $y=u\b_2$, so that
\be\label{Iint4}
I_4(s,\underline \a)= \int_0^{\a^{-1}}\! \d u \, u^{s-1} J_4(u \underline \a)
\, ,\ee
with
\begin{multline}\label{J4int}
J_4(\underline a)= \int_{\substack{x\ge a_1,\, y\ge a_2\\ x+y\le 1-a_3}} {\d x\, \d y\over [x y +(x+y)(1-x-y)]^2} \\
=\int_{\underline x\ge \underline a}\d x_1\, \d x_2\, \d x_3\, {\delta(1-x_1-x_2-x_3)\over [x_1 x_2 + x_1 x_3 + x_2 x_3]^2} \,\cdotp
\end{multline}
This integral can be computed explicitly to yield
\be\label{Jint4}
J_4(\underline a)= \sum_{i\ne j=1}^3
{1-3 a_i\over \sqrt{(1-a_i)(1+3a_i)} }\, \ln{  \sqrt{(1-a_i)(1+3a_i)} +1-a_i-2a_j       \over  \sqrt{(1-a_i)(1+3a_i)} -1+a_i+2a_j } \,\cdotp
\ee
The singularities of (\ref{Iint4}) must come form the region $u\to 0$ and, hence, are related to the behavior of $J_4(u \underline\a)$ for small $u$. Expanding (\ref{Jint4}), we get
\be\label{J4intexp}
J_4(u\underline\a)= -3 \ln u -\sum_{i>j}\ln(\a_i+\a_j) + O(u,u\ln u) \, .
\ee
Of course, this small $\underline a$ behavior can also easily be obtained directly from \eqref{J4int} without computing the integral exactly. Upon inserting \eqref{J4intexp} into \eqref{Iint4} we get
\be\label{Iint4b}
I_4(s,\underline\a)={3\over s^2}-{1\over s}\sum_{i>j}\ln(\a_i+\a_j) \ +\ \text{poles at $s\le -1$}\ \ + \ \text{regular}\, .
\ee
We define the regular part at $s=0$ by
\be\label{Iint4reg}
I_4^{\rm reg}(0,\underline\a)=\lim_{s\to 0}\left[ I_4(s,\underline\a)-  {3\over s^2} + {1\over s}\sum_{i>j}\ln(\a_i+\a_j) \right]\, .
\ee
It is simply given by 
\be\label{Iint5}
I_4^{\rm reg}(0,\underline\a)=\int_0^{\a^{-1}} {\d u\over u} \left(J_4(u\underline\a) +3 \ln u +\sum_{i>j}\ln(\a_i+\a_j) \right) \, .
\ee
One can work out its explicit expression from the above formula, but we will not need it here. 

\subsection{Counterterms and the two-point function at one loop}
\label{apptwopt}

Let us determine here the counterterm coefficients $c_{m}$, $c_{\xi}$ and $c_{\phi}$ from the one-loop two-point function $G^{(2)}(x,y)$. This will also illustrate on a simple example how our formalism can be used to compute $n$-point functions.

Up to order $\hbar$, the two-point function $G^{(2)}(x,y)$ receives contributions from the tree-level  $G(x,y)$, from  the counterterms, as well as from  two one-loop diagrams. There is a one-loop diagram with two internal lines between two cubic vertices  and another one with a single internal line having both ends connected to a quartic vertex,
\begin{multline}\label{completetwopoint}
G^{(2)}(u,v)= G(u,v)
+\int \d^4 x \sqrt{g(x)} G(u,x) \Big(-c_\phi \Delta_x-c_m m^2 -c_\xi \xi R(x)\Big) G(x,v)\\ +{\k_3^2\over 2}\int \d^4 x \sqrt{g(x)}\, \d^4 y \sqrt{g(y)}\, G(u,x) \,\big(G(x,y)\big)^2\, G(y,v)\\
-{\k_4\over 2} \int \d^4 x \sqrt{g(x)}\,  G(u,x)\, G(x,x)\, G(x,v) \, .
\end{multline}
In principle, we should replace all propagators $G$, internal and external, by the regularized propagators $G_{f,\Lambda}$. Replacing the external $G$'s by $G_{f,\lambda}$'s only adds $O(1/\Lambda^2)$ terms to the external $G$'s. Multiplied by a $\Lambda^2$ divergence of a loop, this gives a finite contribution. However, at present, we only want to determine the diverging parts of the counterterm constants and we do not care about such finite contributions. It is then enough to regularize only the  Green's functions appearing in the loops. Let us now compute the different parts of \eqref{completetwopoint}. First, we have
\be\label{DeltaG}
\Delta_x G(x,v) = D_x G(x,v) -\bigl(m^2+\xi R(x)\bigr) G(x,v) 
={\delta(x-v)\over \sqrt{g(x)}} -\bigl(m^2+\xi R(x)\bigr) G(x,v) \, .
\ee
The last term in \eqref{completetwopoint} yields
\begin{multline}\label{twoppointk4}
-{\k_4\over 2} \int \d^4 x \sqrt{g(x)}\,  G(u,x)\, G_{f,\Lambda}(x,x)\, G(x,v) =\\-{\k_4\over 2} \int \d^4 x \sqrt{g(x)}\, G(u,x) G(x,v)  \int\!\d\a \, \vf(\a)\left[ {\L^2\over \a} + a_1(x)  \left(\ln{\Lambda^2\over \m^2 \a}-\g\right)\right]+O(1)\, .
\end{multline}
To evaluate the divergence of the term $\sim \k_3^2$ in \eqref{completetwopoint}, we must compute
\begin{multline}\label{phicubeoneloop}
\int\! \d^4 y \sqrt{g(y)}\,\big(G_{f,\Lambda}(x,y)\big)^2\, G(y,v)
\\= \int\d \a_1 \d\a_2\, \vf(\a_1)\vf(\a_2)  \int\! \d^4 y \sqrt{g(y)}\, \wh K\big({\a_1\over\L^2},x,y\big) \wh K\big({\a_2\over\L^2},x,y\big) \, G(y,v) \, ,
\end{multline}
with $\wh K$ given in \eqref{KhattoK}. A priori, we are not allowed to use the small $t$ expansion of the heat kernel since $t$ is integrated up to infinity. However, it is easy to see that the divergent part of the present diagram only receives contributions from the regions where both $t_i$ are small and we can thus use the small $t$ expansion to compute this divergent part. Moreover, for small $t_i$, only the regions with $y$ close to $x$ contribute to the integral over $y$. This allows us, on the one hand, to expand \be G(y,v)=G(x,v)+(x-y)^i \del_{x_i} G(x,v)+{1\over 2} (x-y)^i(x-y)^j \del_{x_i}\del_{x_j} G(x,v)+\cdots\, ,\ee and, on the other hand, to use the expression for $K$ in Riemann normal coordinates resulting in a trivial Gaussian integral,
\begin{multline}\label{fishdiag}
\int\! \d^4 y \sqrt{g(y)}\,  K(t_1,x,y)  K(t_2,x,y)\, G(y,v)\\
={1\over (4\pi)^2} \left[{G(x,v)\over (t_1+t_2)^2 }+{a_1(x) G(x,v)\over t_1+t_2} - {\Delta_x G(x,v)\ t_1 t_2\over (t_1+t_2)^3}+\cdots\right] \, .
 \end{multline}
The $t_i$ are to be integrated from $\a_i/\L^2$ to some fixed $\L$-independent value (which we take to be $\m^{-2}$) and we see that only the first term contributes to the divergent part
in \eqref{phicubeoneloop},
\begin{multline}\label{phicubeoneloop2}
\int\!\d^4 y \sqrt{g(y)}\,  (G_{f,\L}(x,y))^2 G(y,v)
=\\{G(x,v)\over (4\pi)^2} \int\!\d\a_1 \d\a_2\, \vf(\a_1)\vf(\a_2) \ln{\L^2\over \m^2(\a_1+\a_2)} + O(1)  \, .
\end{multline}
Putting the different pieces together, we find for the diverging part of the  two-point function
\begin{multline}\label{completetwopoint2}
G^{(2)}(u,v)=-c_\phi G(u,v)+\int\! \d^4 x \sqrt{g(x)}\, G(u,x) \bigg\{\big(c_\phi-c_m\big) m^2+ \big( c_\phi-c_\xi \big) \xi R(x)\\
+{\k_3^2\over 2(4\pi)^2}\hskip-1.mm \int\d\a_1 \d\a_2 \vf(\a_1)\vf(\a_2) \ln{\L^2\over \m^2(\a_1+\a_2)}\\
\hskip0.5cm-{\k_4\over 2}\,  \int\d\a \vf(\a)\left[ {\L^2\over \a} + a_1(x)  \left(\ln{\Lambda^2\over \m^2 \a}-\g\right)\right]
\bigg\}\ G(x,v) +O(1)\ .
\end{multline}
Recalling again $a_1=(1/6-\xi)R-m^2$, we see that cancelling the diverging part requires, up to finite shifts,
\begin{align}\label{cphicmcxi}
c_\phi=&0\, , \nonumber\\
m^2 c_m=&-\frac{\k_4}{2(4\pi)^2} \int\d \a\, \vf(\a)\left[   \frac{\Lambda^2}{\a} -m^2 \left(\ln{\Lambda^2\over \m^2 \a}-\g\right)\right] \nonumber\\
&\hskip0.5cm+ {\k_3^2\over 2 (4\pi)^2}  \int\d\a_1\, \d\a_2\, \vf(\a_1)\vf(\a_2) \ln{\L^2\over \m^2(\a_1+\a_2)} \,\cvp 
\nonumber\\
c_\xi\, \xi&=-\frac{\k_4}{2(4\pi)^2}\left({1\over 6}-\xi\right) \int\d \a\, \vf(\a) \left(\ln{\Lambda^2\over \m^2 \a}-\g\right) \, , 
\end{align}
up to terms $O(1)$. Note that we find, in addition to the results obtained in the main text from the computation of the gravitational effective action, that the wave function renormalization vanishes at one loop. If we had evaluated the analogue of \eqref{fishdiag} in $d=6$, the terms $\sim \Delta_x G(x,v)=\delta(x-v)/\sqrt{g(x)}+\cdots$  would also have led to a divergent contribution and one would then conclude that $c_\phi\ne 0$. 
\end{appendix}


%

%
\end{document}